\documentclass{jfm}

\usepackage{amsmath}
\usepackage{graphicx}
\usepackage{natbib}

\usepackage[nameinlink]{cleveref}
\crefname{paragraph}{\S}{\S\S}
\crefformat{section}{\S#2#1#3}
\crefformat{subsection}{\S#2#1#3}
\crefformat{subsubsection}{\S#2#1#3}
\usepackage{subcaption}
\usepackage{float}
\usepackage[outdir=./]{epstopdf}
\usepackage{placeins}
\usepackage{subfloat}
\usepackage{physics}
\usepackage{empheq}
\usepackage[T1]{fontenc}
\usepackage{amsmath,mathtools}
\usepackage[utf8]{inputenc}
\usepackage{amssymb}
\usepackage{bm}
\usepackage{xspace}
\usepackage{subcaption}
\usepackage{float}
\usepackage{multirow}
\usepackage[table]{xcolor}
\usepackage{caption}
\usepackage{tikz}
\allowdisplaybreaks

\begin{document}

\newtheorem{lemma}{Lemma}
\newtheorem{corollary}{Corollary}
\title{Generalised quasilinear approximations of turbulent channel flow: Part 1. Streamwise nonlinear energy transfer}
\author
{
Carlos G. Hern\'andez\aff{1}\corresp{\email{cg1116@imperial.ac.uk}}, 
Qiang Yang\aff{2} \and Yongyun Hwang\aff{1}
}
\affiliation
{
\aff{1}
Department of Aeronautics, Imperial College London, South Kensington, London SW7 2AZ, UK
\aff{2}
State Key Laboratory of Aerodynamics, China Aerodynamics Research and Development Centre, Mianyang 621000, PR China
}
\maketitle
\begin{abstract}
A generalised quasilinear (GQL) approximation (Marston \emph{et al.}, \emph{Phys. Rev. Lett.}, vol. 116, 104502, 2016) is applied to turbulent channel flow at $Re_\tau \simeq 1700$ ($Re_\tau$ is the friction Reynolds number), with emphasis on the energy transfer in the streamwise wavenumber space. The flow is decomposed into low and high streamwise wavenumber groups, the former of which is solved by considering the full nonlinear equations whereas the latter is obtained from the linearised equations around the former. The performance of the GQL approximation is subsequently compared with that of a QL model (Thomas \emph{et al.}, \emph{Phys. Fluids.}, vol. 26, no. 10, 105112, 2014), in which the low-wavenumber group only contains zero streamwise wavenumber. It is found that the QL model exhibits a considerably reduced multi-scale behaviour at the given moderately high Reynolds number. This is improved significantly by the GQL approximation which incorporates only a few more streamwise Fourier modes into the low-wavenumber group, and it reasonably well recovers the distance-from-the-wall scaling in the turbulence statistics and spectra. Finally, it is proposed that the energy transfer from the low to the high-wavenumber group in the GQL approximation, referred to as the `scattering' mechanism, depends on the neutrally stable leading Lyapunov spectrum of the linearised equations for the high wavenumber group. In particular, it is shown that if the threshold wavenumber distinguishing the two groups is sufficiently high, the scattering mechanism can completely be absent due to the linear nature of the equations for the high-wavenumber group. 
\end{abstract}

\section{Introduction}\label{sec:sec1}
It is well known that in canonical wall-bounded turbulent shear flows such as Couette, pipe, channel and boundary-layer flows, linear instability does not arise from the typical mean velocity. This is also true for their laminar base flow at transitional Reynolds numbers. However, even if all infinitesimal disturbances have to eventually decay due to the linear stability of the laminar flow, perturbations can still grow transiently in time and can be amplified by external forcing due to the non-normal nature of the linearized Navier-Stokes equations. This growth has been quantified by the computation of optimal transient growth, optimal harmonic response (resolvent analysis) and stochastic response (controllability gramian) \cite[e.g.][]{trefethen93,butler93,farrell93b,reddy93,schmid94,bamieh01,jovanovic05}, and it forms an important element in the subcritical transition to turbulence. Over the last two decades, a number of studies have also shown that energy-containing motions (coherent structures) in turbulent flows could well be  described with the suitable refinement of the analysis tools for the linearized Navier-Stokes equations, as they share the same amplification mechanism as that of the flow structures observed in transition \cite[e.g.][]{farrell93,kim00,alamo06,cossu09,hwang10,mckeon10,zare17}.

Despite this progress, the linearly stable nature of the typical mean velocity in wall-bounded turbulent flows implies that solely the linearised Navier-Stokes equations are not able to describe sustaining velocity fluctuations. Therefore, some minimal amount of nonlinearity must be retained in the mathematical description for the self-sustaining turbulent fluid motion. Towards this quest, various forms of `quasilinear (QL) approximations' have recently been proposed. Common to all, this approach introduces a decomposition of the given flow into two groups: one in which all nonlinear terms are kept, and the other in which all self-interactions are ignored or suitably modelled. The resulting equations for the first group are unchanged from the original, while those for the second become equivalent to a linearisation around the first group, often with an additional \textit{adhoc} model (e.g. stochastic forcing).
The earliest work may be found from \cite{malkus54}, \cite{malkus56} and \cite{herring63, herring64, herring66} where the `marginal stability' criterion was applied to the second group for the closure of the formulation. Modern variants of the QL framework have been proposed for many different flows with various types of suitable models for the self-interaction term of the second group (e.g. stochastic forcing, eddy viscosity, etc): for example, stochastic structural stability theory (S3T) \citep{farrell07,farrell12}, direct statistical simulation (DSS) \citep{marston08,tobias13prl}, self-consistent approximation for linearly unstable flows \citep{lugo14,lugo16}, minimal quasilinear approximation agumented with an eddy-viscosity model \citep{hwang19,skouloudis}, restricted nonlinear model (RNL) \citep{thomas14,thomas15,farrell17,pausch18,hernandez} and generalised quasilinear approximations (GQL) \citep{marston16,tobias17}. 

The starting point of the present study is the RNL approach, which we shall refer to as `QL model' hereafter. This model decomposes the given velocity into a streamwise mean and the rest. As in a typical QL approximation, the former group considers the full nonlinear equations, whereas the latter group is approximated with the linearised equations about the former. This model has been shown to be successful for a reduced description of turbulence in wall-bounded shear flows especially at low Reynolds numbers \citep{thomas14,thomas15,pausch18}, as it was designed to capture the key dynamics of coherent structures described by the so-called `self-sustaining process' of the flows \citep{hamilton95,waleffe97}: i.e. the two-way interaction between `streamwise elongated' structure of streamwise velocity (streaks) and its `streamwise wavy' instability involving in the generation of cross-streamwise velocities (waves and rolls). In the QL model, the elongated streaks are captured by the streamwise mean, and the streamwise undulating instability is captured by the linearised equations. Despite the success, it remains to be understood whether this approach would be suitable to the flows at high Reynolds numbers without any further refinement, as the fluid motions in such a regime would involve vigorous nonlinear and non-local interactions across a very wide range of length and time scales. 

Recently, \cite{hernandez} examined the wavenumber-space energy transfer of the QL model in uniform shear turbulence to understand its effect on energy cascade. It was found that in the QL model the energy cascade in the streamwise direction was significantly inhibited, resulting in highly elevated spectral energy intensity residing only at the streamwise integral length scale. Only a small number of streamwise Fourier modes therefore remain active and these are driven by the instability of the streamwise-averaged flow, in agreement with the previous findings \cite[]{thomas14,thomas15,farrell16,tobias17}. Importantly, in this study, the QL approximation was found to significantly damage the slow pressure. This consequently inhibits the related pressure strain which transfers the turbulent kinetic energy produced at the streamwie component to the cross-streamwise components, resulting in the anisotropic fluid motions of the QL model across the entire length scales (i.e. from the integral to the Kolmogorov scale). 

In uniform shear turbulence, the existence of a self-sustaining process at single integral length scale  \citep{sekimoto16,yang18} renders it an appealing platform to study the QL model especially in relation to energy cascade. However, in wall-bounded turbulent shear flow, such a self-sustaining process emerges at multiple scales, as the integral length scale of the flow varies continuously with the distance from the wall \citep{flores10,hwang10prl,hwang11,hwang15,hwangbengana16}. Furthermore, there is a growing body of recent evidence suggesting that the interactions between the self-sustaining processes at different length scales are crucial to understand the statistical and dynamical behaviours of the flow \citep{cho18,lee19,doohan19}.
Given the increasing complexities at high Reynolds numbers, it is important to understand what kind of the physical processes are precisely captured by the QL model and whether there is a way to improve the approximation further. Indeed, the recent application of the QL model to a moderate Reynolds number ($Re_\tau \simeq 940$ where $Re_\tau$ is the friction Reynolds number) showed non-negligible differences to direct numerical simulation (DNS) in the mean velocity and turbulence intensities \citep{farrell16}. Importantly, the streamwise wavenumber spectra of the velocity fluctuations of the QL model did not show any robust self-similar scaling with respect to the distance from the wall, expected from the logarithmic mean velocity \citep{hwang_lee_2020}. 

To address this issue, the scope of the present study is to apply the generalised quasilinear (GQL) approximation \citep{marston16,tobias17} to turbulent channel flow at a sufficiently high Reynolds number ($Re_\tau\simeq 1700$). The GQL approximation is much more flexible than the QL model, and it decomposes the flow into two groups, the former of which contains a set of low-wavenumber Fourier modes and the latter are composed of the rest high-wavenumber modes. The former low-wavenumber group is then solved by considering the full nonlinear equations, while the latter high-wavenumber group is obtained from the linearised equations around the former. In one limit where the low-wavenumber group is restricted to be composed of the Fourier modes for streamwise uniform flow, the approximation becomes identical to the QL model. In the other limit where the low-wavenumber group can be set to cover all wavenumbers, it simply becomes a DNS. It is therefore expected that the application of the GQL approximation would improve the statistical and dynamical features of turbulence from the QL model, as has indeed been demonstrated for zonal jets (\citealp{marston16}) and rotating Couette flow (\citealp{tobias17}). 

The GQL approximation was originally proposed to improve the QL model \cite[i.e. DSS;][]{marston08,tobias13prl} in the context of astrophysical and geophysical fluid dynamics \citep{marston16}. As for wall-bounded turbulence, perhaps its true value would more lie in the suppression of particular nonlinear mechanisms in a `controlled' manner, especially given that various forms of large-eddy simulations (LES) have been used successfully over many years in this type of flow. Indeed, the GQL approximation inhibits some specific energy transfer mechanisms to the high-wavenumber group which take place through the nonlinear interactions within the low wavenumber group and within the high wavenumber group (\citealp{marston16}). This feature offers a new opportunity for the study of wall-bounded turbulence, as the GQL approximation can be used as a robust interventional tool to systematically examine nonlinear inter- and intra-scale interactions, probably the most poorly understood processes in wall-bounded turbulence at high Reynolds numbers. 

For this purpose, the following two types of GQL approximations will be considered in the present and the companion study \citep{hernandez_span}, respectively: 1) the low-wavenumber-mode group is given by the plane Fourier modes for $|k_x|\leq k_{x,c}$ ($k_x$ is the streamwise Fourier wavenumber and $k_{x,c}$ the corresponding threshold wavenumber for the decomposition) (present study); 2) the low-wavenumber-mode group is composed of the plane Fourier modes for $|k_z|\leq k_{z,c}$ ($k_z$ is the streamwise Fourier wavenumber and $k_{z,c}$ the corresponding threshold wavenumber for the decomposition) (\citealp{hernandez_span}). The former case is to primarily examine the interactions between the energy-containing streamwise waves, which have been understood to originate from the streak instability and/or transient growth mechanisms in the self-sustaining processes at different length scales \citep{park11,alizard15,cassinelli17,degiovanetti17,lozano2021}, as well as the energy cascade in the streamwise wavenumber space. This is also a direct extension of the previous studies on the QL model \citep{thomas14,thomas15,farrell17,pausch18,hernandez} using the GQL approximation, as the QL model offers the dynamics of such energy-containing streamwise waves in a minimal manner. The latter case is more directly related to the scale interaction and energy transfer between the self-sustaining processes at different integral length scales, the issues which have been studied recently by \cite{cho18} and \cite{doohan21}. The spanwise length scale has been understood to represent the size of coherent structures sustained by the self-sustaining process \citep{hwang15}. It is therefore hoped that the GQL approximation offers a new perspective for these issues, while confirming the previous findings obtained by `non-intrusive' analysis \cite[e.g.][]{cho18,doohan21}. 

This paper, which forms the first part of the two companion papers, is organized as follows. The GQL model is introduced in \cref{sec:sec2}, where its spectral energy budget is formulated. In \cref{sec:sec3}, the statistics and spectra of the GQL model for channel flow are compared to those of a wall-resolved LES. The energy-budget and pressure-strain spectra are also presented here with a detailed analysis to explain the statistical features of the GQL model. Further discussion is given in \cref{sec:sec4} especially on the multi-scale behaviours of the QL and GQL models and on the mechanism of energy transfer to the high-wavenumber group in the QL/GQL models. Finally, the paper concludes in \cref{sec:sec5} with some remarks.

\section{Problem formulation} \label{sec:sec2}
\subsection{Generalised quasilinear approximation}
We consider a pressure-driven plane channel flow, where the density and kinematic viscosity of the fluid are given by $\rho$ and $\nu$, respectively. The time is denoted by $t$, and the space is denoted by $\mathbf{x}=(x,y,z)$ with $x$, $y$ and $z$ being the streamwise, wall-normal and spanwise directions, respectively. The lower and upper walls of the channel are located at $y=\pm h$. For the GQL approximation, 
the velocity $\bf u$ is decomposed into two groups using a discrete Fourier transform in the streamwise and spanwise directions:
\begin{subequations}\label{eq:2.0}
\begin{equation}\label{eq:2.3z}
\bold{u}=\bold{U}_l+\bold{u}_h,
\end{equation}
where 
\begin{equation}\label{eq:2.0b}
    \bold{U}_l=\sum_{n=-M_{z,F}}^{M_{z,F}}\sum_{m=-M_{x,F}}^{M_{x,F}} \hat{\bold{u}}_{m,n} e^{i(m k_{x,0}x+n k_{z,0}z)}, 
\end{equation}
\end{subequations}
and $\mathbf{u}_h$ is given from (\ref{eq:2.3z}). Here, $\hat{\bold{u}}_{m,n}$ is the discrete Fourier mode of the velocity,  $k_{x,0}$ and $k_{z,0}$ are the fundamental streamwise and spanwise wavenumbers for the given horizontal computational domain (see \S\ref{sec:sec23} for further computational details), and $M_{x,F}$ and $M_{z,F}$ define the threshold streamwise and spanwise wavenumbers for the decomposition such that $k_{x,c}=M_{x,F}k_{x,0}$ and $k_{z,c}=M_{z,F}k_{z,0}$. 

With the decomposition in (\ref{eq:2.0}), two related projection operators are defined such that
\begin{equation}\label{eq:2.4}
    \mathcal{P}_l[\bold{u}]\equiv\bold{U}_l,~~\mathcal{P}_h[\bold{u}]\equiv\bold{u}-\bold{U}_l=\bold{u}_h,
\end{equation}
and they satisfy the following properties: 
\begin{subequations}\label{eq:2.5}
\begin{equation}\label{eq:2.5a}
\mathcal{P}_l[\cdot]+\mathcal{P}_h[\cdot]=\mathcal{I}[\cdot],
\end{equation}
\begin{equation}
\mathcal{P}_l[\mathcal{P}_l[\cdot]]=\mathcal{P}_l[\cdot],~ \mathcal{P}_h[\mathcal{P}_h[\cdot]]=\mathcal{P}_h[\cdot],
\end{equation}
\begin{equation}
\mathcal{P}_l[\mathcal{P}_h[\cdot]]=\mathcal{P}_h[\mathcal{P}_l[\cdot]]=\mathbf{0},
\end{equation}
\end{subequations}
where $\mathcal{I}[\cdot]$ is the identity operator. We note that the projection operators are linear like the Fourier transform, implying that their application to linear terms does not yield any change in their original form. Using the definition and the properties listed in (\ref{eq:2.4}) and (\ref{eq:2.5}), the Navier-Stokes equations are first projected onto the $\mathcal{P}_l$ (or low wavenumber) and $\mathcal{P}_h$ (or high wavenumber) subspaces. The subsequent linearisation of the equations for $\mathbf{u}_h$ about $\mathbf{U}_l$ leads to the GQL system of interest, i.e.
\begin{subequations}\label{eq:full}
\begin{align}\label{eq:aa}
&\pdv{\bold{U}_l}{t} + \mathcal{P}_{l}[(\bold{U}_{l} \cdot \nabla) \bold{U}_{l}]  =-\frac{1}{\rho}\bold{\nabla} P_l + \nu \bold{\nabla}^2  \bold{U}_l +  \mathcal{P}_{l}[ \bold{\nabla} \cdot \bm{\tau}_{SGS} ] \nonumber\\ &-\mathcal{P}_{l}\left[\left(\bold{U}_{l}\cdot\nabla\right)\bold{u}_{h}\right]-\mathcal{P}_{l}\left[\left(\bold{u}_{h}\cdot\nabla\right)\bold{U}_{l}\right]-\mathcal{P}_{l}\left[\left(\textbf{u}_{h}\cdot\nabla\right)\textbf{u}_{h}\right],
\end{align}
and 
\begin{equation}\label{eq:bb}
\pdv{\bold{u}_h}{t} + \mathcal{P}_h[(\bold{u}_h \cdot \bold{\nabla}) \bold{U}_l] + \mathcal{P}_h[(\bold{U}_l \cdot \bold{\nabla}) \bold{u}_h] =-\frac{1}{\rho}\bold{\nabla} p_h +  \nu \bold{\nabla}^2  \bold{u}_h +  \mathcal{P}_{h}[ \bold{\nabla} \cdot \bm{\tau}_{SGS} ] , 
\end{equation}
\end{subequations}
where $P_l$ and $p_h$ are defined to enforce $\bold{\nabla} \cdot \bold{U}_l=0$ and $\bold{\nabla} \cdot \bold{u}_h=0$, respectively, with $p=P_l+p_h$. The numerical simulations in this study are performed using LES. Therefore, the subgrid-scale stress (SGS) tensor, given by $\bm{\mathbf{\tau}}_{SGS}= - ( \bold{u} \bold{u} - \overline{\bold{u} \bold{u}})$, appears in (\ref{eq:full}), where the overbar $(\overline{\cdot})$ denotes the application of the grid filter given by the numerical discretisation of the equations (see \S\ref{sec:sec23}). The SGS tensor for the present LES here employs a mixing-length type model
\begin{equation}
\bm{\mathbf{\tau}}_{SGS}= \nu_t (\nabla \bold{u}+ \nabla \bold{u}^T),
\end{equation}
where the eddy viscosity $\nu_t$ was employed from the model proposed by \cite{vreman}. This model was developed to closely match the theoretically predicted algebraic properties of subgrid-scale dissipation from a database of flows. In particular, the dissipation by the eddy viscosity vanishes in the near-wall region, not requiring any special treatment such as utilisation of a wall-damping function. The model has shown a performance comparable to that of the standard dynamic Smagorinsky model \cite[e.g.][]{germano91}, and the value for the model constant used in this work is $C=0.03$ \cite[see also][for the details]{vreman}. 

The GQL approximation here can also be simplified to the QL approximation if $M_{x,F}=0$ and $M_{z,F}=N_{z,F}$ ($N_{z,F}$ is the total number of the spanwise Fourier modes used for simulation). In this case, the first two terms in the second line of (\ref{eq:aa}) disappear, resulting in the equations identical to those in the previous studies \citep{thomas14,thomas15,farrell17,pausch18,hernandez}. On the contrary, if $M_{x,F}=N_{x,F}$ and $M_{z,F}=N_{z,F}$ are set ($N_{x,F}$ is the total number of the streamwise Fourier modes used for simulation), $\bold{u}_h=\bold{0}$, $p_h=0$ and $\mathcal{P}_{h}[ \cdot ]=\bold{0}$. Therefore, (\ref{eq:aa}) becomes identical to the equations used for the LES in this study. 


\subsection{Reynolds decomposition}\label{sec:sec21}
To analyse the turbulence statistics of the GQL and the original full system, we consider the Reynolds decomposition of the velocity $\mathbf{u}=(u,v,w)$:
\begin{equation}\label{eq:2.1}
    \mathbf{u}=\mathbf{U}+\mathbf{u}',
\end{equation}
in which $\mathbf{U}(\equiv\langle \mathbf{u} \rangle_{x,z,t})=(U(y),0,0)$ is the mean velocity with $\langle\cdot\rangle_{x,z,t}$ being an average in $t$-, $x$- and $z$-directions. 
The equations for turbulent fluctuations are given by
\begin{equation}\label{eq:2.2b}
\pdv{\bold{u}'}{t} + (\bold{U} \cdot \bold{\nabla}) \bold{u}' + (\bold{u}' \cdot \bold{\nabla}) \bold{U} =-\frac{1}{\rho}\bold{\nabla} p' + \nu \bold{\nabla}^2 \bold{u}' + \bold{\nabla} \cdot \bm{\mathbf{\tau}}_{SGS}' -(\bold{u}' \cdot \bold{\nabla} )\bold{u}'+\langle(\bold{u}' \cdot \bold{\nabla} )\bold{u}'\rangle_{x,z,t}.
\end{equation}
For the GQL approximation to (\ref{eq:2.2b}), the turbulent velocity fluctuation is further decomposed into the low and high wavenumber components as in (\ref{eq:2.3z}):
\begin{equation}\label{eq:2.3}
\bold{u}'=\bold{u}_l+\bold{u}_h.
\end{equation}
Using the definition and the properties listed in (\ref{eq:2.4}) and (\ref{eq:2.5}), the projection of the equations for turbulent fluctuation onto the $\mathcal{P}_l$ and $\mathcal{P}_h$ subspaces leads to the following momentum equations:
\begin{subequations}\label{eq:2.6}
\begin{align}\label{eq:2.6a}
&\frac{\partial \bold{u}_{l}}{\partial t}+\left( \textbf{u}_{l}\cdot\nabla\right)\textbf{U}+\left(\textbf{U}\cdot\nabla\right)\textbf{u}_{l}=-\frac{1}{\rho}\nabla p_{l} + \nu \bold{\nabla}^2  \bold{u}_l +  \mathcal{P}_{l}[ \bold{\nabla} \cdot \bm{\tau}_{SGS}' ] \nonumber \\
&-\mathcal{P}_{l}\left[\left(\bold{u}_{l}\cdot\nabla\right)\bold{u}_{l}\right]-\mathcal{P}_{l}\left[\left(\bold{u}_{l}\cdot\nabla\right)\bold{u}_{h}\right]-\mathcal{P}_{l}\left[\left(\bold{u}_{h}\cdot\nabla\right)\bold{u}_{l}\right]-\mathcal{P}_{l}\left[\left(\textbf{u}_{h}\cdot\nabla\right)\textbf{u}_{h}\right] \nonumber \\
&+\langle(\bold{u}_l \cdot \bold{\nabla})\bold{u}_l\rangle_{x,z,t}
+\langle(\bold{u}_l \cdot \bold{\nabla})\bold{u}_h\rangle_{x,z,t}
+\langle(\bold{u}_h \cdot \bold{\nabla} )\bold{u}_l\rangle_{x,z,t}
+\langle(\bold{u}_h \cdot \bold{\nabla} )\bold{u}_h\rangle_{x,z,t}
\end{align}
and
\begin{align}\label{eq:2.6b}
&\frac{\partial \bold{u}_{h}}{\partial t}+\mathcal{P}_{h}\left[\left(\textbf{u}_{h}\cdot\nabla\right)(\bold{U}+\bold{u}_l)\right]+\mathcal{P}_{h}\left[\left((\bold{U}+\bold{u}_l)\cdot\nabla\right)\textbf{u}_{h}\right]=-\frac{1}{\rho}\nabla p_{h}+  \nu \bold{\nabla}^2  \bold{u}_h \nonumber \\
&+  \mathcal{P}_{h}[ \bold{\nabla} \cdot \bm{\tau}_{SGS}' ] -\mathcal{P}_{h}\left[\left(\textbf{u}_{l}\cdot\nabla\right)\textbf{u}_{l}\right]-\mathcal{P}_{h}\left[\left(\textbf{u}_{h}\cdot\nabla\right)\textbf{u}_{h}\right],
\end{align}
\end{subequations}
where $p_l$ and $p_h$ are defined to enforce $\bold{\nabla} \cdot \bold{u}_l=0$ and $\bold{\nabla} \cdot \bold{u}_h=0$, respectively, with $p'=p_l+p_h$. For the GQL approximation in (\ref{eq:full}), the self-interaction terms $\mathcal{P}_{h}\left[\left(\textbf{u}_{l}\cdot\nabla\right)\textbf{u}_{l}\right]$ and $\mathcal{P}_{h}\left[\left(\textbf{u}_{h}\cdot\nabla\right)\textbf{u}_{h}\right]$ in (\ref{eq:2.6b}) are ignored. 

\subsection{Spectral energetics}\label{sec:sec22}

To study the effect of the GQL approximation on the inter- and intra-scale energy transfer of the given flow, here we consider the energy transfer in the Fourier space as was recently studied by \cite{cho18}. For this formulation, it is more convenient to consider a one-dimensional continuous Fourier transform than the discrete version in (\ref{eq:2.0b}):
\begin{equation}\label{ft}
u_j^{\prime}(t,r)=\int_{-\infty}^{\infty} \widehat{u_j^{\prime}}(t,k) e^{\mathrm{i} k r} \mathrm{d}k
\end{equation}
for $j=1,2,3$, where $\widehat{\cdot}$ denotes the Fourier-transformed coefficient, $(u_1^{\prime},u_2^{\prime},u_3^{\prime})=(u^{\prime},v^{\prime},w^{\prime})$, $r(=x~\mathrm{or}~z)$ is the streamwise or spanwise coordinate, and $k(=k_x~\mathrm{or}~k_z)$ the corresponding wavenumber. We then take the Fourier transformation (\ref{ft}) to  (\ref{eq:2.2b}), and multiply it by the complex conjugate of $\widehat{u_i'}(k)$. 
By taking an average in time and the planar direction along which the Fourier transform is not taken (denoted by $r^\perp$), the following energy balance in the Fourier space for the GQL approximation is obtained:
\begin{align}\label{eq:2.9}
&\underbrace{\left\langle\frac{\partial{\widehat{e}(k)}}{\partial{t} } \right\rangle_{r^\perp,t}}_{=0} = \underbrace{\left\langle \Real \left\{-{\widehat{u^{\prime}}}^* (k)\widehat{v^{\prime}}(k) \: \frac{\textup{d} U}{\textup{d} y} \right\} \right\rangle_{r^\perp,t}}_{\widehat{P}\left(y, k \right)}
+\underbrace{\left\langle - \nu \frac{\partial{{\widehat{u_{i}^{\prime}}}^*(k)} }{\partial{x_{j}} } \frac{\partial{{\widehat{u_{i}^{\prime}}}(k)} }{\partial{x_{j}} } \: \right\rangle_{r^\perp,t}}_{\widehat{\varepsilon}\left(y, k \right)} \nonumber\\
&+\underbrace{\left\langle \Real \left\{- {\widehat{u_{i}^{\prime}}}^*(k) \left(\frac{\partial }{\partial{x_{j}} }  \left( \widehat{ \tau_{ij,SGS}^{\prime}}(k) \right) \right)\: \right\} \right\rangle_{r^\perp,t}}_{\widehat{\varepsilon}_{SGS}\left(y, k \right)} +\underbrace{\left\langle \Real \left\{\frac{\textup{d}}{\textup{d} y} \left ( \frac{\widehat{p^{\prime}}(k) {\widehat{v^{\prime}}}^*(k)}{\rho} \right) \right\} \right\rangle_{r^\perp,t}}_{\widehat{T}_{p}\left(y, k \right)} \nonumber \\
&+ \underbrace{\left\langle \Real \left\{-{\widehat{u_{i}^{\prime}}}^*(k) \left(\frac{\partial }{\partial{x_{j}} }  \left( \widehat{ u_{i}^{\prime}u_{j}^{\prime} }(k) -\mathcal{P}_h\left[\widehat{ u_{h,i}^{\prime}u_{h,j}^{\prime} }(k)\right]
-\mathcal{P}_h\left[\widehat{ u_{l,i}^{\prime}u_{l,j}^{\prime} }(k)\right] \right) \right)\: \right\} \right\rangle_{r^\perp,t}}_{\widehat{T}_{turb}\left(y, k\right)}\nonumber \\
&+\underbrace{\left\langle \nu \frac{\textup{d}^2{\widehat{e}(k)}}{\textup{d}{y}^2 } \right\rangle_{r^\perp,t}}_{\widehat{T}_{\nu}\left(y, k \right)},
\end{align}
where $(x_1,x_2,x_3)=(x,y,z)$, $\widehat{e}(k)=( |{\widehat{u'}(k)} |^2 + |{\widehat{v'}(k)} |^2 +|{\widehat{w'}(k)} |^2 )/2$, the superscript $(\cdot)^*$ indicates the complex conjugate, and $\Real \left \{\hspace{1mm}\cdot\hspace{1mm}\right \}$ the real part. In (\ref{eq:2.9}) the left-hand side is the rate of each streamwise/spanwise Fourier mode of TKE, which should vanish in a statistically steady flow. The terms on the right-hand side are the rate of turbulence production $\widehat{P}(y,k)$, viscous dissipation $\widehat{\varepsilon}(y,k)$, SGS dissipation $\widehat{\varepsilon}_{SGS}(y,k)$, pressure transport $\widehat{T}_p(y,k)$, turbulent transport $\widehat{T}_{turb}(y,k)$ and viscous transport $\widehat{T}_{\nu}(y,k)$, respectively. 

Equation (\ref{eq:2.9}) can be further split into each component for the componentwise TKE budget. In this case, the energy transport by pressure strain appears:
\begin{align}
&\widehat{\Pi}_x(y, k)=\left \langle \Real \left \{ \frac{\widehat{p'} (k)}{\rho} \pdv{{\widehat{u'}}^* (k)}{x} \right \} \right \rangle_{r^\perp,t}, \quad \widehat{\Pi}_y(y, k)= \left \langle \Real \left \{\frac{\widehat{p'} (k)}{\rho} \pdv{{\widehat{v'}}^*(k)}{y} \right \} \right \rangle_{r^\perp,t} , \nonumber\\ 
&\widehat{\Pi}_z(y, k)=\left \langle \Real \left \{\frac{\widehat{p'} (k)}{\rho} \pdv{{\widehat{w'}}^* (k)}{z} \right \} \right \rangle_{r^\perp,t},
\end{align}
where $\widehat{\Pi}_x$, $\widehat{\Pi}_y$ and $\widehat{\Pi}_z$ are one-dimensional spectra of the streamwise, wall-normal and spanwise components of pressure strain, respectively. As discussed in detail in previous studies \citep[][]{mizuno16,cho18,lee19,hernandez}, the pressure-strain terms play an important role in the TKE distribution to the individual velocity components. In parallel shear flows, the turbulent production (source term) only takes place in the streamwise component, but not in the wall-normal nor in the spanwise components. The pressure strain terms subsequently transfer the energy produced in the streamwise component to the other two components, as can be understood from the relation
\begin{equation}
\widehat{\Pi}_x(y,k)+\widehat{\Pi}_y(y,k)+\widehat{\Pi}_z(y,k)=0
\end{equation}
due to the continuity equation. If the isotropy of fluid motions at dissipation scale is assumed \cite[]{kolmogorov41}, this implies that the pressure-strain terms must mediate the conversion of highly anisotropic large scale into isotropic small scale during the energy cascade.
\begin{table}
  \begin{center}
\def~{\hphantom{0}}
  \begin{tabular}{lccccccccccc}
\multicolumn{1}{c}{\multirow{2}{*}{Case}} &
\multicolumn{1}{c}{\multirow{2}{*}{$Re$}} &
\multicolumn{1}{c}{\multirow{2}{*}{$Re_{\tau}$}} & \multicolumn{1}{c}{\multirow{2}{*}{$L_x/h$}}  &  \multicolumn{1}{c}{\multirow{2}{*}{$L_z/h$}}   & 
\multicolumn{1}{c}{\multirow{2}{*}{$\Delta_x^+$}}  & \multicolumn{1}{c}{\multirow{2}{*}{$\Delta_z^+$}} & 
\multicolumn{1}{c}{\multirow{2}{*}{$M_{x, F}$}} &
\multicolumn{1}{c}{\multirow{2}{*}{$\lambda_{x,c}/h$}} &
\multicolumn{1}{c}{\multirow{2}{*}{$\lambda_{x,c}^+$}} &
\multicolumn{1}{c}{\multirow{2}{*}{$N_{x} \times N_y \times N_z$}} \\ \\[3pt]
\multicolumn{1}{c}{LES}  & 55555 & \multicolumn{1}{c}{1673} &  \multicolumn{1}{c}{$\pi$} & \multicolumn{1}{c}{$\pi/2$}  &  \multicolumn{1}{c}{61.6} & \multicolumn{1}{c}{30.8} &  \multicolumn{1}{c}{42}&  \multicolumn{1}{c}{-}&  \multicolumn{1}{c}{-} &\multicolumn{1}{c}{$128 \times 129 \times 128$} \\ [2pt]
\multicolumn{1}{c}{QL} & 55555  & \multicolumn{1}{c}{1501}  & \multicolumn{1}{c}{$\pi$} & \multicolumn{1}{c}{$\pi/2$}  &  \multicolumn{1}{c}{55.2} & \multicolumn{1}{c}{27.6} & \multicolumn{1}{c}{0} &  \multicolumn{1}{c}{0}&  \multicolumn{1}{c}{0}& \multicolumn{1}{c}{$128 \times 129 \times 128$}   \\ [2pt]
\multicolumn{1}{c}{GQL1} & 55555  & \multicolumn{1}{c}{1751} &   \multicolumn{1}{c}{$\pi$} & \multicolumn{1}{c}{$\pi/2$}  &   \multicolumn{1}{c}{64.4} & \multicolumn{1}{c}{32.2} & \multicolumn{1}{c}{1} &  \multicolumn{1}{c}{3.14}&  \multicolumn{1}{c}{4713}& \multicolumn{1}{c}{$128 \times 129 \times 128$} \\ [2pt]
\multicolumn{1}{c}{GQL5} & 55555  & \multicolumn{1}{c}{1792} & \multicolumn{1}{c}{$\pi$} & \multicolumn{1}{c}{$\pi/2$}  &   \multicolumn{1}{c}{66.0} & \multicolumn{1}{c}{33.0} & \multicolumn{1}{c}{5} &  \multicolumn{1}{c}{0.63}&  \multicolumn{1}{c}{1126}& \multicolumn{1}{c}{$128 \times 129 \times 128$} \\ [2pt]
\multicolumn{1}{c}{GQL25} & 55555  & \multicolumn{1}{c}{1733} & \multicolumn{1}{c}{$\pi$} & \multicolumn{1}{c}{$\pi/2$}  &   \multicolumn{1}{c}{63.8} & \multicolumn{1}{c}{31.9} & \multicolumn{1}{c}{25}&  \multicolumn{1}{c}{0.13}&  \multicolumn{1}{c}{218} & \multicolumn{1}{c}{$128 \times 129 \times 128$} \\ [2pt]
\end{tabular}
\caption{Simulation parameters in the present study. $L_x/h$, $L_y/h$ and $L_z/h$ indicate the domain size in the $x$-, $y$- and $z$ directions, respectively. Here, $Re=U_0 h/\nu$ and $Re_\tau=u_\tau h /\nu$, where $U_0$ and $u_\tau$ are the centerline velocity of the corresponding laminar base flow and the wall shear (or friction) velocity, respectively. The grid spacings in the $x$- and $z$-directions are $\Delta_x^+$ and $\Delta_z^+$ (after aliasing). $\lambda_{x,c}$ is the threshold streamwise wavelength. $N_x$, $N_y$, $N_z$ are the number of grid points in the $x$-, $y$- and $z$-directions, respectively. 
}
\label{tab:tab}
\end{center}
\end{table}

\subsection{Numerical simulations}\label{sec:sec23}

The LESs for the full and the GQL equations are carried out by imposing constant mass flux across the channel. The numerical solver used in this study is \texttt{diablo} (\citealp{bewley14}), the use of which has been verified by a number of previous studies \cite[e.g.][]{yang18,doohan19,hernandez}. In this solver, the streamwise and spanwise directions are discretized using Fourier series with $2/3$ rule for dealiasing, and the wall-normal direction is discretized using the second-order central difference. The time integration is conducted semi-implicitly based on the fractional-step method (\citealp{kim85}). All the viscous terms are implicitly advanced with the second-order Crank–Nicolson method, while the rest of the nonlinear advection terms are explicitly integrated with a low-storage third-order Runge–Kutta method. 
The present LES has previously been validated over a range of Reynolds numbers from $Re_\tau \approx 1000$ to $Re_\tau \approx 4000$ \citep{degiovanetti16,degiovanetti17}.

Table \ref{tab:tab} summarizes the parameters for the simulations performed in this study. The Reynolds number for all the considered cases is $Re=U_0 h /\nu=55555$ ($U_0$ is the centreline velocity of the laminar base flow of each simulation). The computations are carried out in the minimal unit for large-scale self-sustaining process  with $L_x/h=\pi$, $L_z/h=\pi/2$ \citep{hwang10prl,hwangbengana16}, which are identical to those in \cite{farrell17}. We note that the fundamental streamwise wavenumber $k_{x,0}(\equiv 2\pi/L_x)$ in this domain primarily resolves the streak instability (or transent growth) wave emerging from the large-scale self-sustaining process \cite[]{degiovanetti17}. 
For the QL and GQL approximations, the number of streamwise Fourier number used in the $\mathcal{P}_l$-subspace group of the velocity is varied from the minimal ($M_{x,F}=0$) to the maximum number used for the original full LES ($M_{x,F}=42$), while $M_{z,F}$ is kept with the one in the full LES. The threshold streamwise wavelength for the decomposition of the velocity into the two groups in (\ref{eq:2.0}) is given by $\lambda_{x,c}=2\pi/k_{x,c}$ with $k_{x,c}=2\pi M_{x,F}/L_x$. 


\section{Results}\label{sec:sec3}

\subsection{Turbulence statistics and spectra}\label{sec:sec31}

\begin{figure*}
\centering
\begin{subfigure}[b]{0.45\textwidth}
\includegraphics[width=\textwidth]{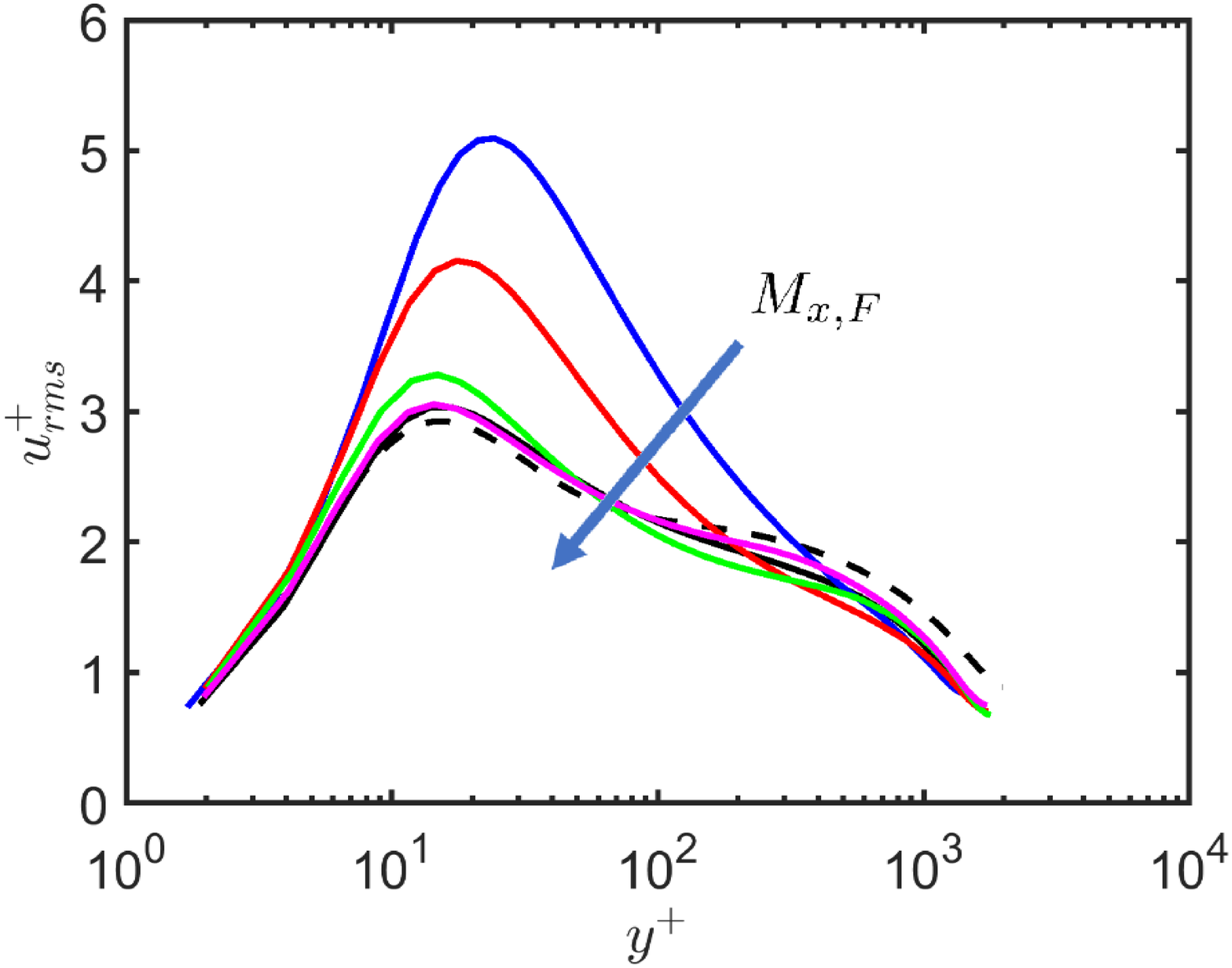}
\caption{$u_{rms}^+(y^+)$}
\label{fig:uu}
\end{subfigure}
\begin{subfigure}[b]{0.45\textwidth}
\includegraphics[width=\textwidth]{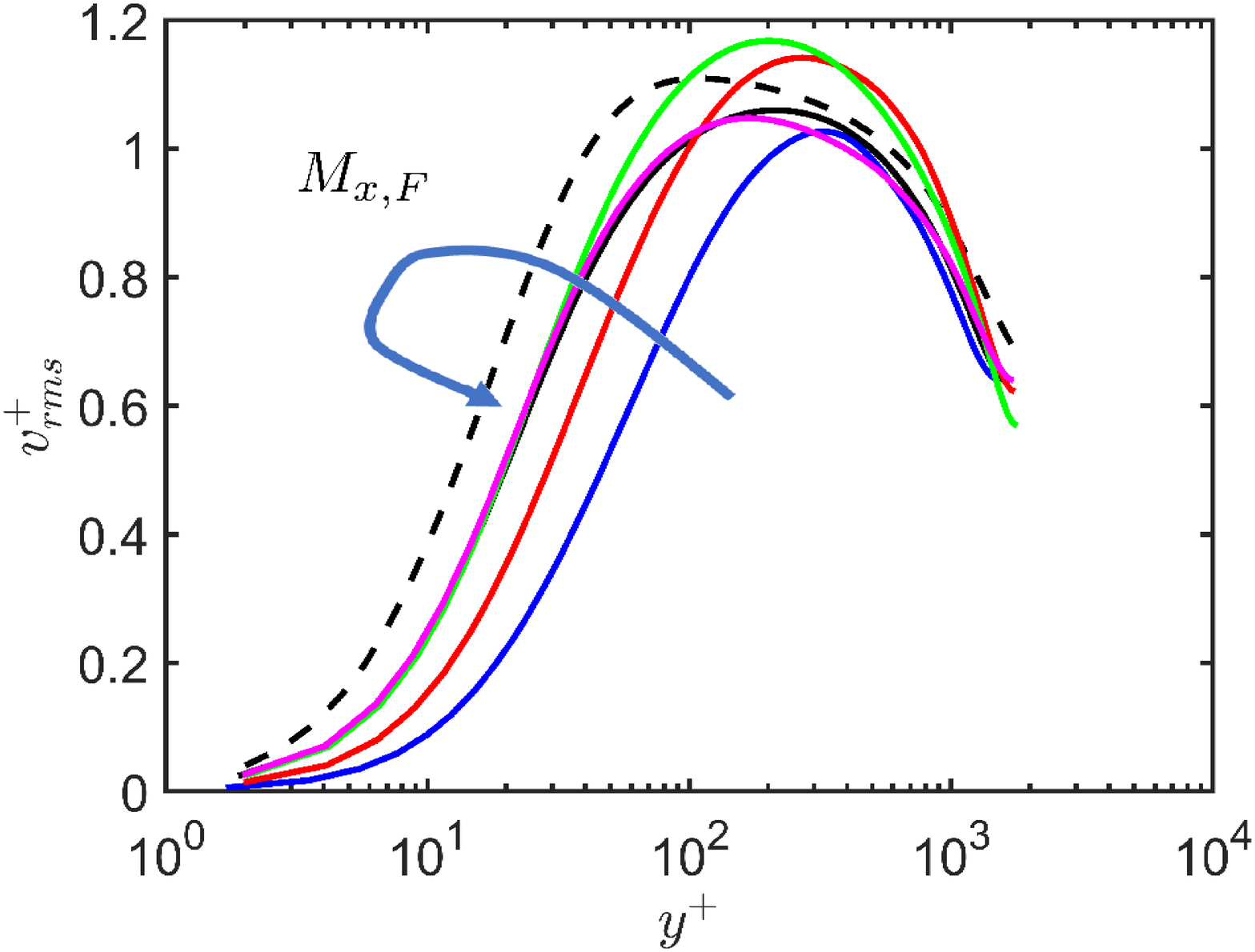}
\caption{$v_{rms}^+(y^+)$}
\label{fig:vv}
\end{subfigure}
\begin{subfigure}[b]{0.45\textwidth}
\includegraphics[width=\textwidth]{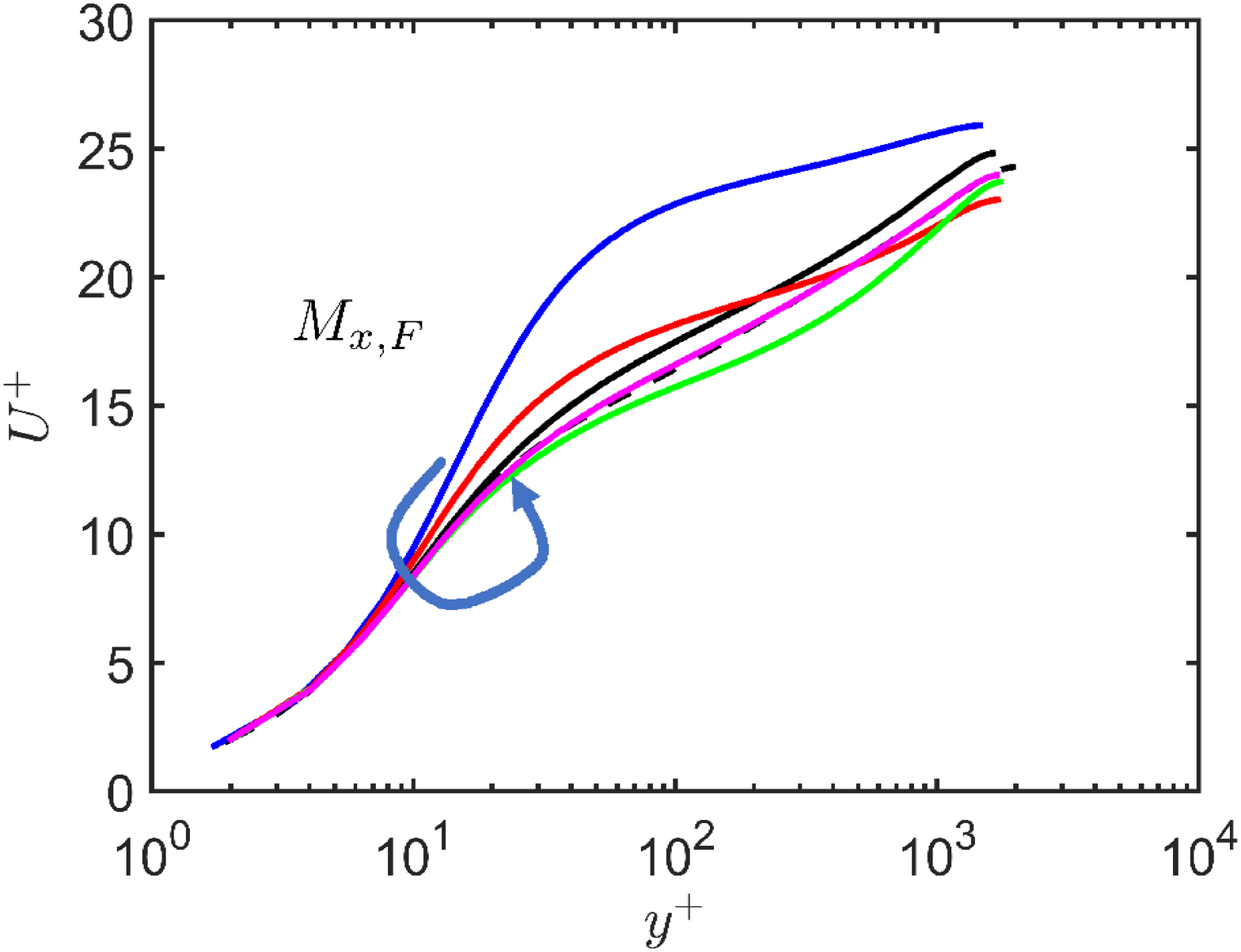}
\caption{$U^+(y^+)$}
\label{fig:ww}
\end{subfigure}
\begin{subfigure}[b]{0.45\textwidth}
\includegraphics[width=\textwidth]{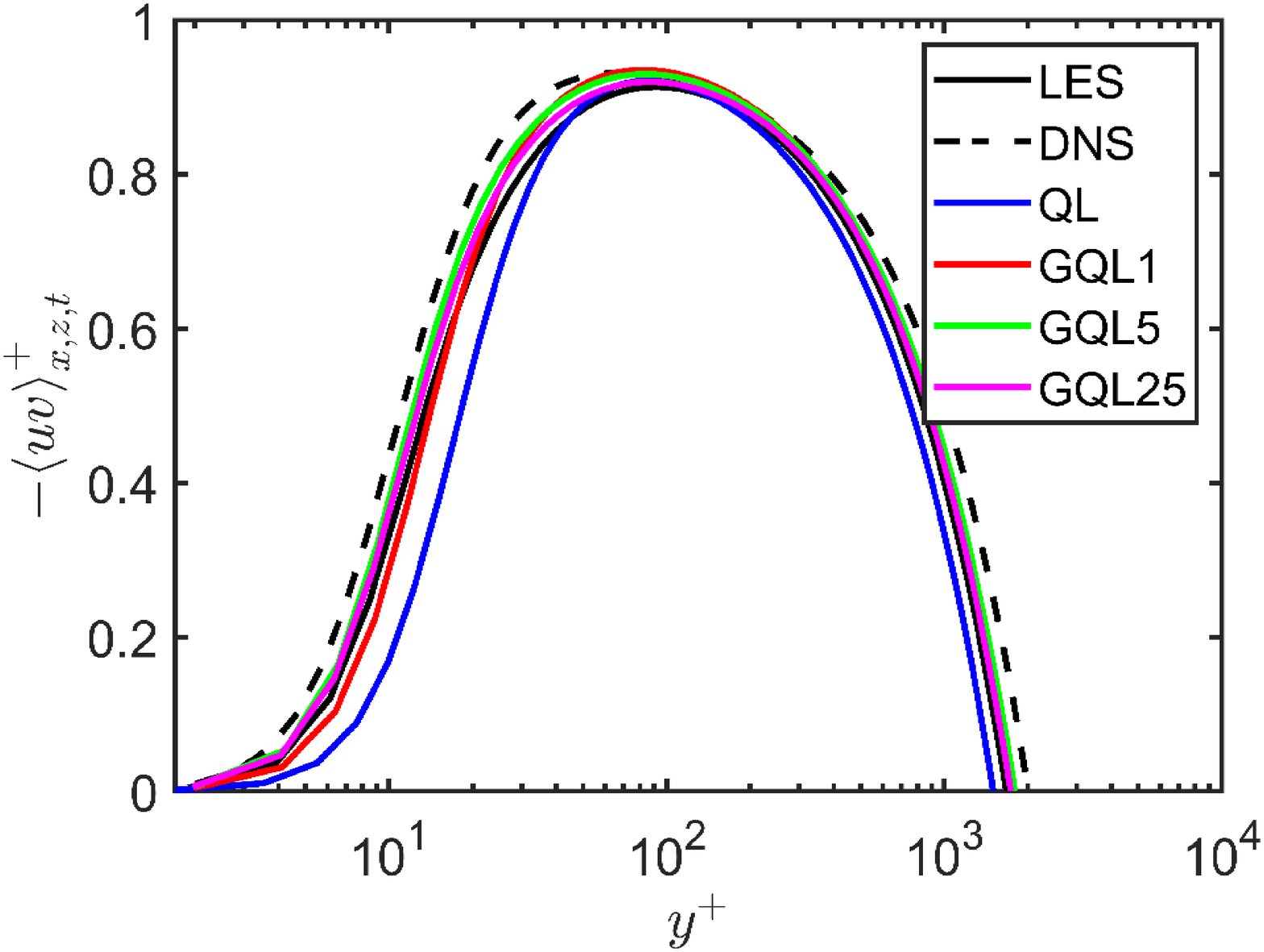}
\caption{$ \langle u^{\prime} v^{\prime}\rangle_{x,z,t} ^+(y^+)$}
\label{fig:uv}
\end{subfigure}
\caption{First- and second-order turbulence statistics for the present LES, DNS at $Re_\tau=2003$ (\citealp{hoyas08}), QL, GQL1, GQL5 and GQL25 cases: (a) $u_{rms}^+(y^+)$; (b) $v_{rms}^+(y^+)$; (c) $U^+(y^+)$; (d) $ \langle u^{\prime} v^{\prime}\rangle_{x,z,t}^+(y^+)$.}
\label{fig:stat}
\end{figure*}

The first- and second-order turbulence statistics as a function of the wall-normal direction $y^+$ are plotted in figure \ref{fig:stat} (the superscript $(\cdot)^+$ denotes normalisation by viscous inner scale). The DNS statistics from \cite{hoyas08} at $Re_\tau=2003$, plotted with dashed line, agree generally well with the ones of the present LES. The statistics of the QL case (i.e. $\lambda_{x,c}=\infty$) are the most anisotropic in comparison to the reference LES. In particular, $u_{rms}^+$ (figure \ref{fig:stat}a) and $U^+$ (figure \ref{fig:stat}c) are larger than those of the reference LES, whilst $v_{rms}^+$ (figure \ref{fig:stat}b) , $w_{rms}^+$ (not shown) and $\langle u^{\prime} v^{\prime}\rangle_{x,z,t}^+$ are smaller. This behaviour agrees well with that of \cite{thomas14} and \cite{farrell16}. By further incorporating the next immediate Fourier streamwise mode into the $\mathcal{P}_l$-subspace group (i.e. GQL1 case), we observe that the magnitude of the near-wall streamwise velocity peak is considerably reduced, whilist $v_{rms}^+$, $w_{rms}^+$ and $\langle u^{\prime} v^{\prime}\rangle_{x,z,t}^+$ become slightly larger than those of the reference LES. Importantly, the inner-scaled mean velocity $U^+$ now provides a much better approximation to that of the reference LES than the QL case, even though only one more streamwise Fourier mode is incorporated into the $\mathcal{P}_l$-subspace group. We note that the QL case shows a substantial depletion of $-\langle u^{\prime} v^{\prime}\rangle_{x,z,t}^+$ in the near-wall region ($y^+\leq 30$) compared to that of the other GQL and LES cases (figure \ref{fig:stat}d). From the following integral form of the mean streamwise momentum equation,
\begin{equation}
    U^+(y^+)=y^+-\frac{(y^+)^2}{2Re_\tau}+\int_0^{y^+} \langle u^{\prime} v^{\prime}\rangle_{x,z,t}^+~dy^+,
\end{equation}
where the vanishing near-wall contribution of the SGS model is ignored, the smaller value in $-\langle u^{\prime} v^{\prime}\rangle_{x,z,t}^+$ should lead to a large value of $U^+$ in the near-wall region, consistent with the mean velocity in figure \ref{fig:stat}(c). As the more streamwise Fourier modes are included in the $\mathcal{P}_l$-subspace group, all the turbulence statistics converge to those of the full LES (GQL1, GQL5 and GQL25 cases in figure \ref{fig:stat}). Finally, the statistics of the GQL25 case give the closest match to those of the reference LES, as expected.

\begin{figure}
\vspace{-0.8cm}
\begin{minipage}{\textwidth}
\centering
\begin{subfigure}[b]{0.42\textwidth}
  \includegraphics[width=\textwidth]{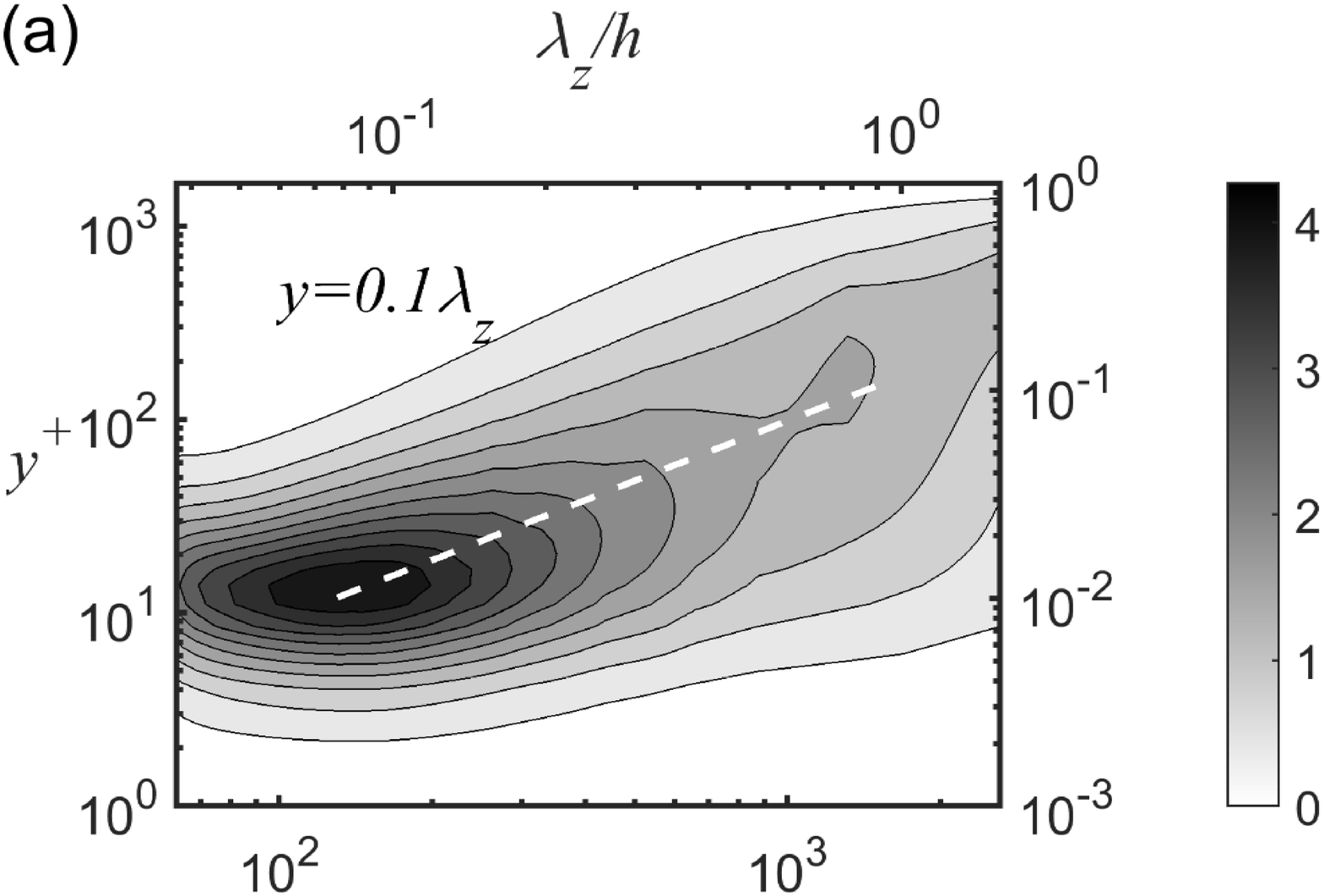}
\label{1}
\vspace{-0.8cm}
\end{subfigure}
\begin{subfigure}[b]{0.42\textwidth}
  \includegraphics[width=\textwidth]{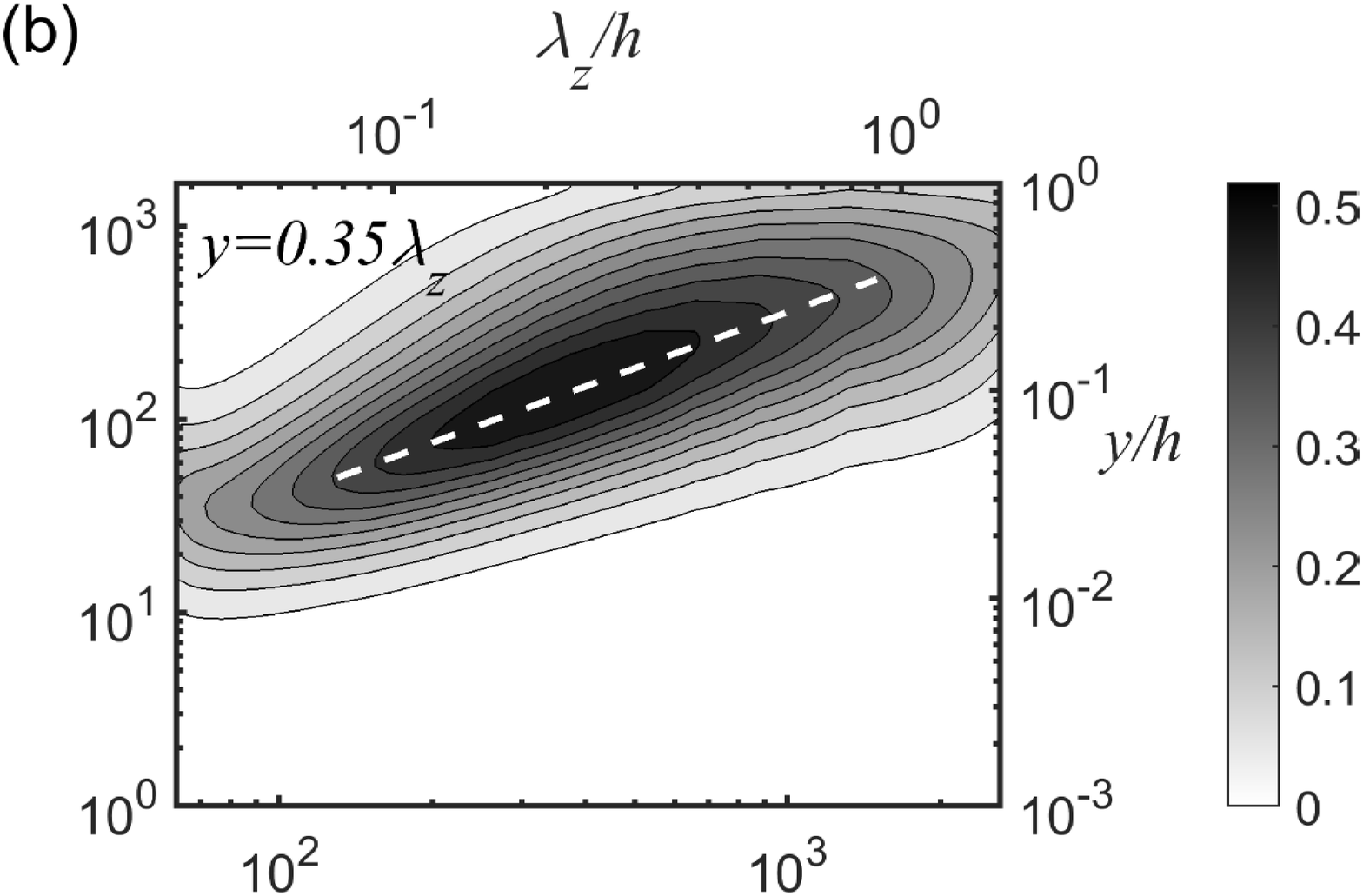}
\label{2}
\vspace{-0.8cm}
\end{subfigure}
\vspace{-0.8cm}
\begin{subfigure}[b]{0.42\textwidth}
  \includegraphics[width=\textwidth]{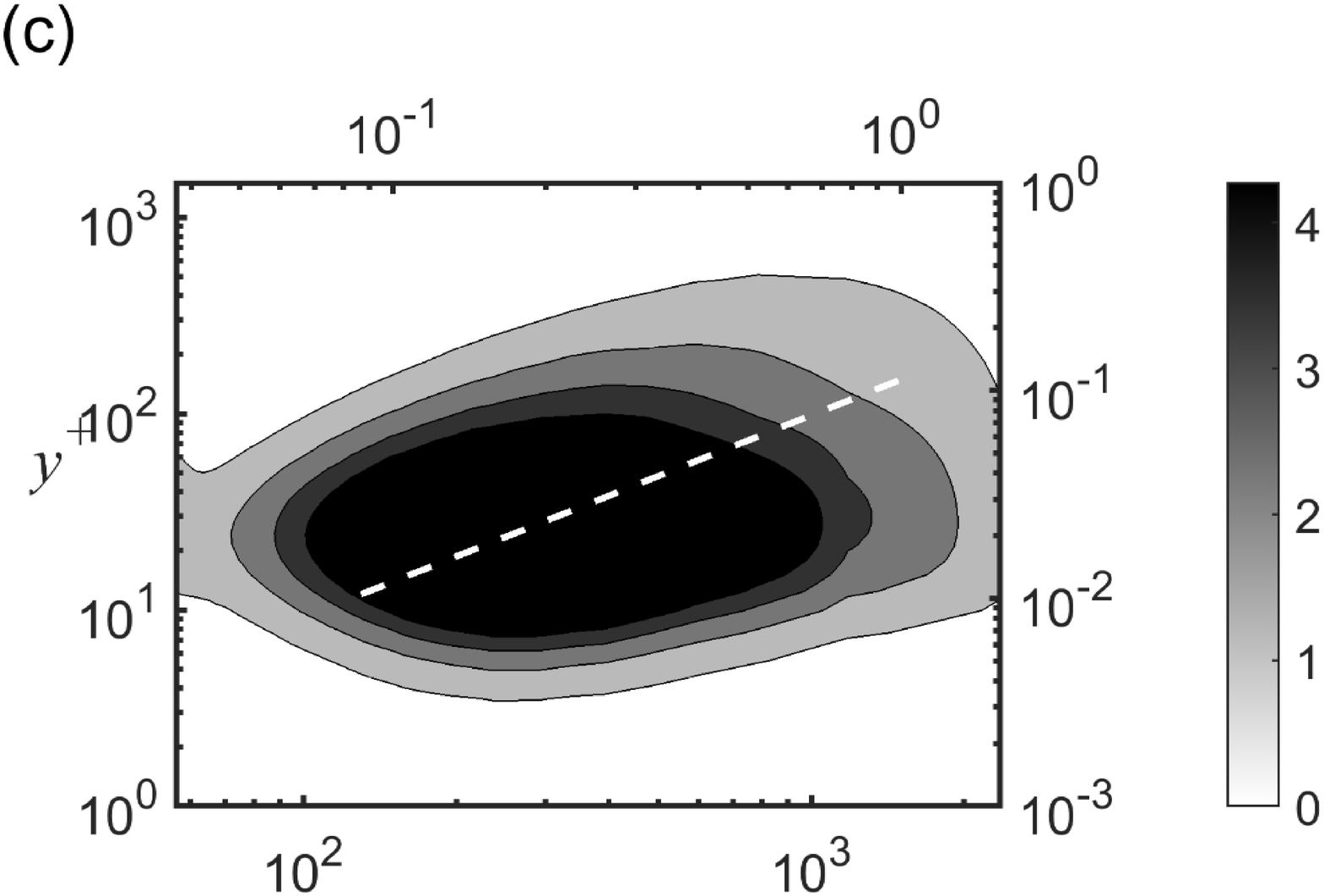}
  \label{3}
\end{subfigure}
\begin{subfigure}[b]{0.42\textwidth}
  \includegraphics[width=\textwidth]{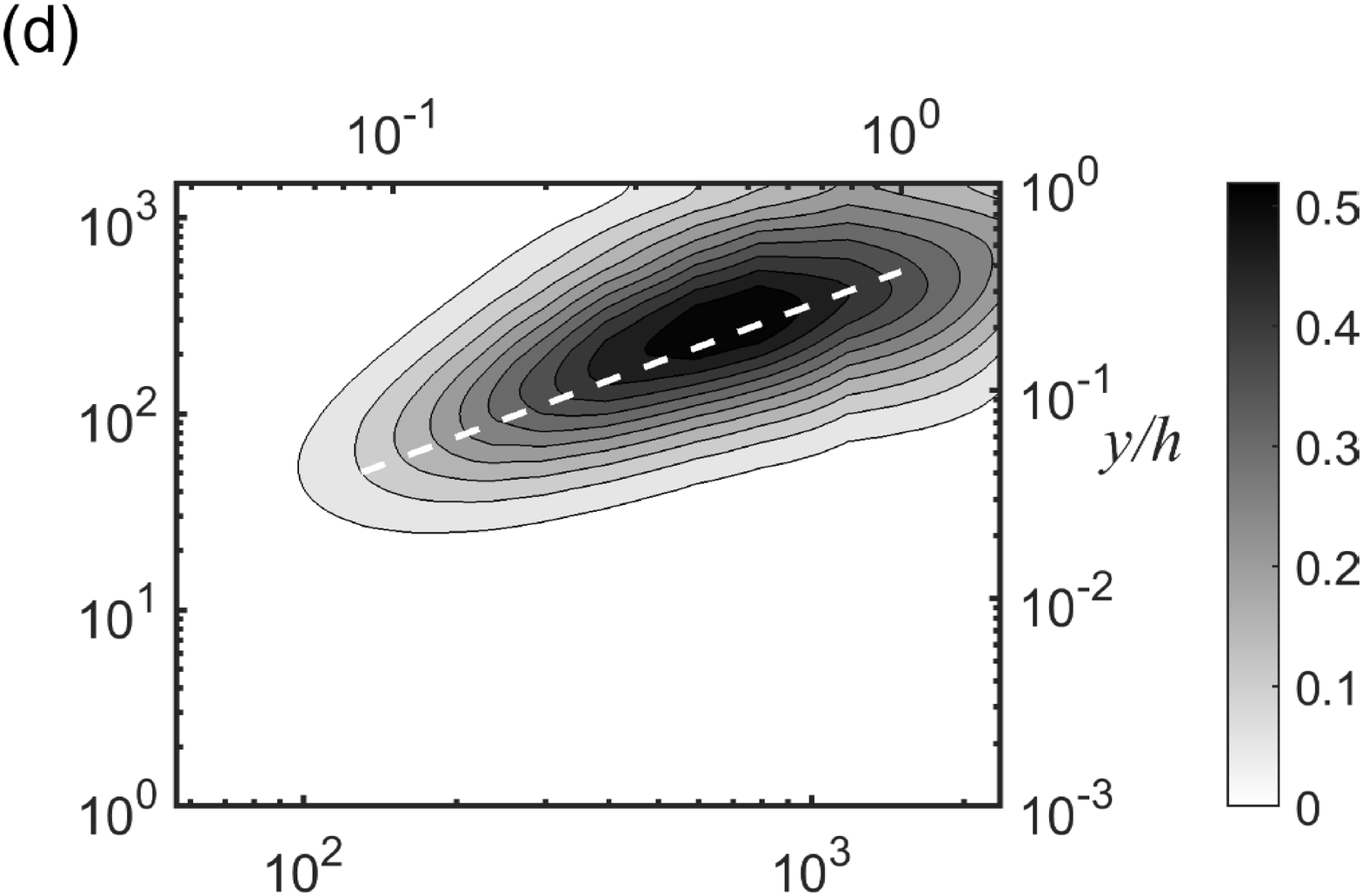}
  \label{4}
\end{subfigure}
\vspace{-0.8cm}
\begin{subfigure}[b]{0.42\textwidth}
  \includegraphics[width=\textwidth]{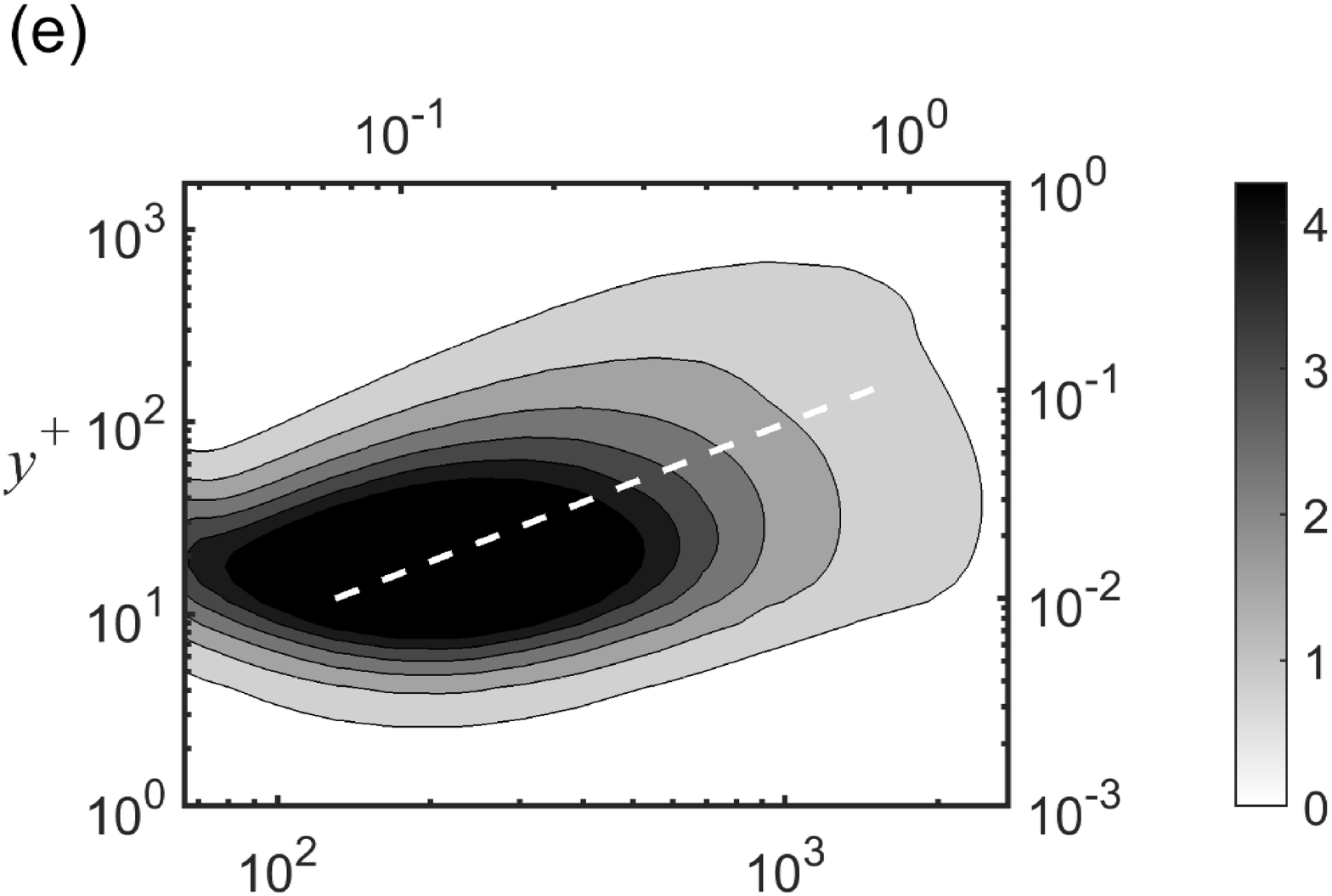}
  \label{5}
\end{subfigure}
\begin{subfigure}[b]{0.42\textwidth}
  \includegraphics[width=\textwidth]{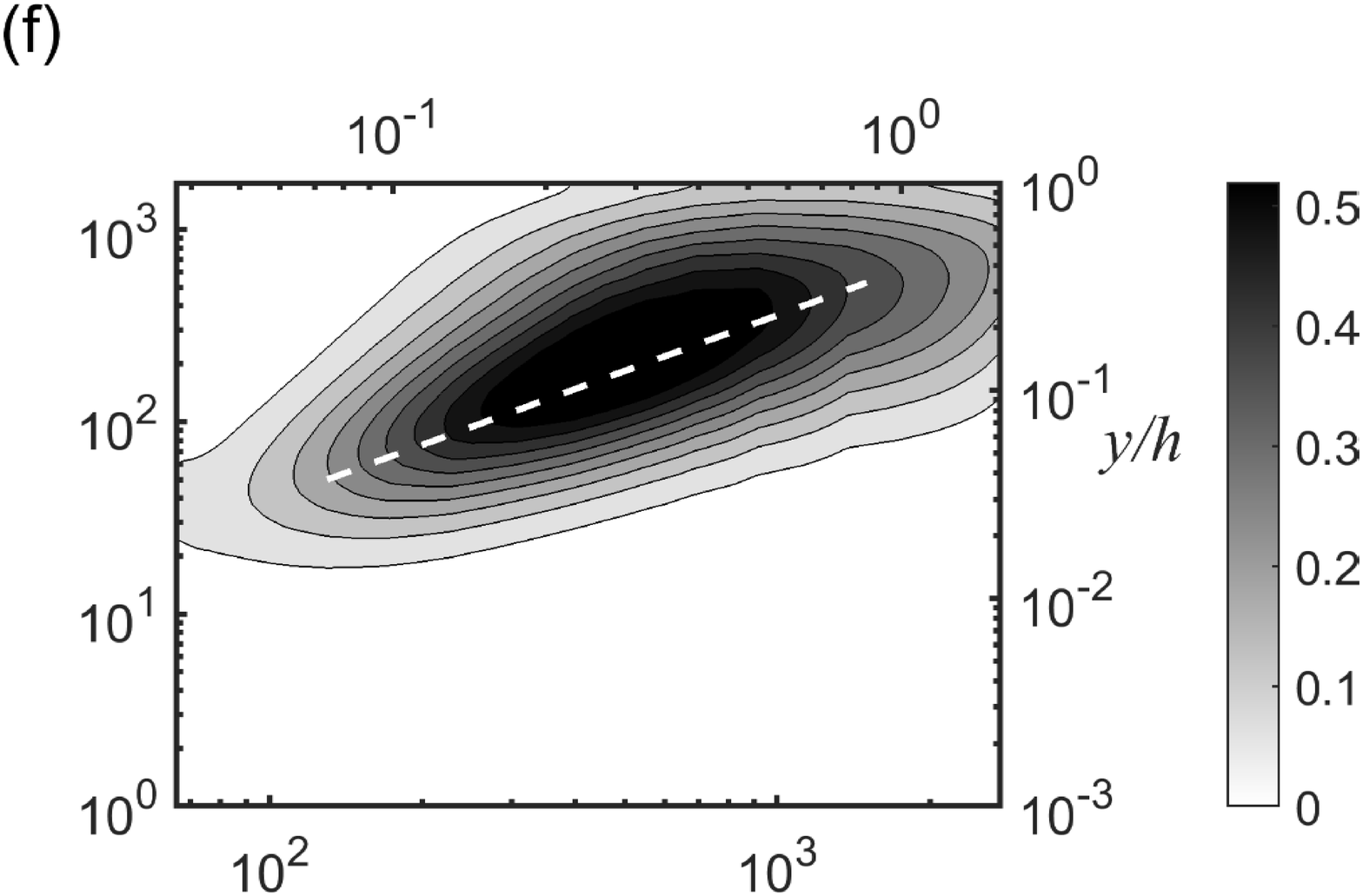}
  \label{6}
\end{subfigure}
\vspace{-0.8cm}
\begin{subfigure}[b]{0.42\textwidth}
  \includegraphics[width=\textwidth]{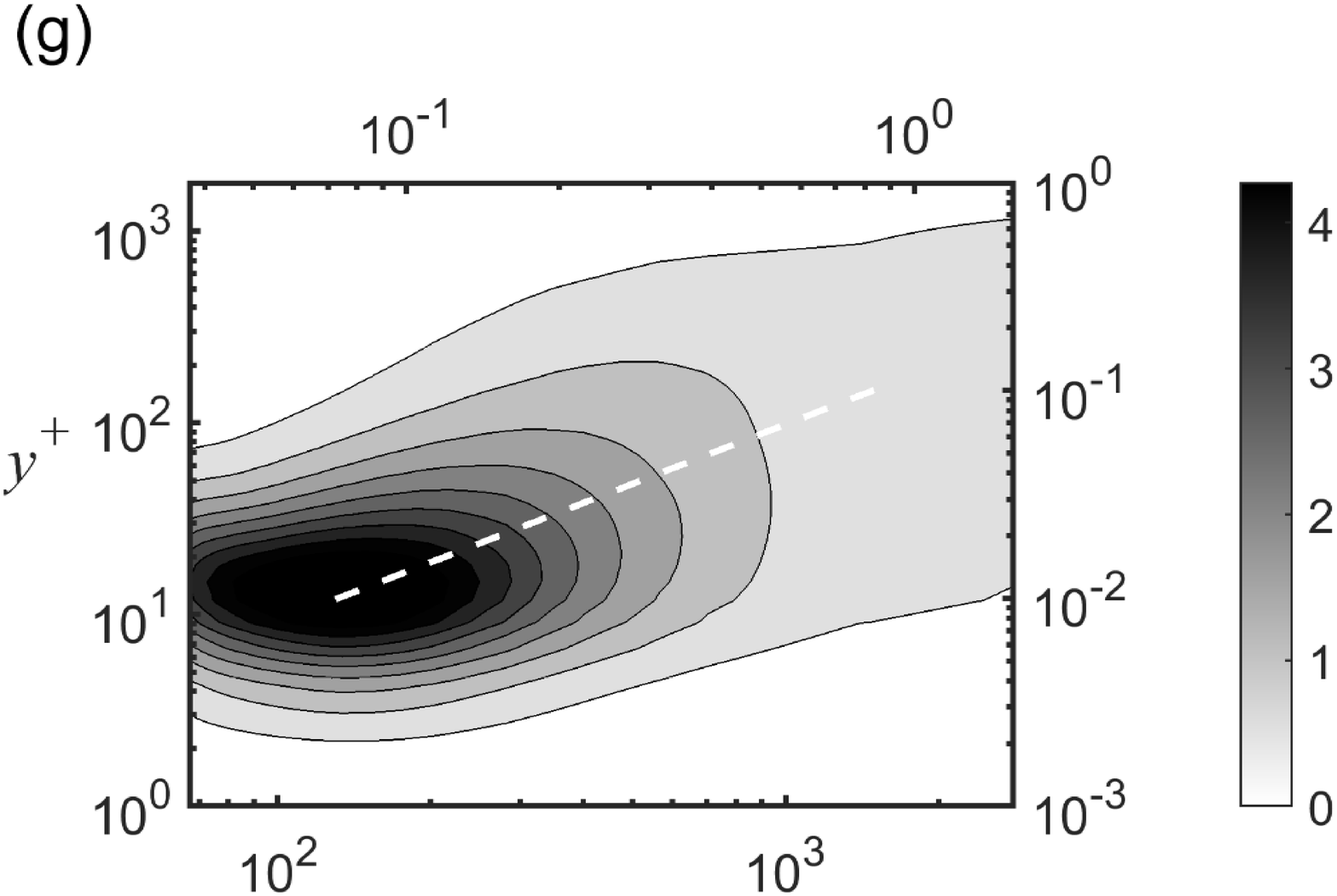}
  \label{5}
\end{subfigure}
\begin{subfigure}[b]{0.42\textwidth}
  \includegraphics[width=\textwidth]{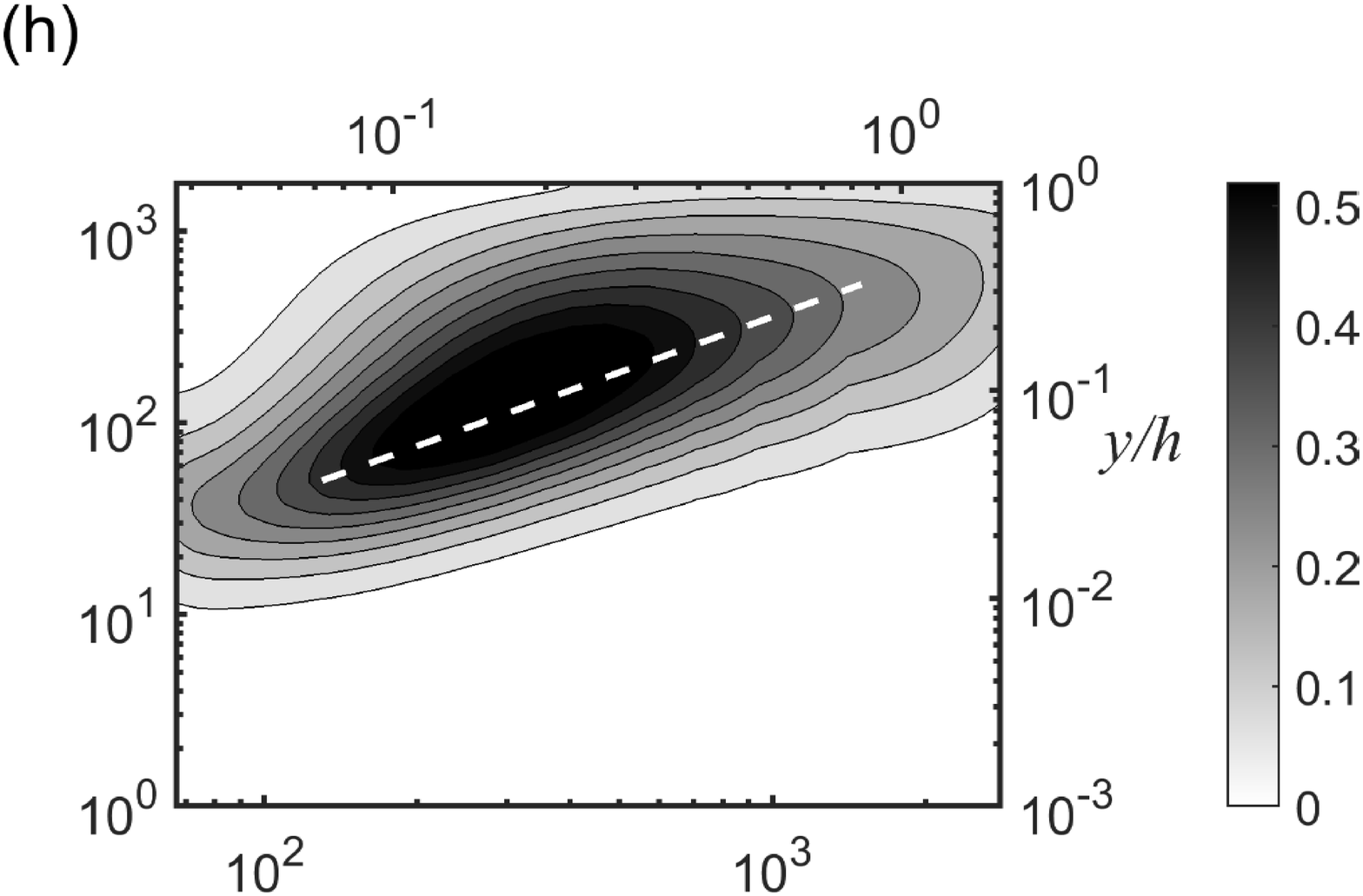}
  \label{6}
\end{subfigure}
\begin{subfigure}[b]{0.42\textwidth}
  \includegraphics[width=\textwidth]{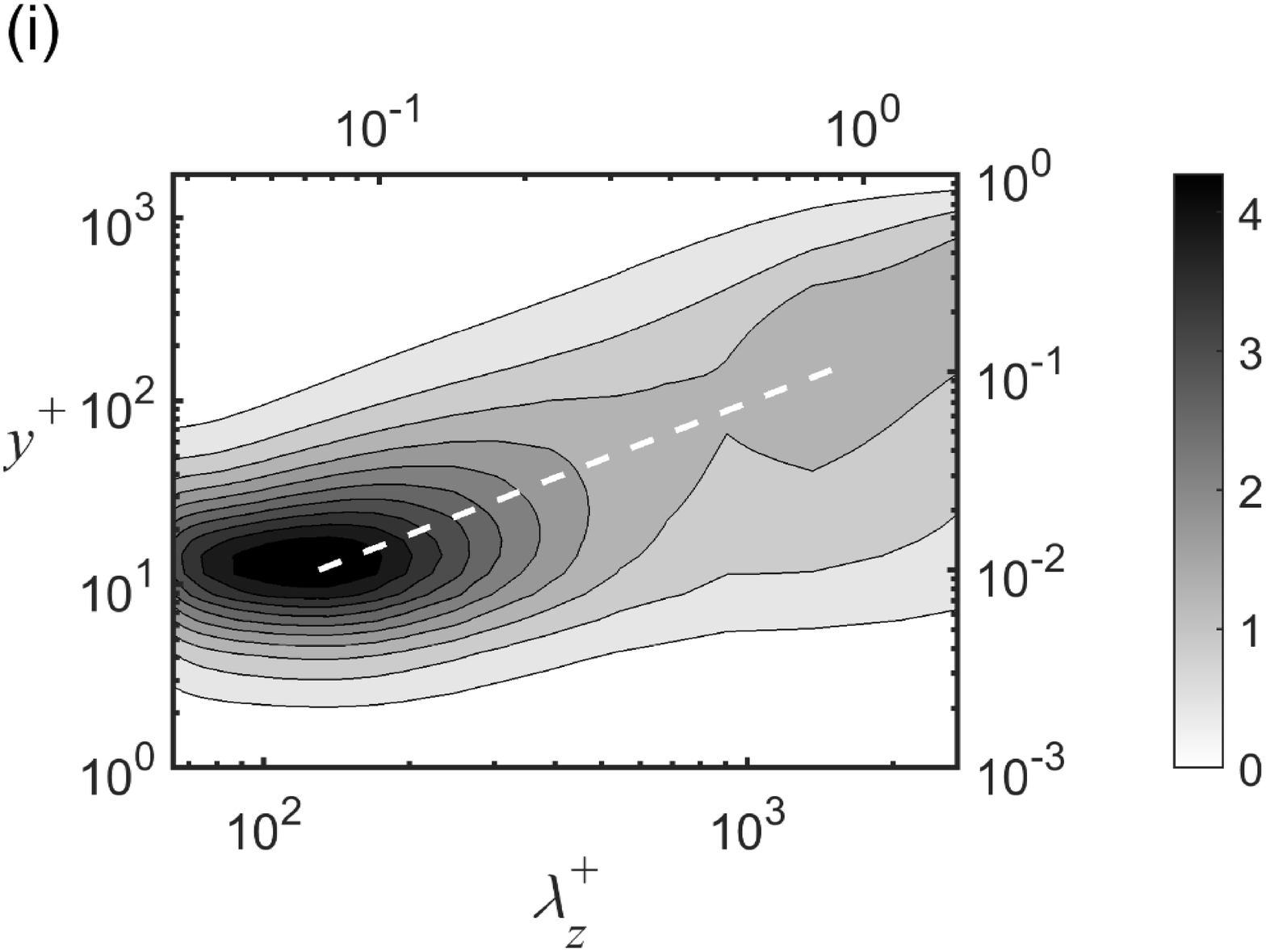}
\end{subfigure}
\begin{subfigure}[b]{0.42\textwidth}
  \includegraphics[width=\textwidth]{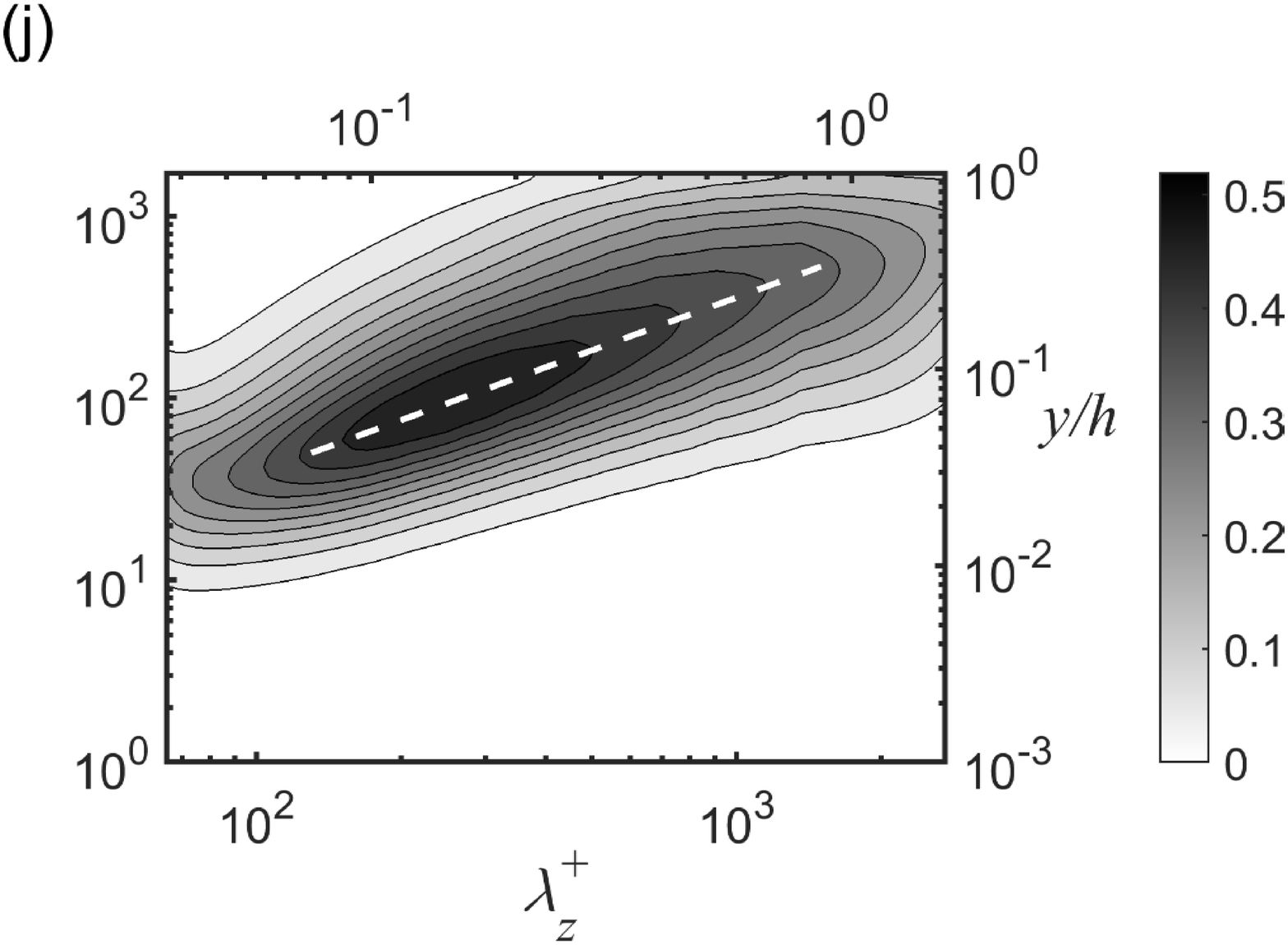}
\end{subfigure}
\end{minipage}
\caption{Premultiplied spanwise wavenumber spectra of streamwise $k_z^+ \Phi_{uu}^+(y^+,\lambda_z^+)$ (left column) and wall-normal $k_z^+ \Phi_{vv}^+(y^+,\lambda_z^+)$ (right column) velocity for (a,b) LES, (c,d) QL, (e,f) GQL1, (g,h) GQL5 and (i,j) GQL25 cases.}
\label{fig:zspectra}
\end{figure}

Figure \ref{fig:zspectra} compares the premultiplied spanwise wavenumber spectra of streamwise (left column) and wall-normal (right column) velocities of the reference LES with those of the QL and GQL cases. The spectra of the LES show the typical features of energy-containing motions in turbulent channel flow \cite[e.g.][]{hwang15}: the spanwise wavenumber spectra of streamwise velocity are aligned along a linear ridge $y \approx 0.1 \lambda_z$ throughout the logarithmic region, indicating the linear growth of the spanwise integral length scale with the distance from the wall. At the bottom end of this ridge ($y^+\approx 10$ and $\lambda_z^+\approx 100$), we find a local maximum with the spanwise spacing of the near-wall streaks (\citealp{kline67}). At the top end ($y=0.1h$ and $\lambda_z=1h$), we find another local maximum whose spanwise wavelength is reminiscent of the spanwise length scale of large- and very-large-scale motions (\citealp{kovasznay70,delalamo03}). The existence of the aforementioned linearly scaling ridge connecting the two local maxima has been understood as a refutation of the attached eddy hypothesis (\citealp{townsend76}). It was also shown that the size of each of the attached eddies is characterised by its spanwise length scale, as was demonstrated in \cite{hwang15}. Similarly, the spanwise wavenumber spectra of wall-normal velocity are found to be aligned along a linear ridge $y \approx 0.35 \lambda_z$ (figure \ref{fig:zspectra}).

As indicated earlier for the statistics, the QL case shows an increased energy in some part of the streamwise velocity spectra compared to the LES case. This is particularly notable at $\lambda_z^+ \approx 300$ $(\lambda_z/h \approx 0.2)$, while the spectra at the other wavelengths show reduced energy (figure \ref{fig:zspectra}c). On the contrary, the energy in the spanwise wavenumber spectra of wall-normal velocity is found to be decreased at most of the spanwise wavelengths, and little energy is seen for $\lambda_z^+ \lesssim 100$ (figure \ref{fig:zspectra}d). The reduced energy in the spectra of both streamwise and wall-normal velocities for $\lambda_z^+ \lesssim 100$ implies that there will be a signficant reduction in the Reynolds shear stress, since the Reynolds shear stress is simply a correlation between the streamwise and wall-normal velocities. This explains the lack of Reynolds shear stress of the QL case in the near-wall region (figure \ref{fig:stat}d). As more streamwise Fourier modes are included in the $\mathcal{P}_l$-subspace group (i.e. GQL1, GQL5 and GQL25), the peak location in the streamwise velocity spectra gradually moves towards $(\lambda_z^+,y^+) \approx (100,10)$ along the linear ridge $y=0.1\lambda_z$, while the outer part of the spectra also becomes more energetic. In the case of the wall-normal velocity spectra, this leads the spectra to span more towards the wall along the linear ridge $y=0.35 \lambda_z$. Finally, the spectra of the GQL25 are found to be fairly similar to those of the reference LES.

\begin{figure}
\vspace{-0.8cm}
\begin{minipage}{\textwidth}
\centering
\begin{subfigure}[b]{0.42\textwidth}
  \includegraphics[width=\textwidth]{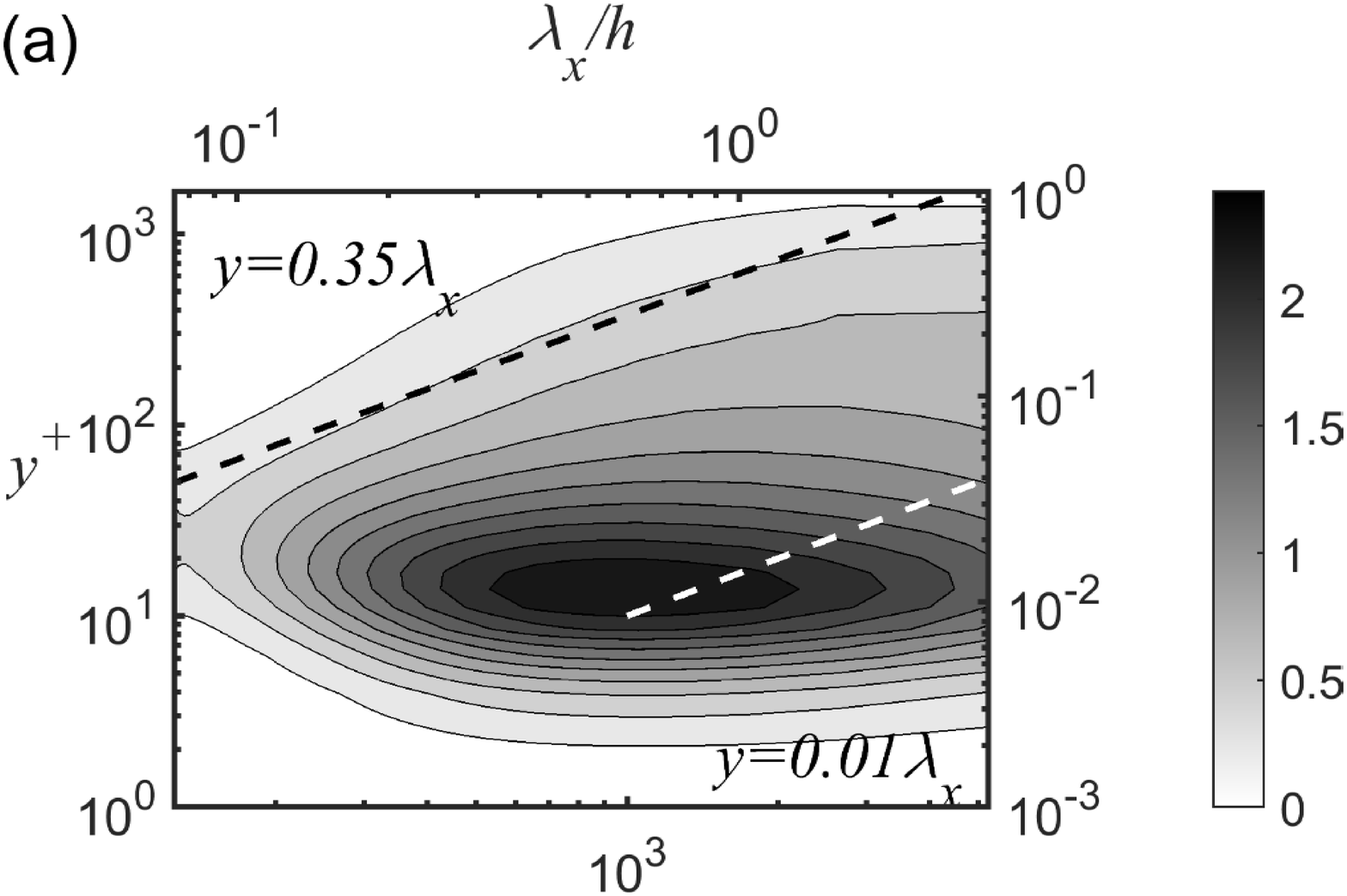}
\label{1}
\vspace{-0.8cm}
\end{subfigure}
\begin{subfigure}[b]{0.42\textwidth}
  \includegraphics[width=\textwidth]{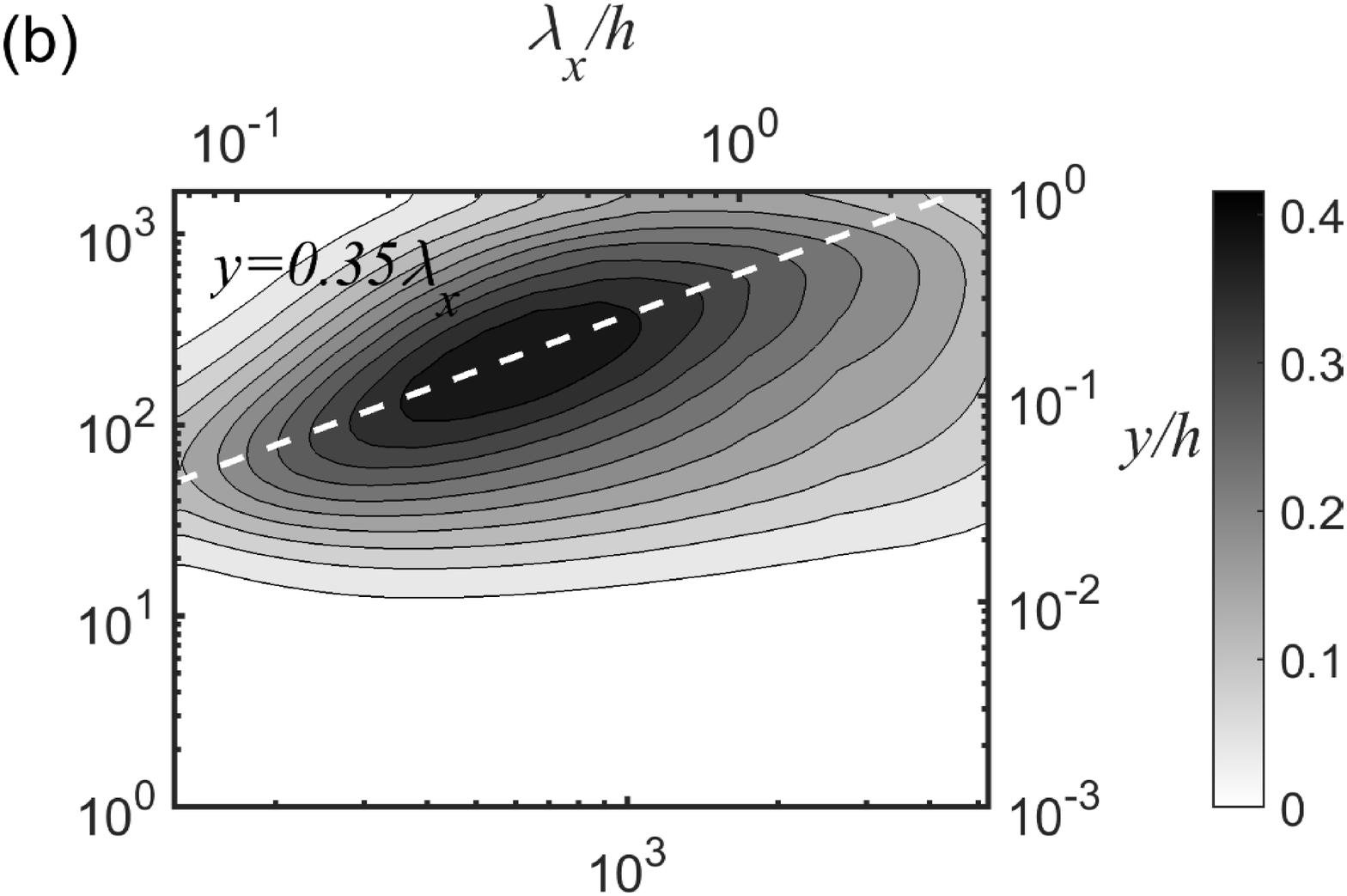}
\label{2}
\vspace{-0.8cm}
\end{subfigure}
\vspace{-0.8cm}
\begin{subfigure}[b]{0.42\textwidth}
  \includegraphics[width=\textwidth]{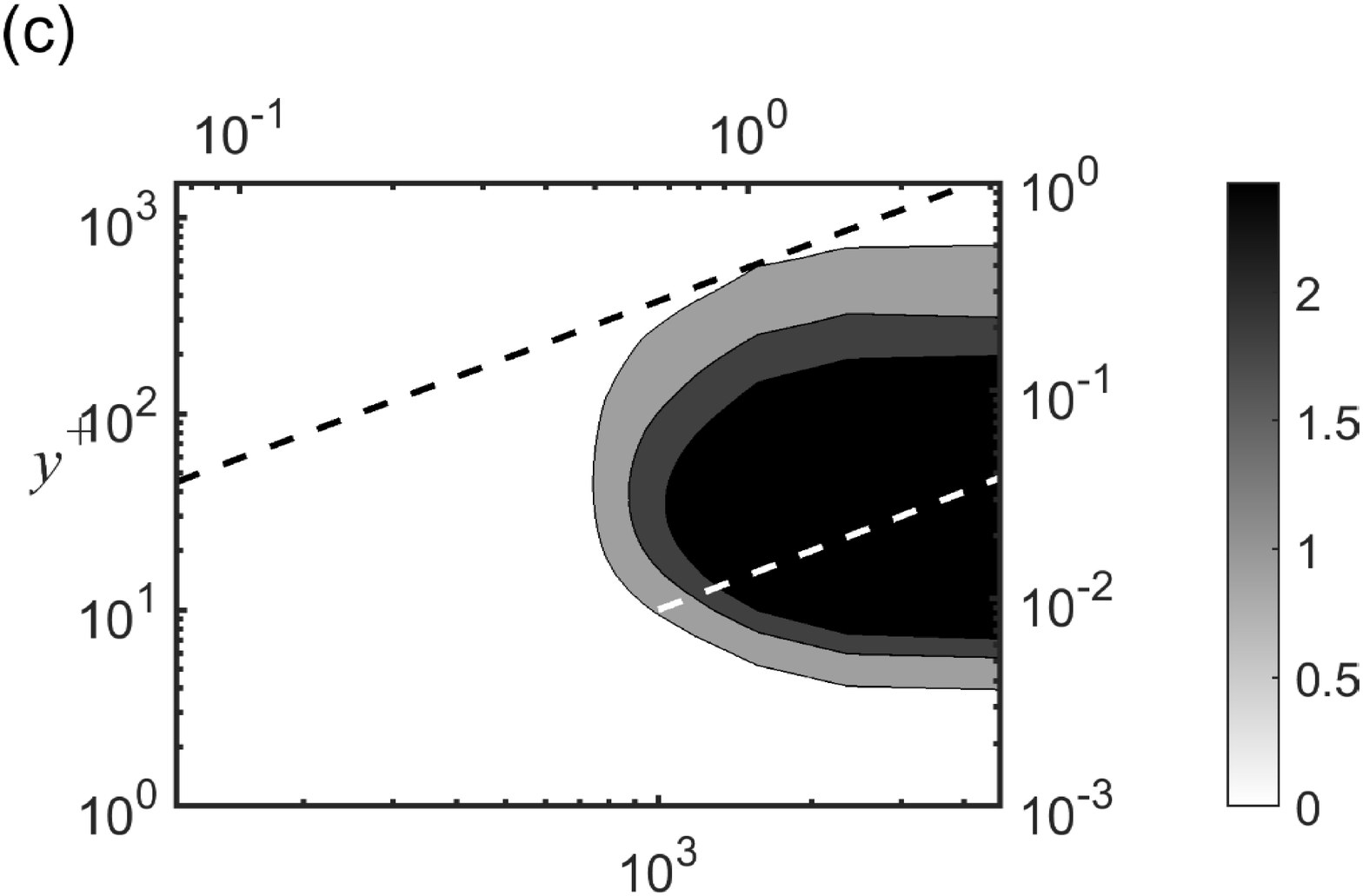}
  \label{3}
\end{subfigure}
\begin{subfigure}[b]{0.42\textwidth}
  \includegraphics[width=\textwidth]{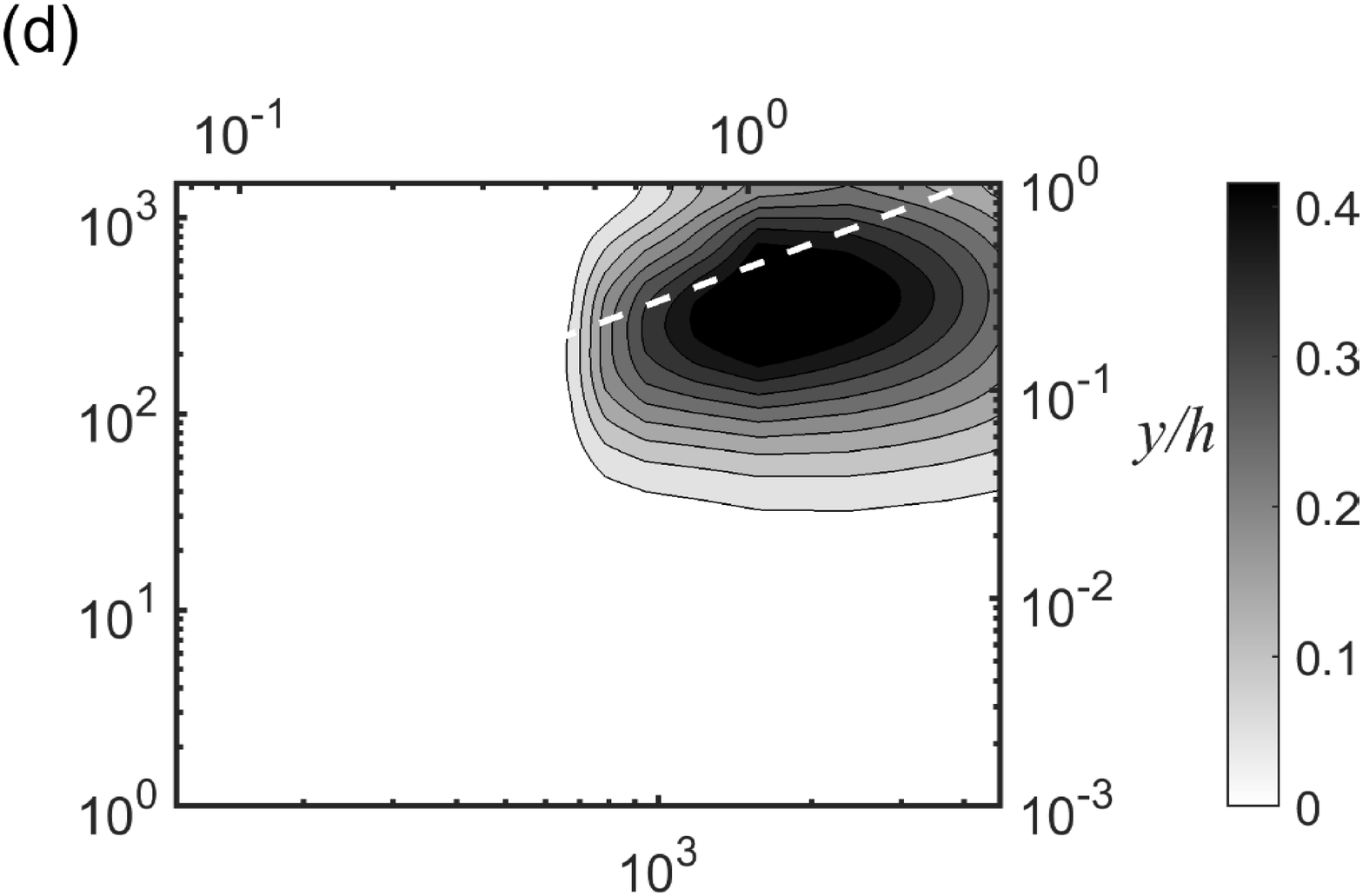}
  \label{4}
\end{subfigure}
\vspace{-0.8cm}
\begin{subfigure}[b]{0.42\textwidth}
  \includegraphics[width=\textwidth]{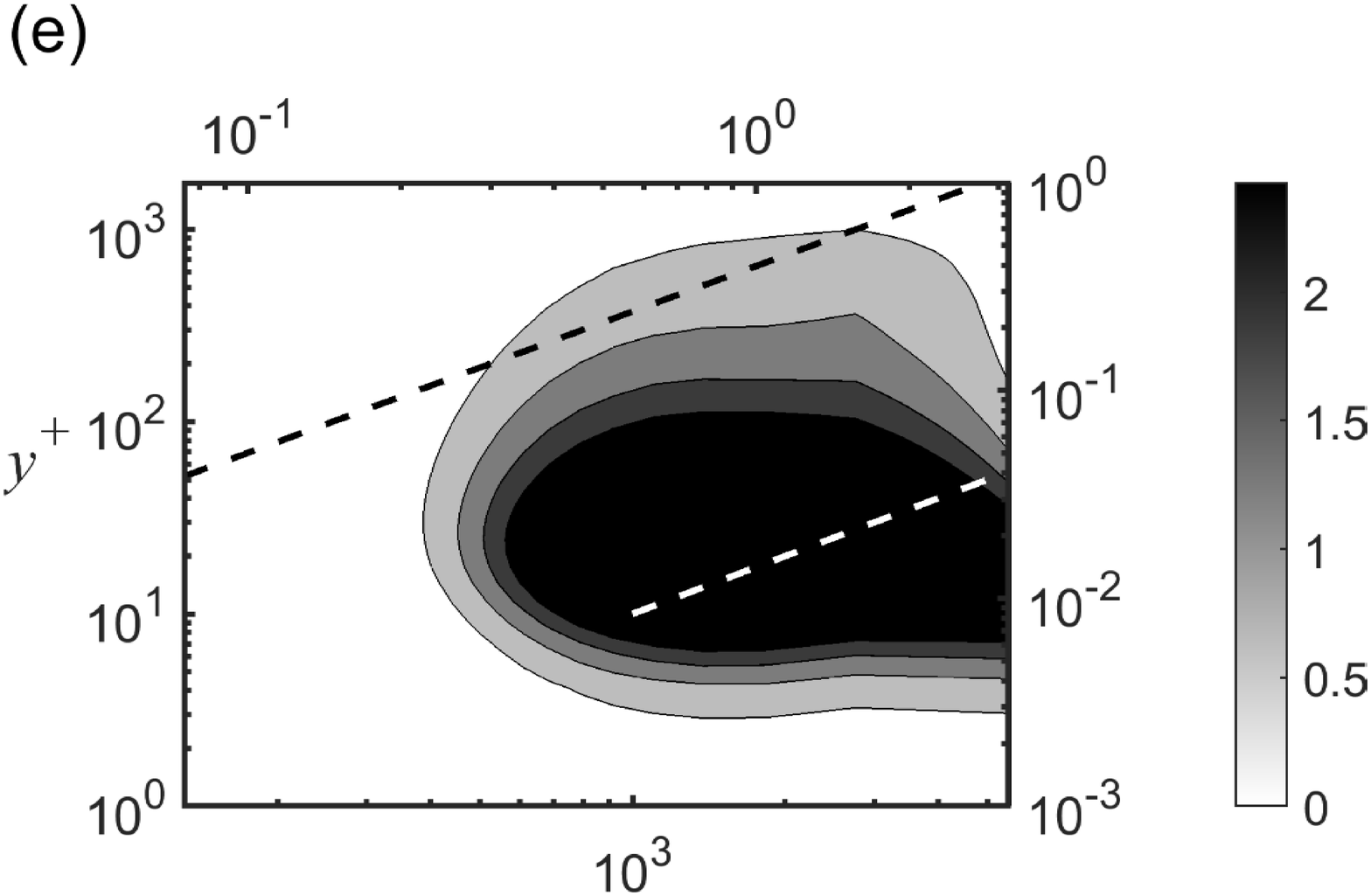}
  \label{5}
\end{subfigure}
\begin{subfigure}[b]{0.42\textwidth}
  \includegraphics[width=\textwidth]{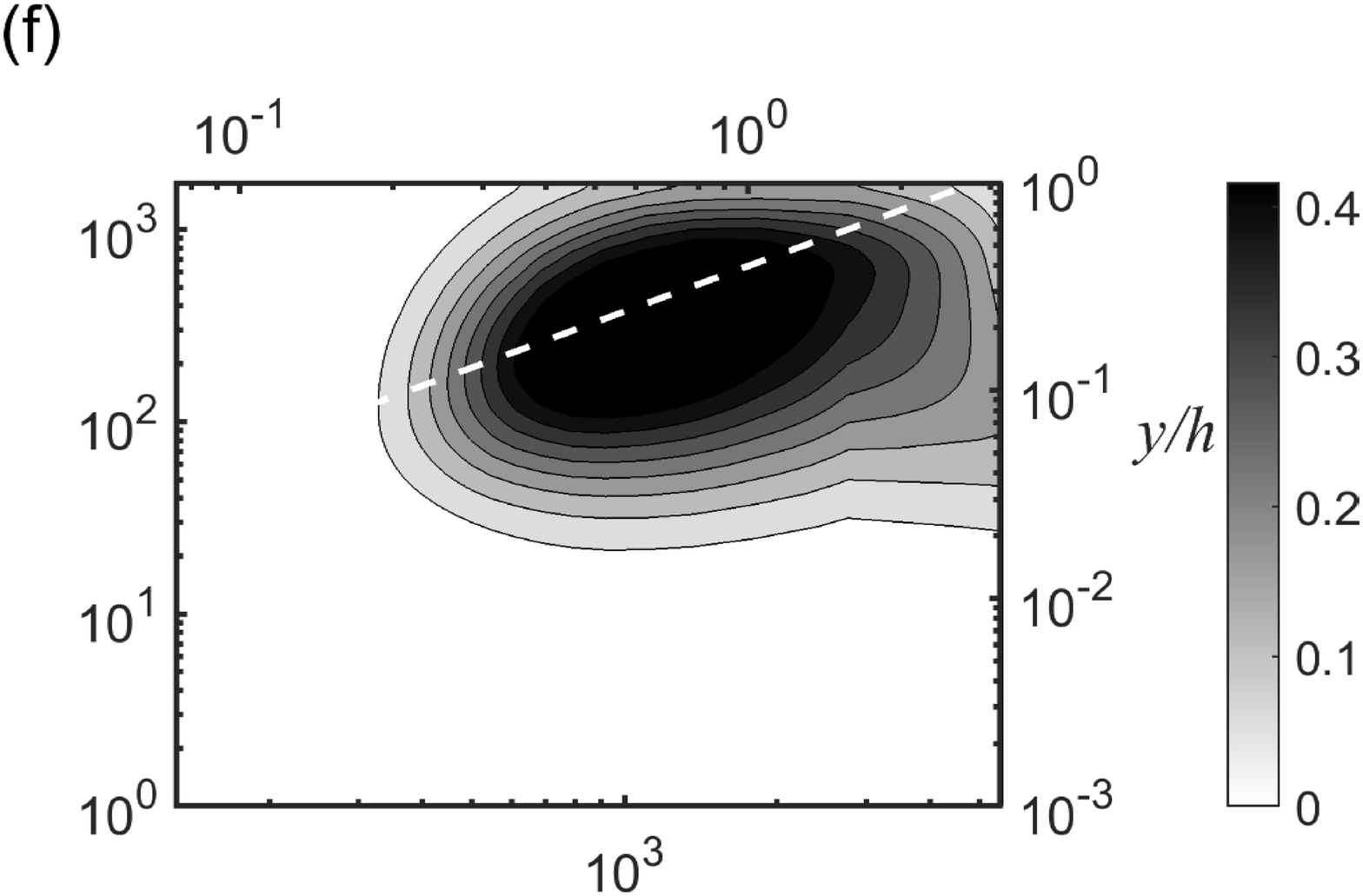}
  \label{6}
\end{subfigure}
\vspace{-0.8cm}
\begin{subfigure}[b]{0.42\textwidth}
  \includegraphics[width=\textwidth]{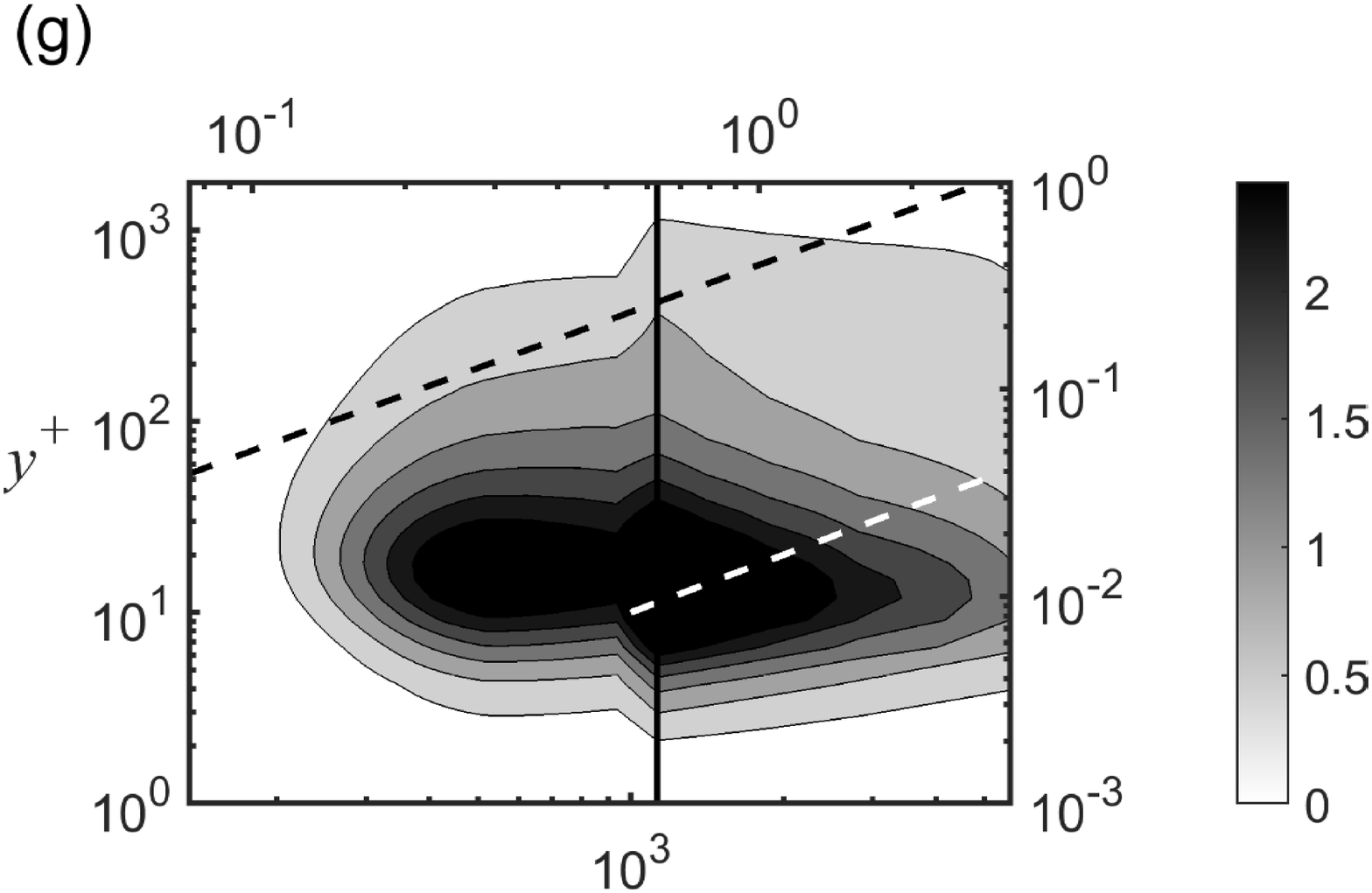}
  \label{5}
\end{subfigure}
\begin{subfigure}[b]{0.42\textwidth}
  \includegraphics[width=\textwidth]{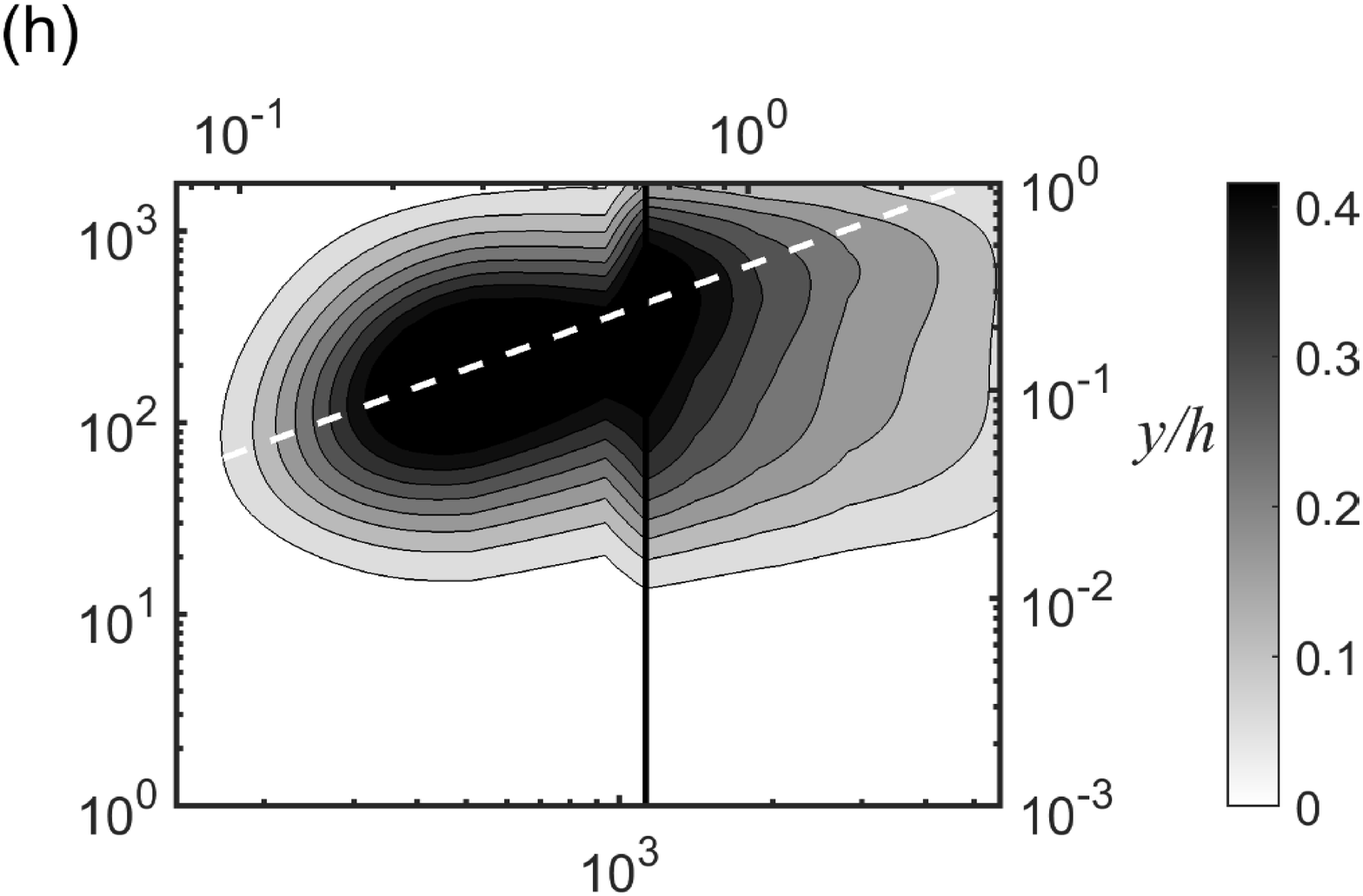}
  \label{6}
\end{subfigure}
\begin{subfigure}[b]{0.42\textwidth}
  \includegraphics[width=\textwidth]{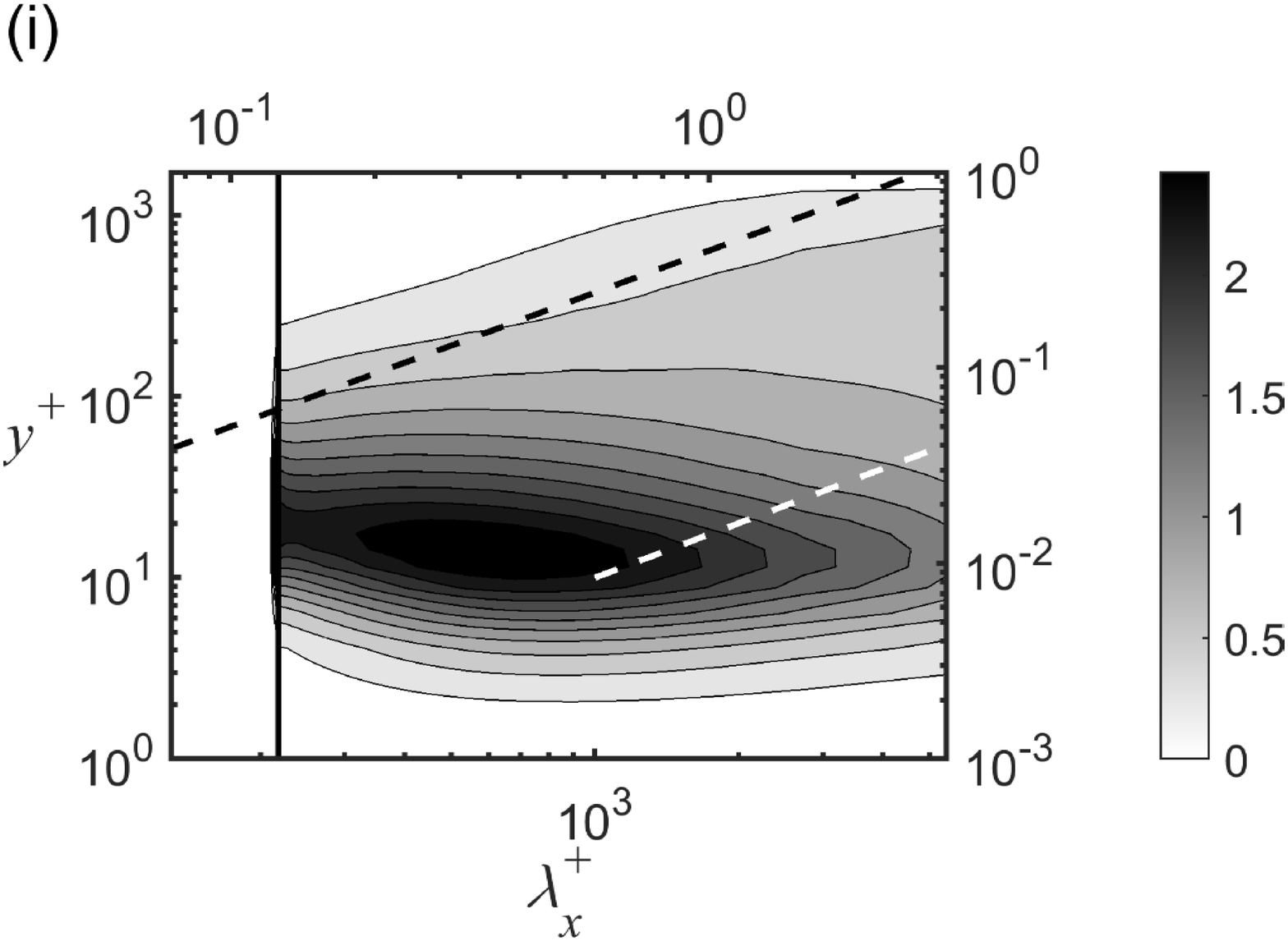}
\end{subfigure}
\begin{subfigure}[b]{0.42\textwidth}
  \includegraphics[width=\textwidth]{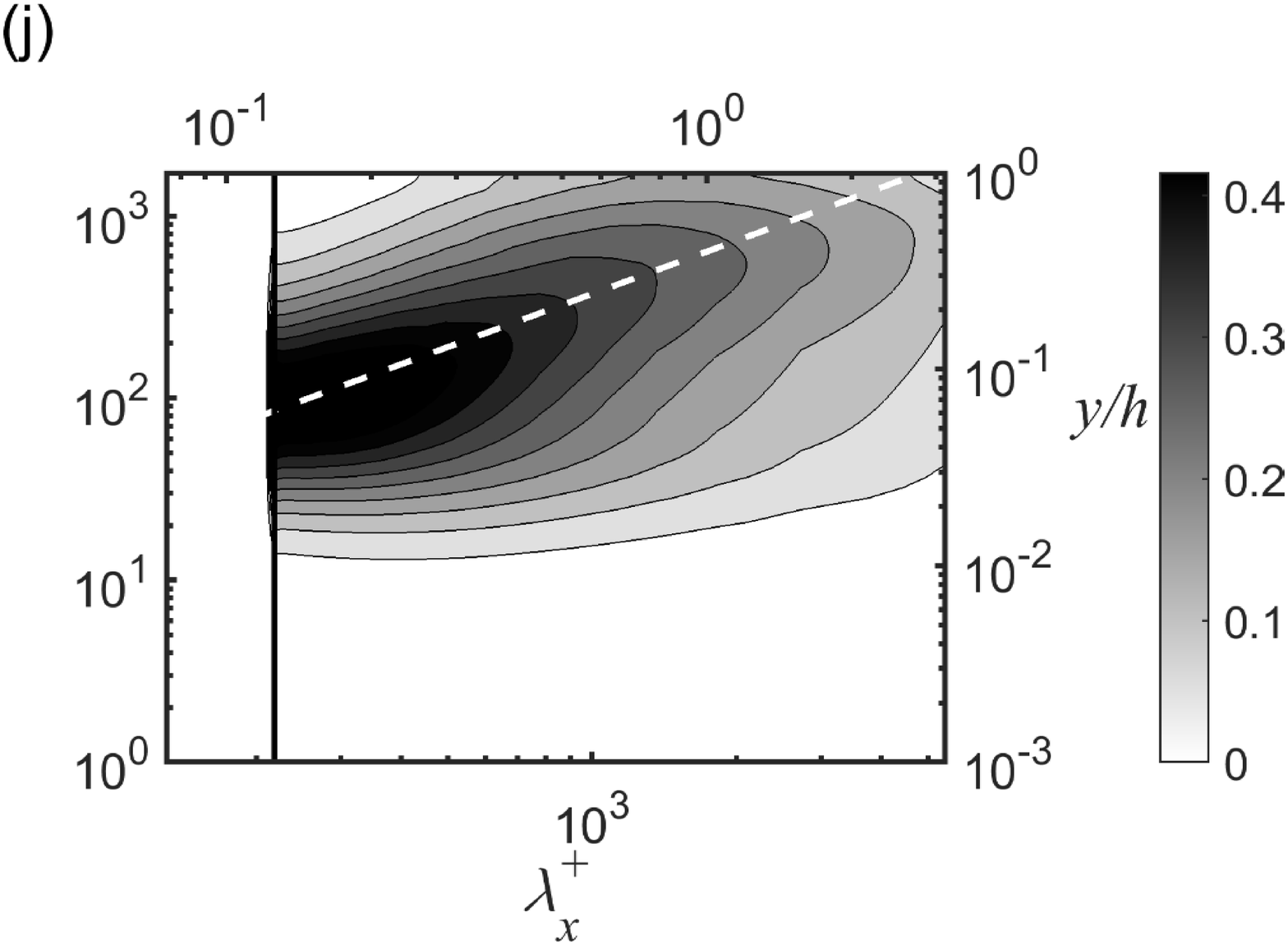}
\end{subfigure}
\end{minipage}
\caption{Premultiplied streamwise wavenumber spectra of streamwise $k_x^+ \Phi_{uu}^+(y^+,\lambda_x^+)$ (left column) and wall-normal $k_x^+ \Phi_{vv}^+(y^+,\lambda_x^+)$ (right column) velocity for (a,b) LES, (c,d) QL, (e,f) GQL1, (g,h) GQL5 and (i,j) GQL25 cases. Here, the vertical line represents the streamwise cut-off wavelength ($\lambda_{x,c}$) dividing the $\mathcal{P}_h$- (left) and $\mathcal{P}_l$-subspace  (right) regions.}
\label{fig:xspectra}
\end{figure}

The premultiplied streamwise wavenumber spectra of streamwise and wall-normal velocities are shown in figure \ref{fig:xspectra} for the LES and GQL cases. The streamwise wavenumber spectra of the LES case also show the typical features of energy-containing motions in turbulent channel flow. \cite{hwang15} showed that the energy-containing motions at a given spanwise length scale are composed of two components: a long streaky structure mainly carrying streamwise turbulent kinetic energy and a short and tall vortex packet carrying turbulent kinetic energy at all the velocity components. This bimodal structure guides the observation of the different features of the spectra: the linear ridges $y \approx 0.35 \lambda_x$ (upper line) and $y \approx 0.01 \lambda_x$ (lower line) are plotted in figure \ref{fig:xspectra}. The streamwise velocity spectra appear to be very energetic along $y \approx 0.01 \lambda_x$, the ridge corresponding to the aforementioned long streaky structure (figure \ref{fig:xspectra}a), although this behaviour is not very clearly seen for large $\lambda_x$ due to the small computational domain of the present simulations in the streamwise direction -- note that, in this case, a significant amount of spectral intensity is carried by zero streamwise wavenumbers due to the streaks extending over the entire streamwise domain \cite[i.e. $\lambda_x = \infty$; see][for the cases with a sufficiently large streamwise domain]{hwang15}. The spectra also show a non-negligible amount of energy along $y \approx 0.35 \lambda_x$, corresponding to the vortex packet (\citealp{hwang15}), and the wall-normal velocity spectra are very well aligned with this ridge.

Compared to the reference LES, the QL case shows that the energy in both streamwise and wall-normal velocity spectra is concentrated in the streamwise Fourier modes at $\lambda_x^+ \gtrsim 700$ and exhibits poor correlation with the two linear ridges, $y \approx 0.35 \lambda_x$ and  $y \approx 0.01 \lambda_x$, as was shown in \cite{farrell16}. This behaviour is also consistent with the previous observations \cite[e.g.][]{thomas14,thomas15,farrell16,tobias17,hernandez}, where only a reasonably small number of the streamwise Fourier modes are active in the QL case. It has been claimed that this feature is the basis for the significantly reduced computational cost of the QL model, as many Fourier modes would not be needed for the QL model in the streamwise direction. However, it is evident that this feature could conversely become a non-trivial limitation of the QL model especially at high Reynolds numbers, as it destroys the fundamental scaling behaviour of the streamwise wavenumber spectra in the logarithmic region. As the streamwise Fourier modes are further incorporated into the $\mathcal{P}_l$-subspace group through the GQL approximation (figures \ref{fig:xspectra}e-j; i.e. GQL1, GQL5 and GQL25), the two spectra begin to show more energy at smaller $\lambda_x$ along the linear ridges. As a consequence, in the GQL1 and GQL5 cases, the near-wall region is better resolved and the spectra reach out to smaller scales down to $\lambda_x^+ \approx 200$. The spectra of the GQL25 case greatly resemble those of the LES case but for a detail. It is interesting to note that the spectra do not extend below $\lambda_{x,c}$ in this case, implying that the velocity field in the $\mathcal{P}_h$-subspace group yields the trivial solution. This issue has been found to be intricately linked to the nature of the GQL approximation, and it will be discussed in detail in \S\ref{sec:sec43}.


\subsection{Spectral energy transfer}\label{sec:sec32}
\begin{figure}
\begin{minipage}{\textwidth}
\centering
\begin{subfigure}[b]{0.42\textwidth}
  \includegraphics[width=\textwidth]{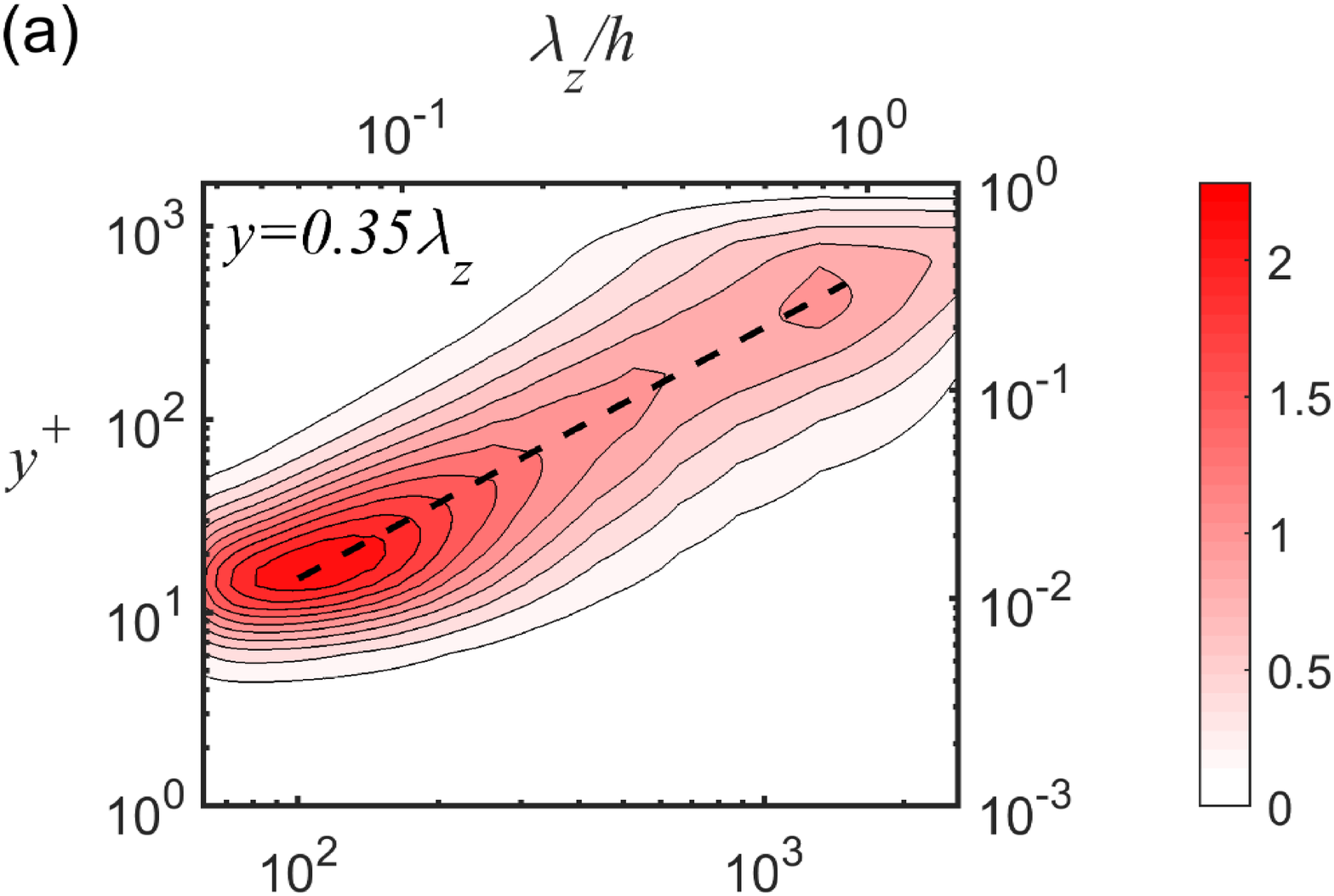}
\label{1}
\end{subfigure}
\vspace{-0.7cm}
\begin{subfigure}[b]{0.42\textwidth}
  \includegraphics[width=\textwidth]{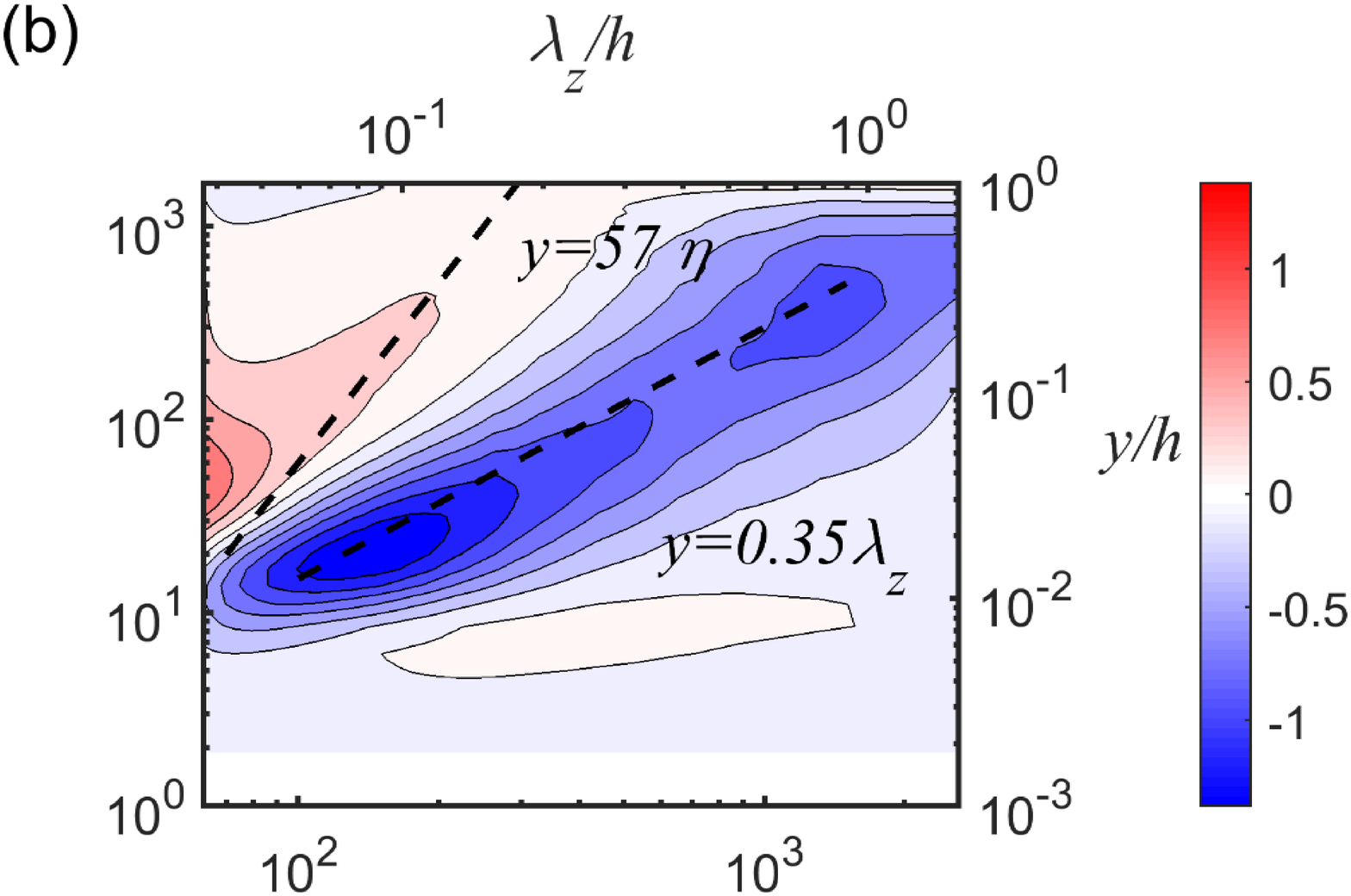}
\label{2}
\end{subfigure}
\begin{subfigure}[b]{0.42\textwidth}
  \includegraphics[width=\textwidth]{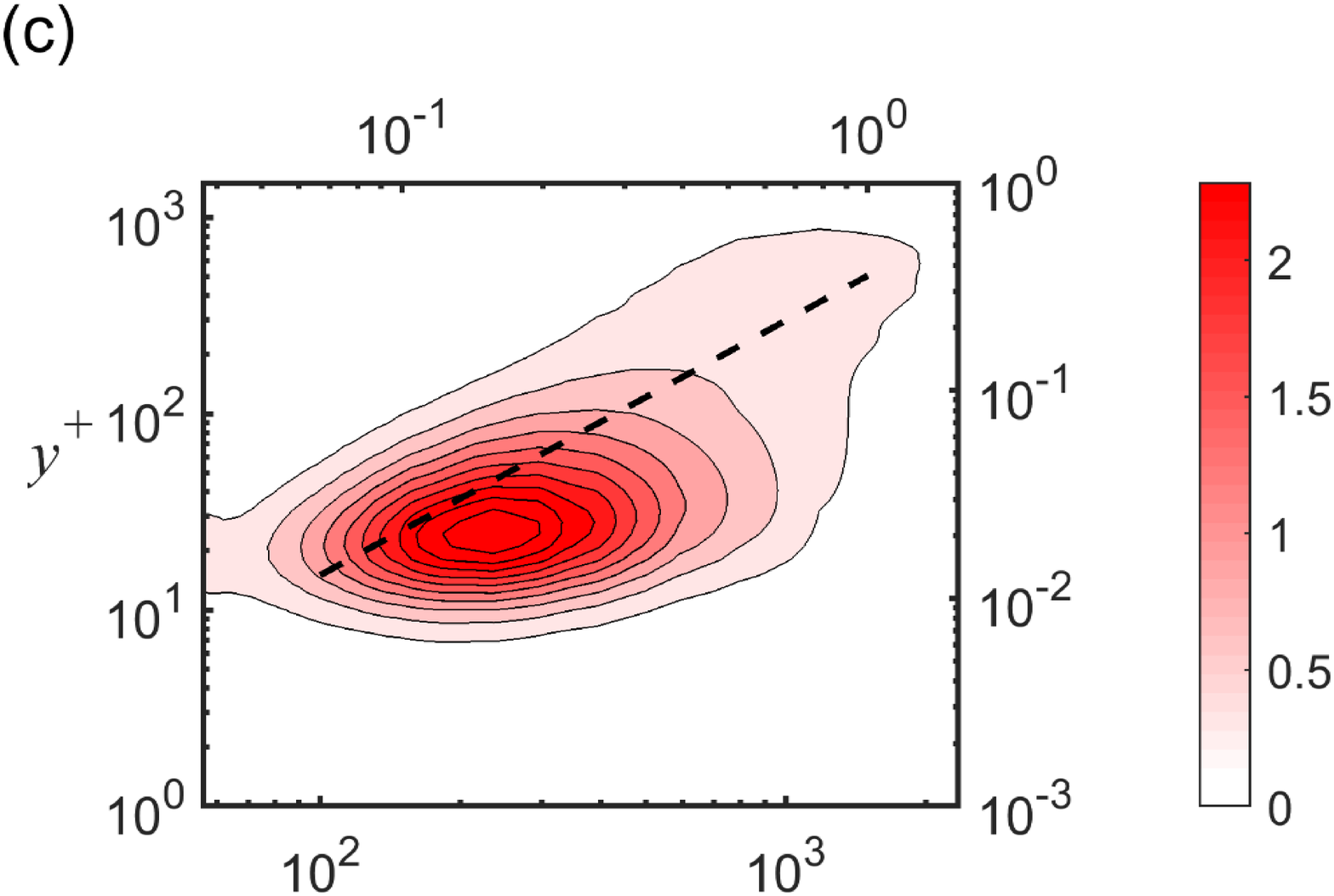}
  \label{3}
\end{subfigure}
\vspace{-0.7cm}
\begin{subfigure}[b]{0.42\textwidth}
  \includegraphics[width=\textwidth]{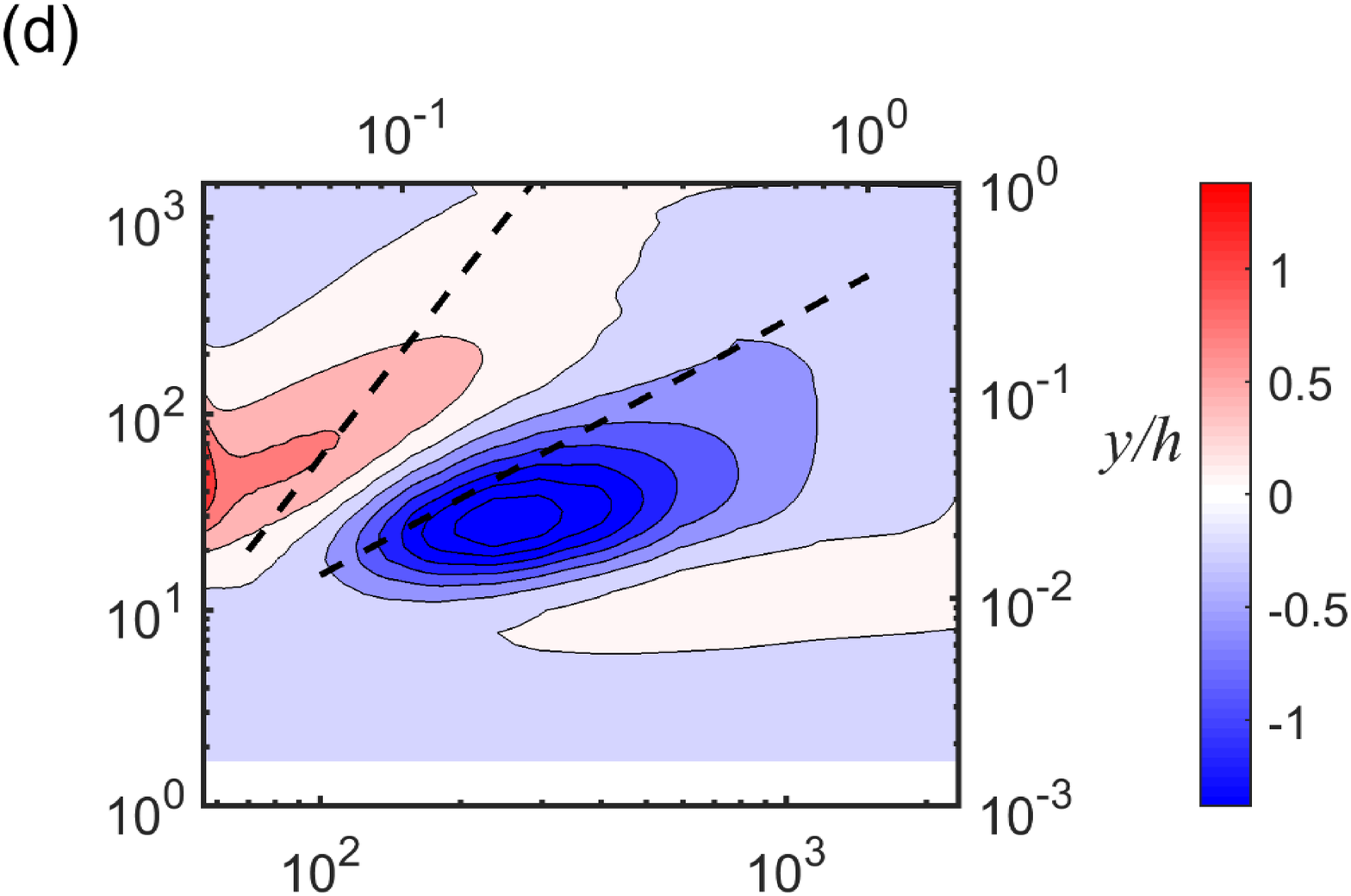}
  \label{4}
\end{subfigure}
\begin{subfigure}[b]{0.42\textwidth}
  \includegraphics[width=\textwidth]{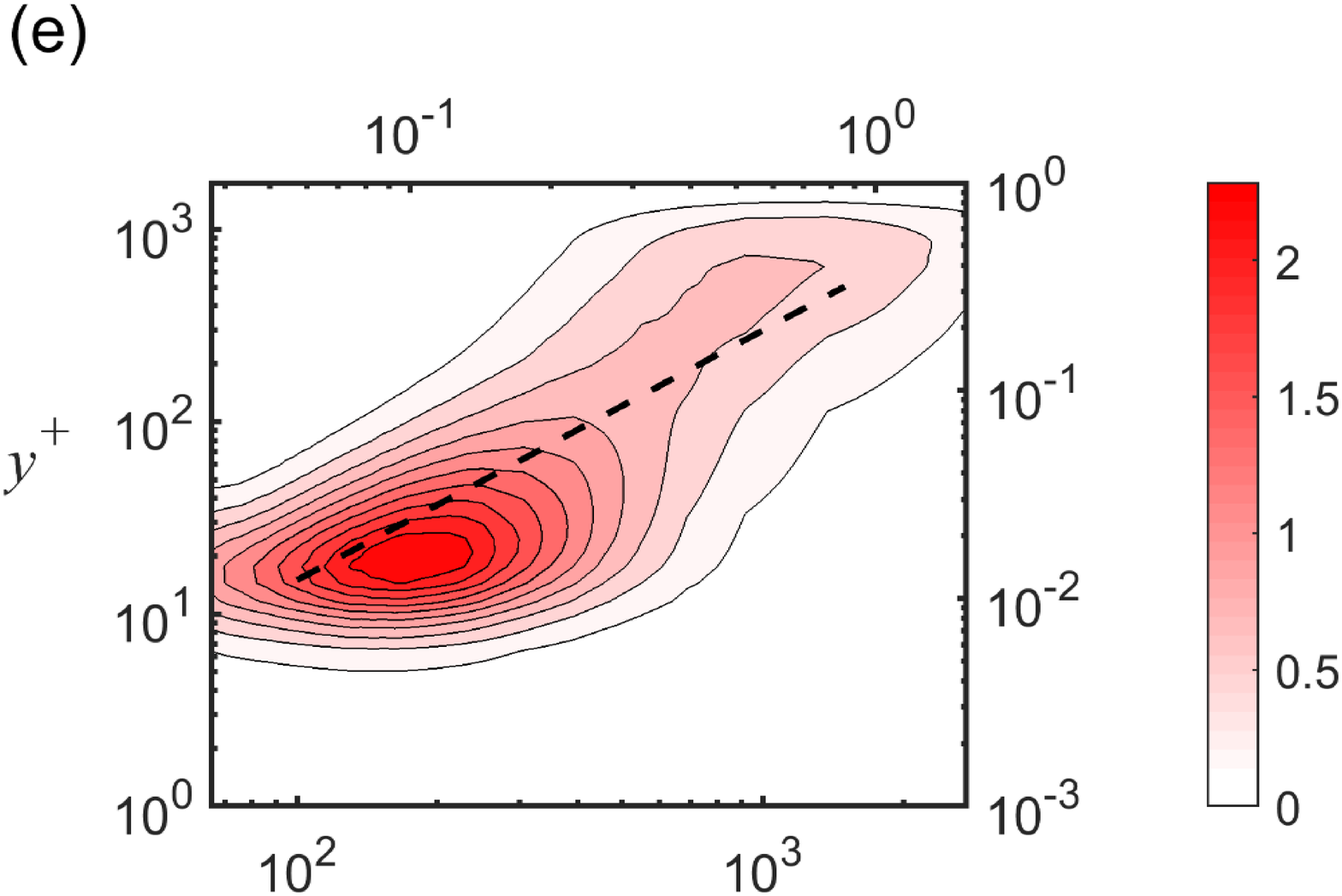}
  \label{5}
\end{subfigure}
\vspace{-0.7cm}
\begin{subfigure}[b]{0.42\textwidth}
  \includegraphics[width=\textwidth]{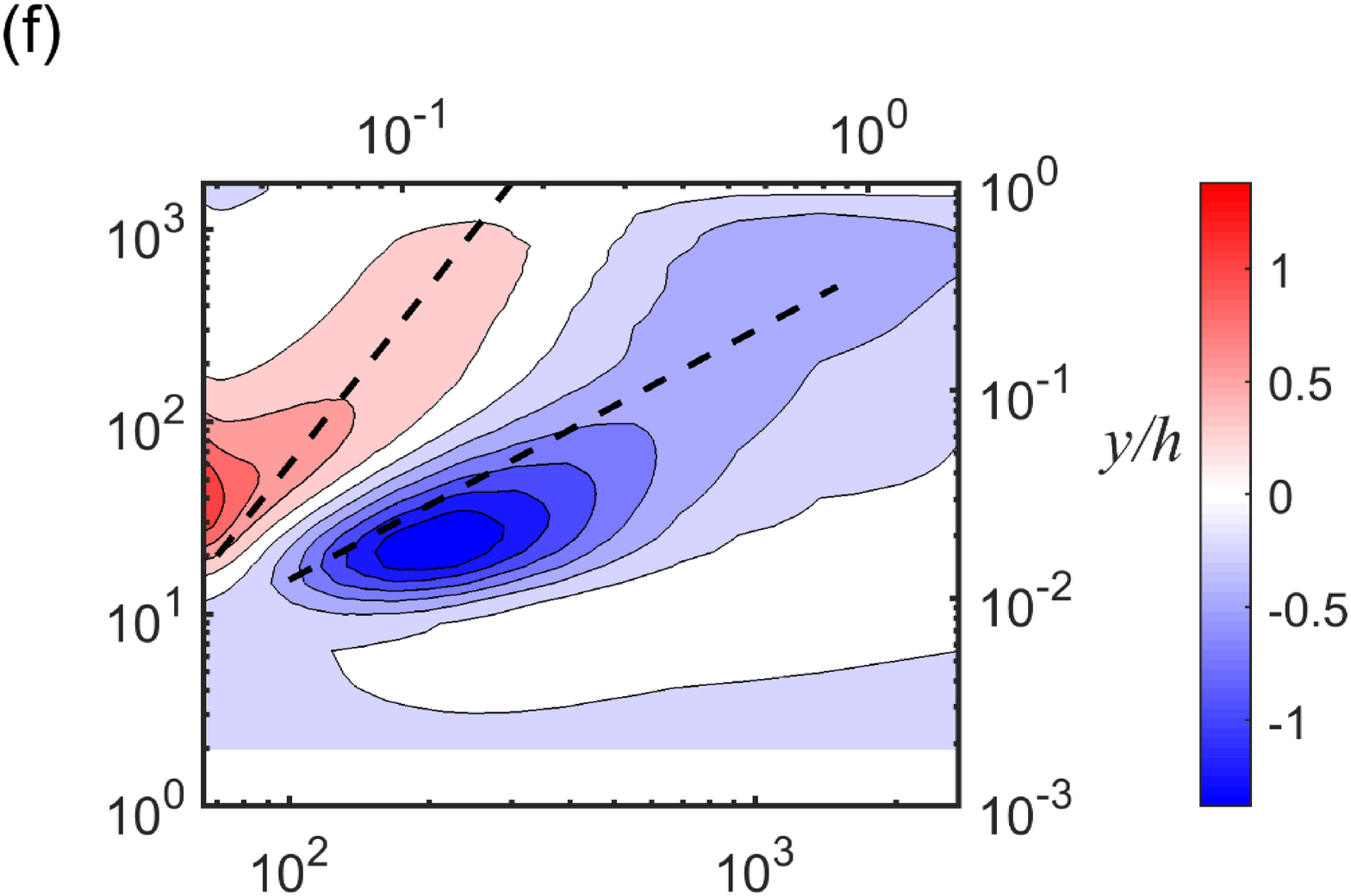}
  \label{6}
\end{subfigure}
\begin{subfigure}[b]{0.42\textwidth}
  \includegraphics[width=\textwidth]{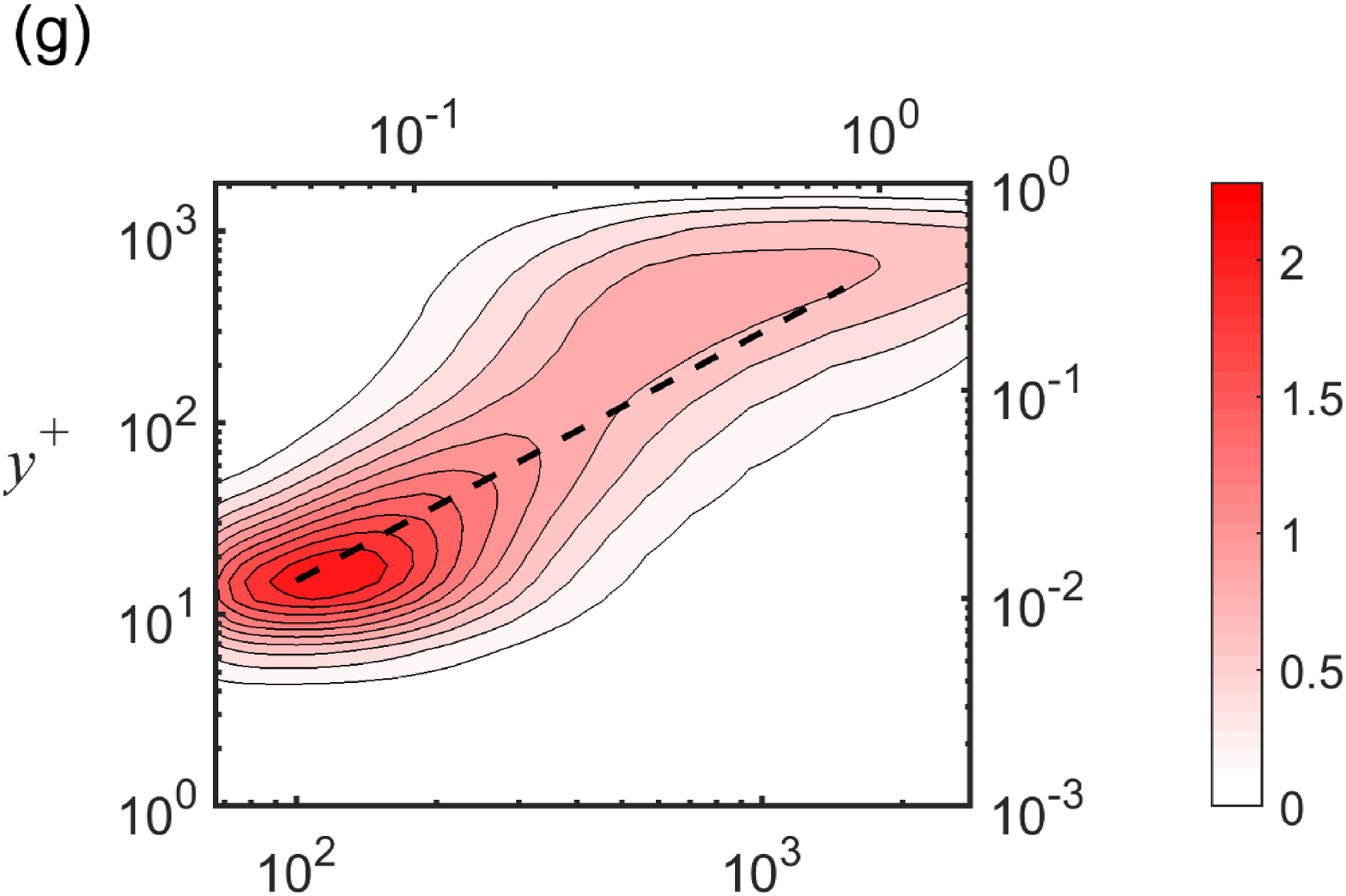}
  \label{5}
\end{subfigure}
\vspace{-0.7cm}
\begin{subfigure}[b]{0.42\textwidth}
  \includegraphics[width=\textwidth]{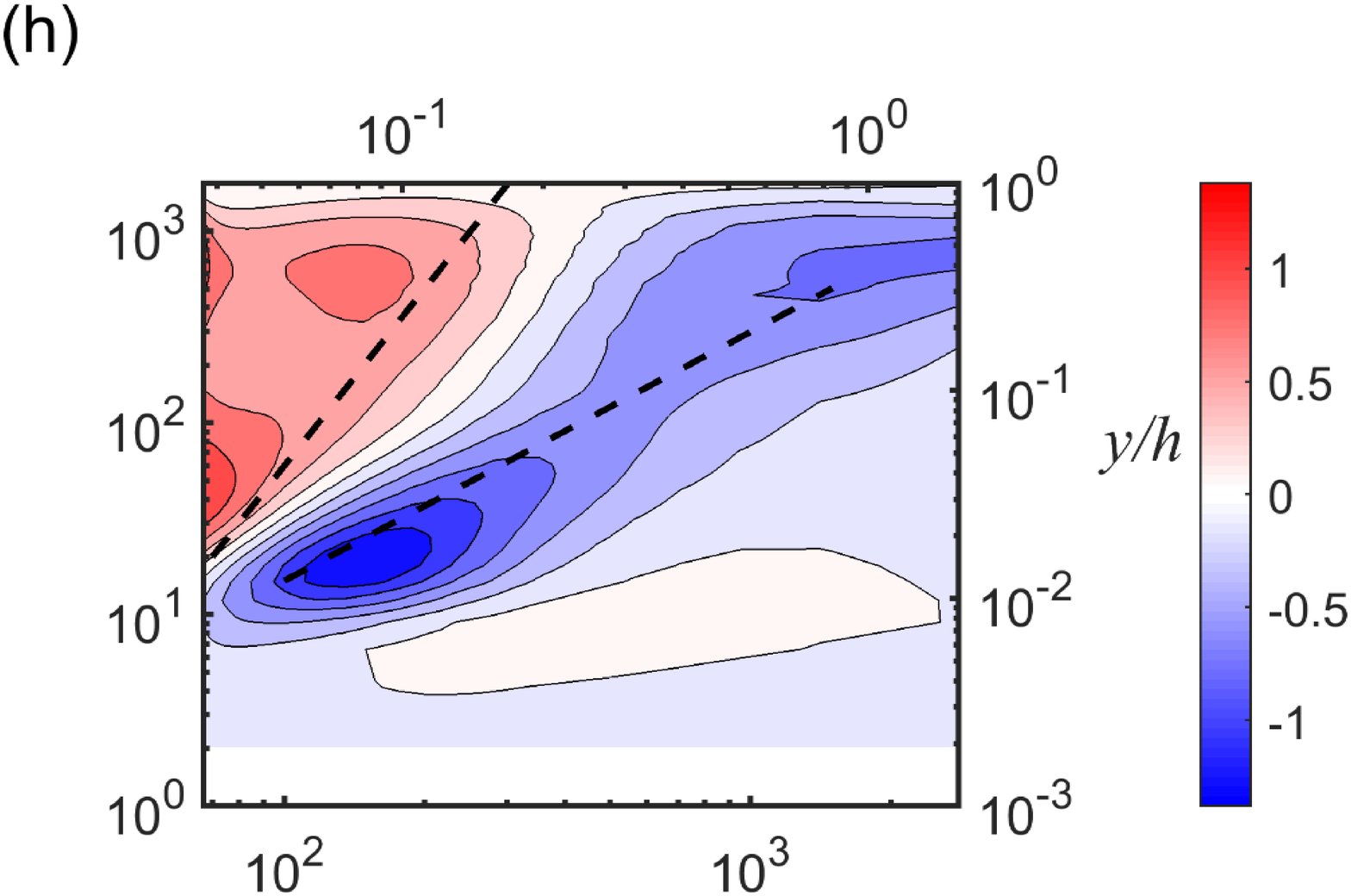}
  \label{6}
\end{subfigure}
\begin{subfigure}[b]{0.42\textwidth}
  \includegraphics[width=\textwidth]{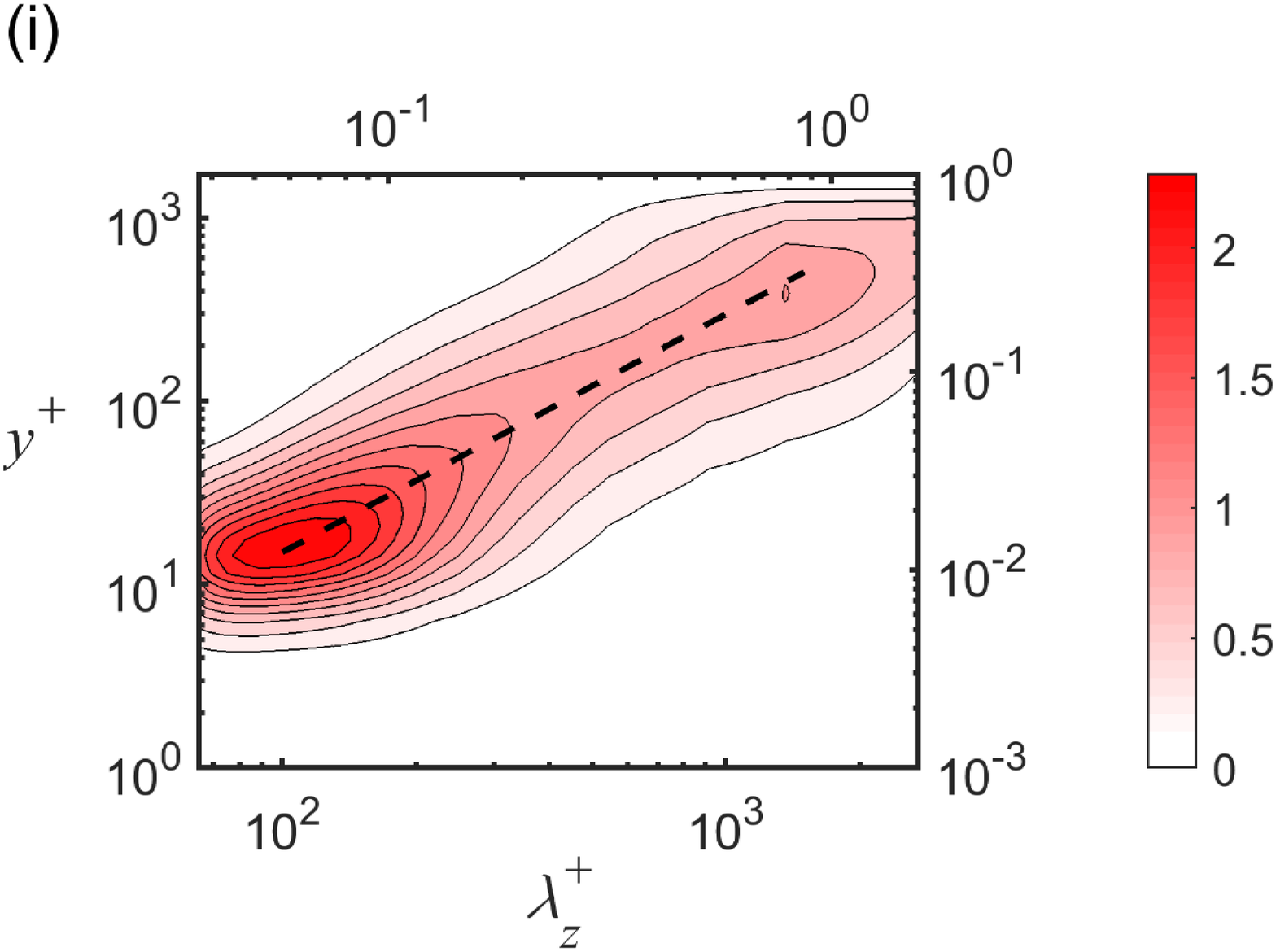}
\end{subfigure}
\begin{subfigure}[b]{0.42\textwidth}
  \includegraphics[width=\textwidth]{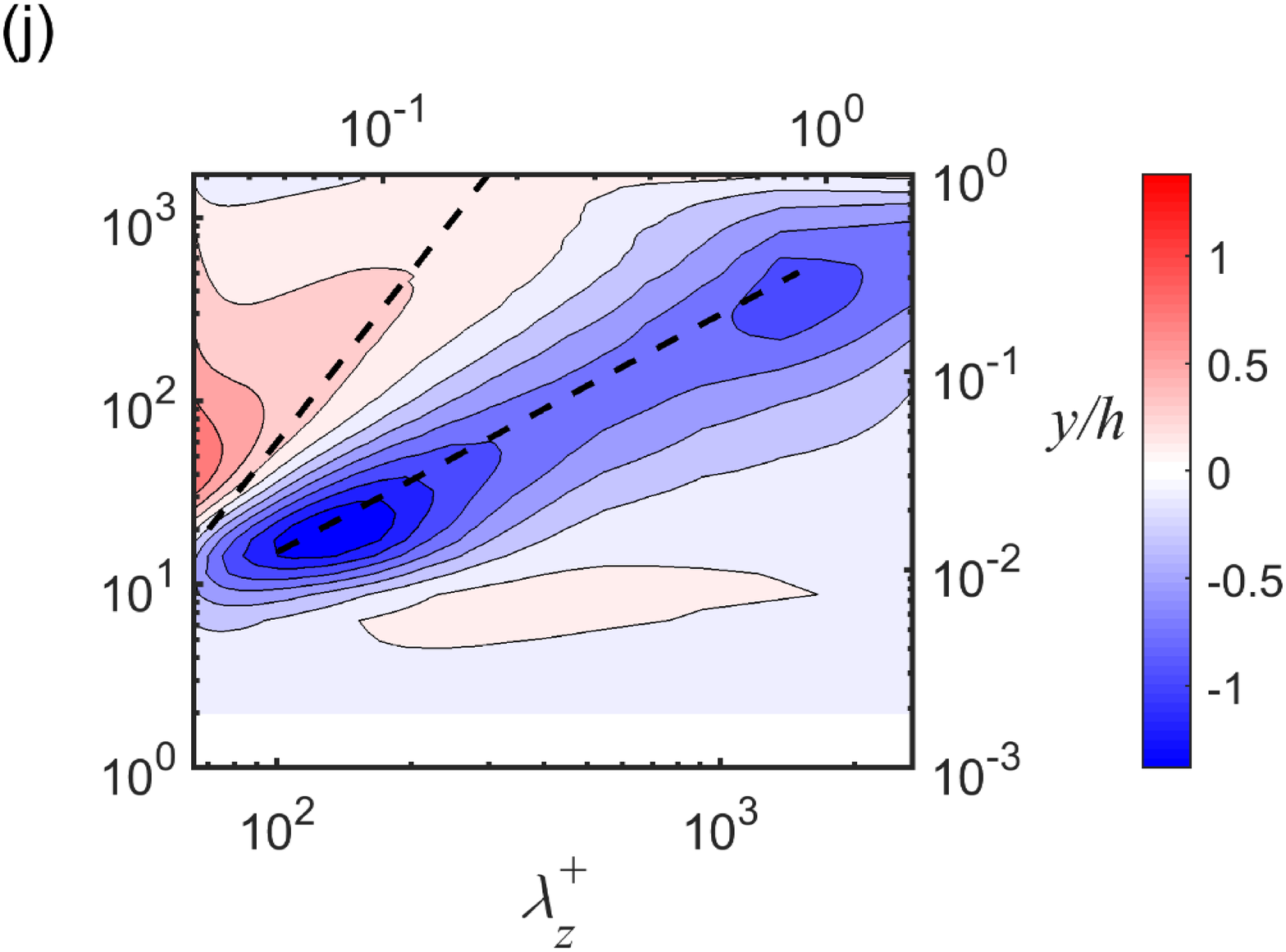}
\end{subfigure}
\end{minipage}
\caption{Premultiplied spanwise wavenumber spectra of production $k_z^+ y^+ \widehat{P}^+(y^+,\lambda_z^+)$ (left column) and turbulent transport $k_z^+ y^+ \widehat{T}_{turb}^+(y^+,\lambda_z^+)$ (right column) for (a,b) LES, c,d) QL, (e,f) GQL1,  (g,h) GQL5 and  (i,j) GQL25 cases.}
\label{fig:zenergy}
\end{figure}

Now, we study the spectral energy transfer in the LES, QL and GQL cases. The energy transfer in the spectral space has recently been examined in detail in the recent studies of \cite{cho18} and \cite{lee19}, where pressure and viscous transport spectra are shown to be negligibly small in the log and outer regions. Also, given that the present study is based on LES, the analysis in this section will be focused on the production and turbulent transport spectra only. The premultiplied one-dimensional spanwise wavenumber spectra of the production and turbulent transport of each case are plotted in figure \ref{fig:zenergy}. The spectra of the LES case (figures \ref{fig:zenergy}a,b) show the typical features of turbulent channel flow (\citealp{cho18}). The turbulent production spectra in figure \ref{fig:zenergy}(a) are almost uniformly distributed along the ridge $y \approx 0.35 \lambda_z$, especially over the range of the spanwise wavelength corresponding to the log layer ($300 \delta_\nu \lesssim \lambda_z \lesssim 1 h$, where $\delta_\nu = \nu/u_\tau$). The production also shows a spectral energy peak at $\lambda_z^+ = 100$ and $y^+ = 15$. The premultiplied turbulent transport spectra \ref{fig:zenergy}(b) show postive and negative regions due to the energy conservative nature of the nonlinear terms in the Navier-Stokes equations (\citealp{cho18}). The negative turbulent transport (blue contours in figure \ref{fig:zenergy}b) is almost balanced with the (positive) production in the logarithmic and outer layers, and the positive turbulent transport appears along a ridge indicating the Kolmogorov scale (i.e. $y \approx 57 \eta$ where $\eta$ is the Kolmogorov scale). Finally, the turbulent transport spectra are weakly positive in the region very close to the wall ($y^+<10$) over a wide range of the spanwise wavelength scales ($200 \delta_\nu \lesssim \lambda_z \lesssim 1 h$) and this is due to an inverse energy transfer from small to large scales \cite[]{cho18,doohan21}.


The spectra of the QL case show increased intensity and a production peak which has switched to $\lambda_z^+ \approx 300$. The range of spanwise wavelengths covered by the production spectra is further reduced, featuring little spectral intensity at small scales ($\lambda_z^+ \lesssim 100$) and a significant lack of it at large scales ($\lambda_z/h \gtrsim 1$). The linear scaling of $\lambda_z^+$ with $y^+$ seems to have almost disappeared. Furthermore, the region of negative turbulent transport has shrunk to spanwise wavelengths in the range of $\lambda_z^+ \approx 100-1000$. The near-wall region of weakly positive turbulent transport is found to grow to larger spanwise scales in the QL model, despite the lack of spectral intensity in the outer region. This suggests that the near-wall positive turbulent transport at large $\lambda_z$ in the original full LES case is at least not from the large-scale structures, since the QL model exhibits considerably weak energy and production spectra. Given the more energetic near-wall production of the QL model at $\lambda_z^+ \approx 300$, this behaviour is consistent with the explanation given by \cite{cho18}, who showed that this near-wall positive turbulent transport originate from smaller scale by visualising the related triadic wave interactions, but in contradiction to the argument in \cite{kawata21} that the spanwise interscale transport is mainly related to the individual dynamics of each scale. Once again, by allowing more streamwise mode to interact nonlinearly, the resulting spectra start to recover the original range of spanwise wavelengths, the scaling with $y$ and the original position of the peak in the GQL1, GQL5 and GQL25 cases. The spectra of the GQL25 case provide an excellent match with those of LES.

\begin{figure}
\begin{minipage}{\textwidth}
\centering
\begin{subfigure}[b]{0.42\textwidth}
  \includegraphics[width=\textwidth]{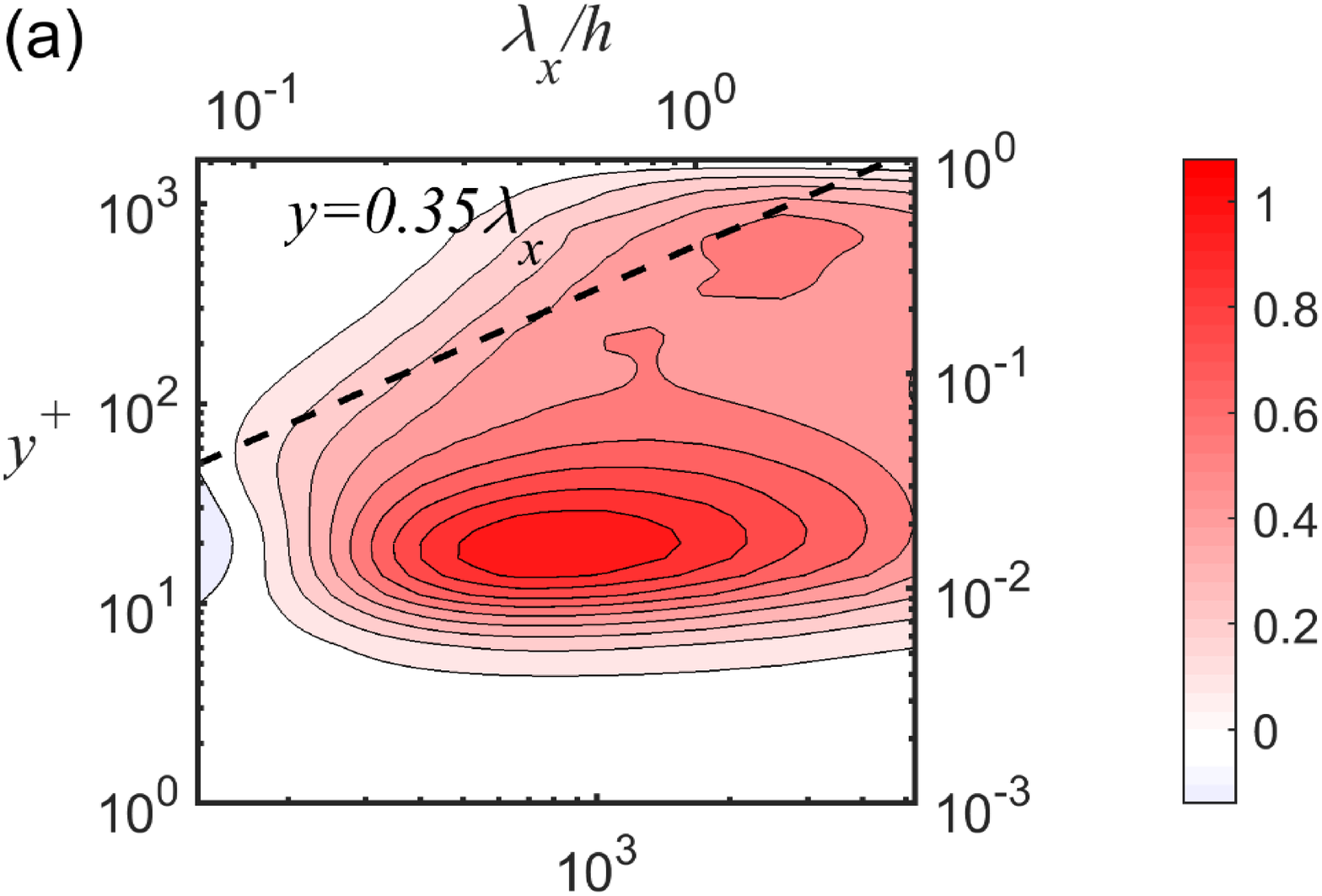}
\label{1}
\end{subfigure}
\vspace{-0.7cm}
\begin{subfigure}[b]{0.42\textwidth}
  \includegraphics[width=\textwidth]{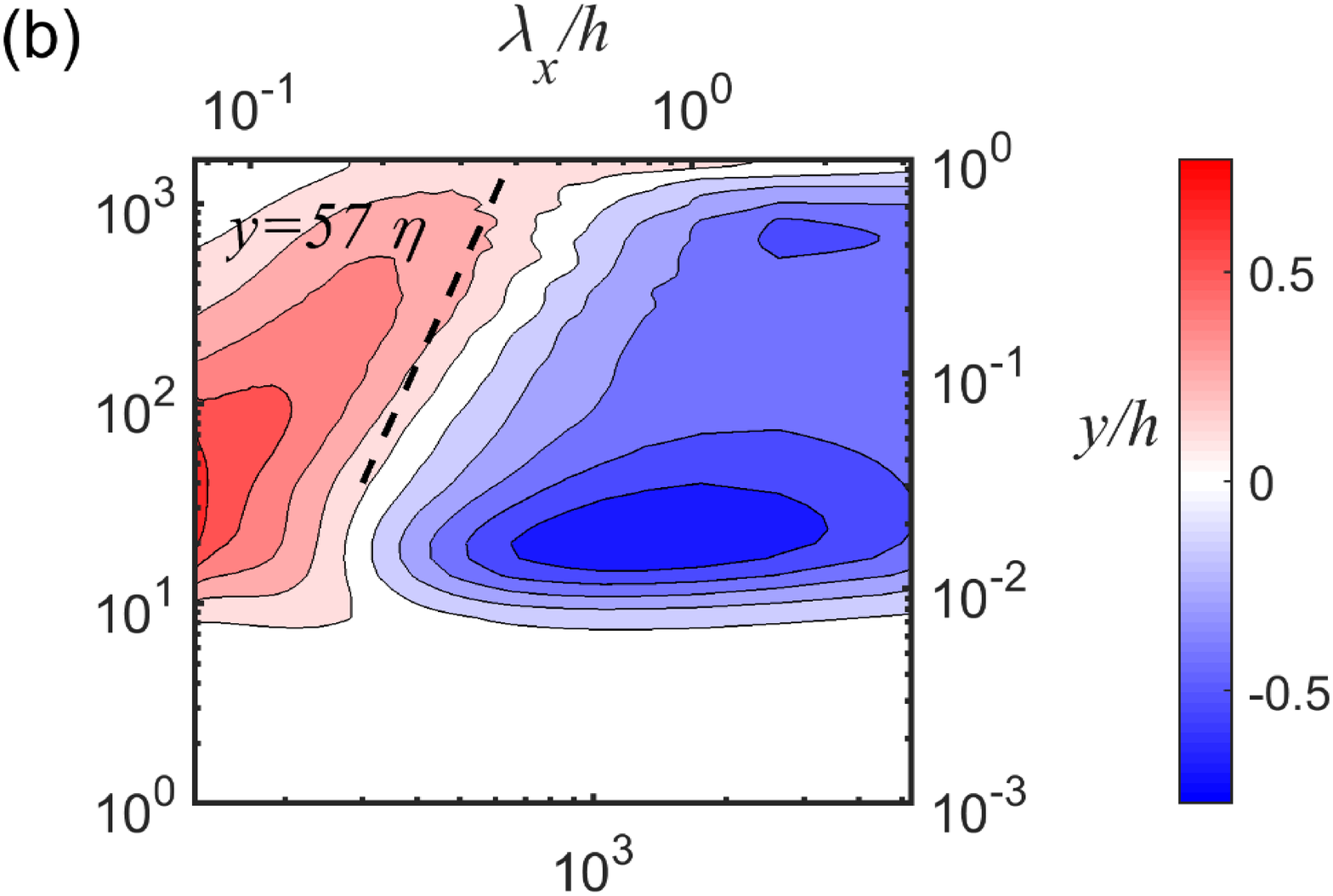}
\label{2}
\end{subfigure}
\begin{subfigure}[b]{0.42\textwidth}
  \includegraphics[width=\textwidth]{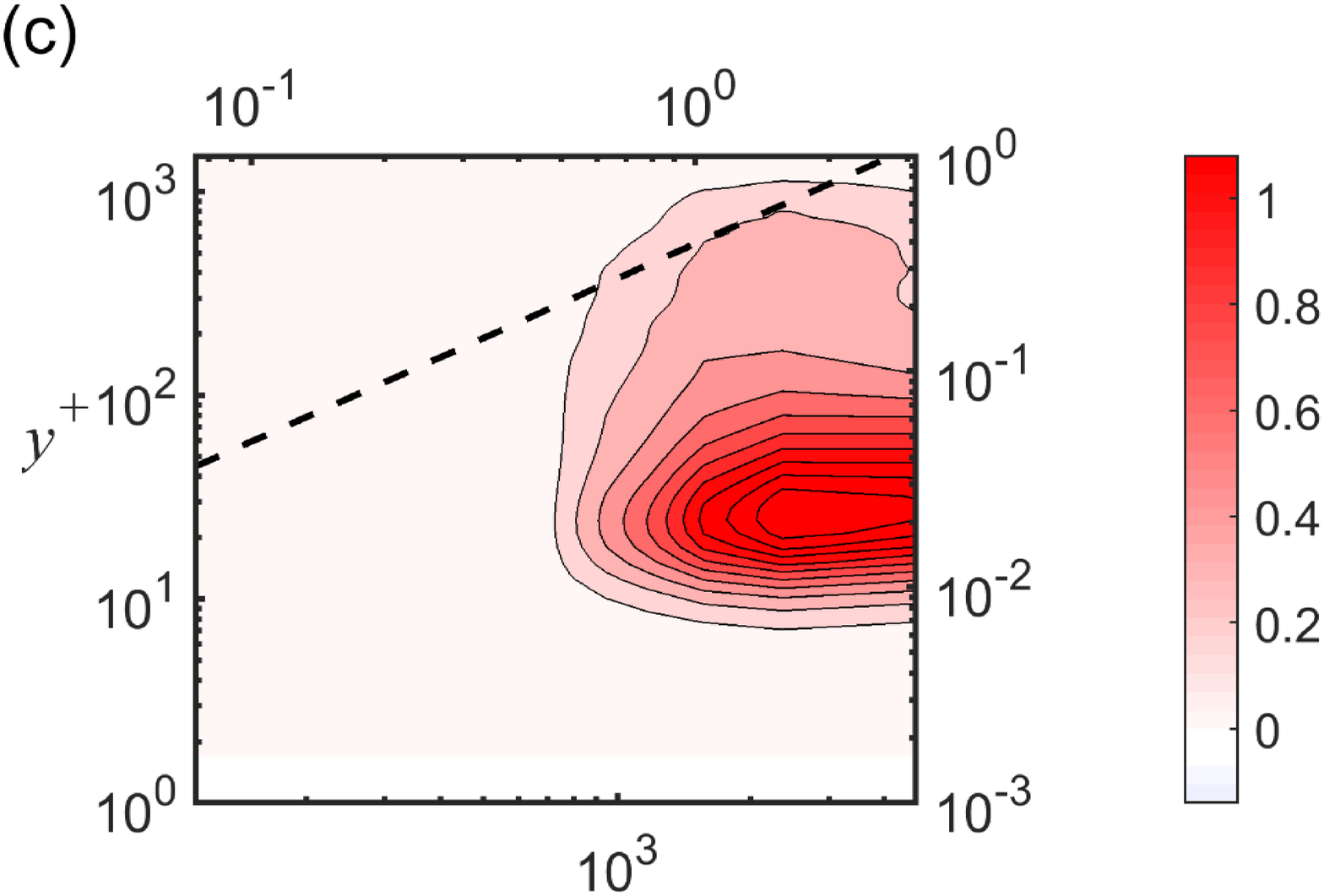}
\label{fig:prodx}
\end{subfigure}
\vspace{-0.7cm}
\begin{subfigure}[b]{0.42\textwidth}
  \includegraphics[width=\textwidth]{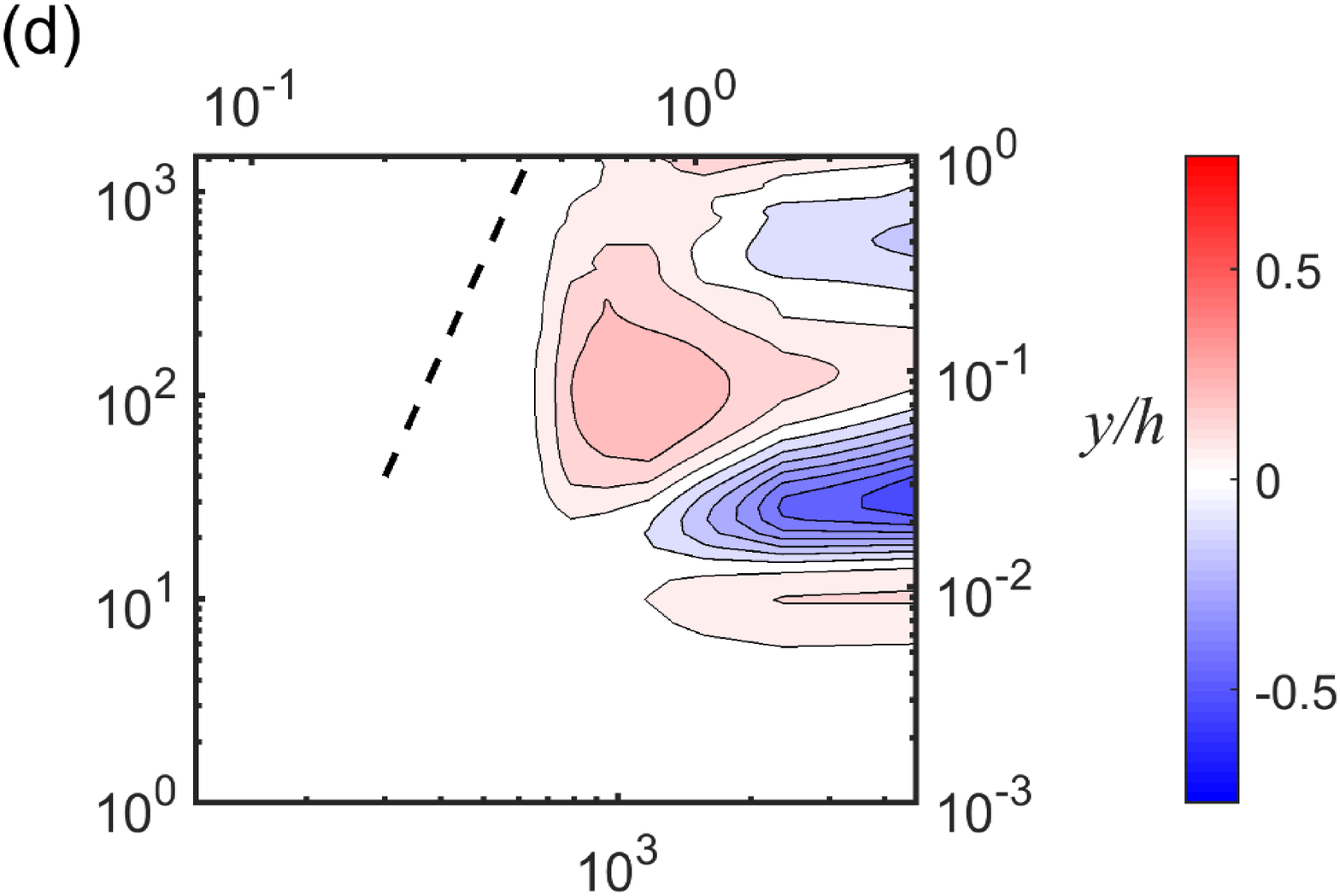}
  \label{4}
\end{subfigure}
\begin{subfigure}[b]{0.42\textwidth}
  \includegraphics[width=\textwidth]{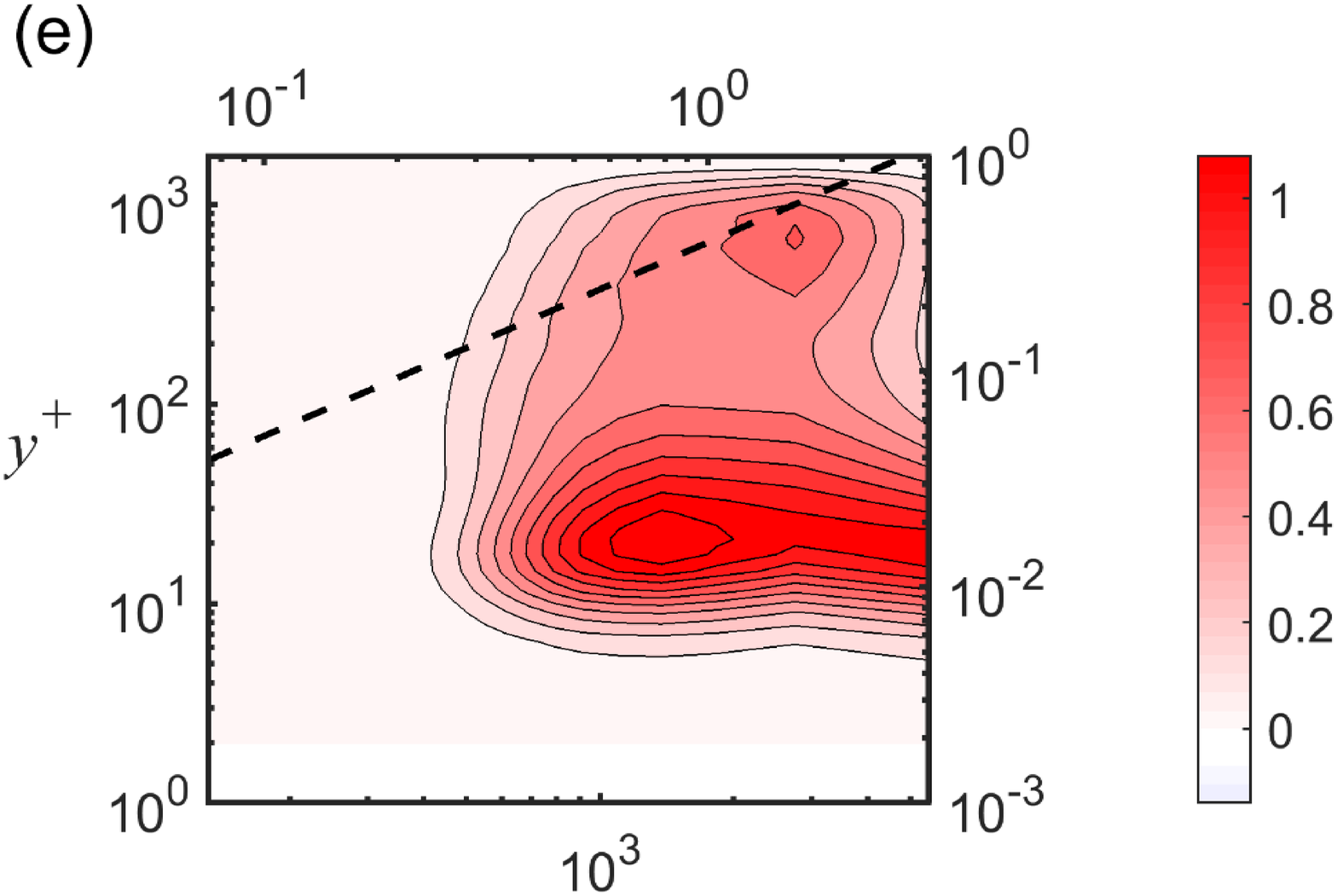}
  \label{5}
\end{subfigure}
\vspace{-0.7cm}
\begin{subfigure}[b]{0.42\textwidth}
  \includegraphics[width=\textwidth]{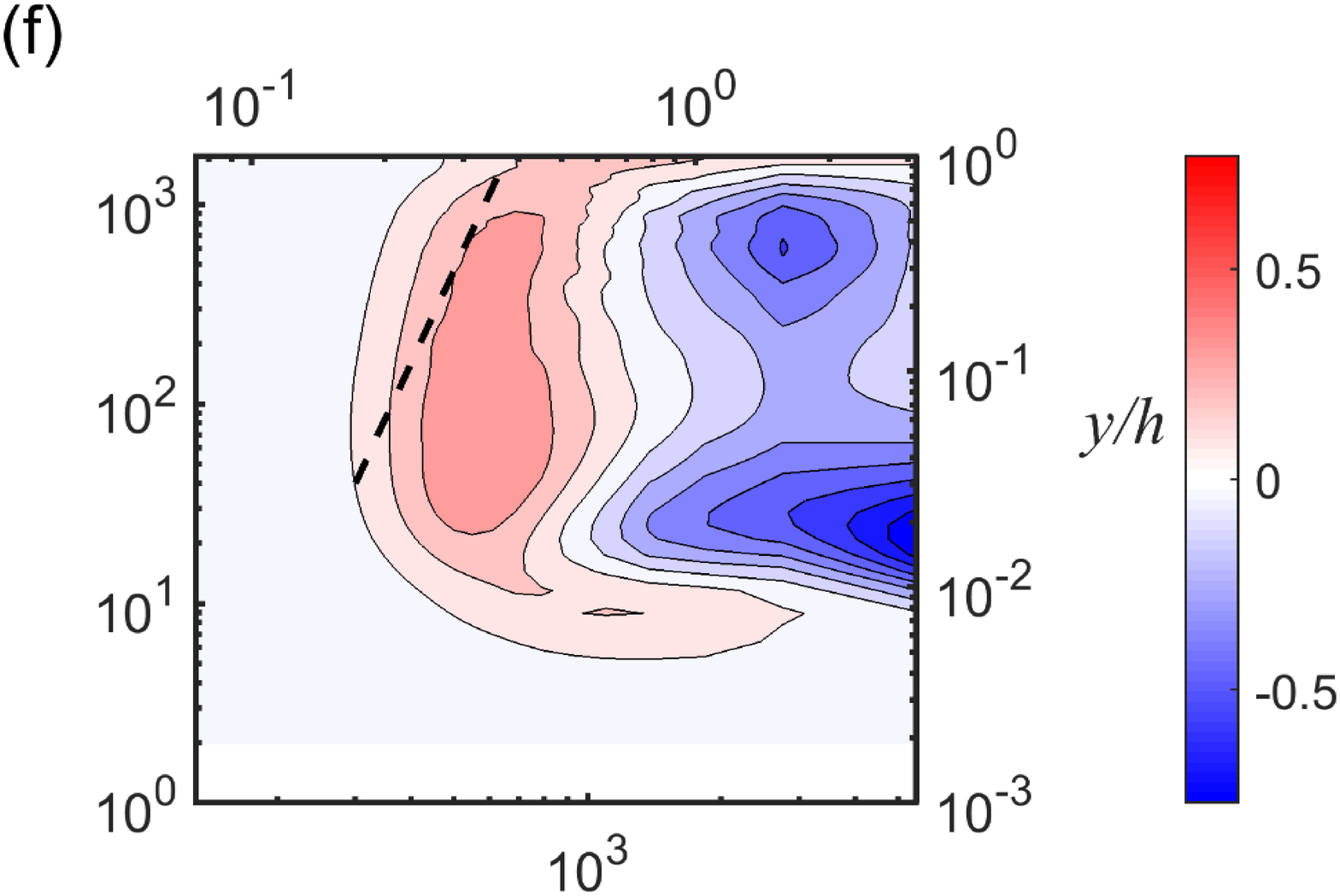}
  \label{6}
\end{subfigure}
\begin{subfigure}[b]{0.42\textwidth}
  \includegraphics[width=\textwidth]{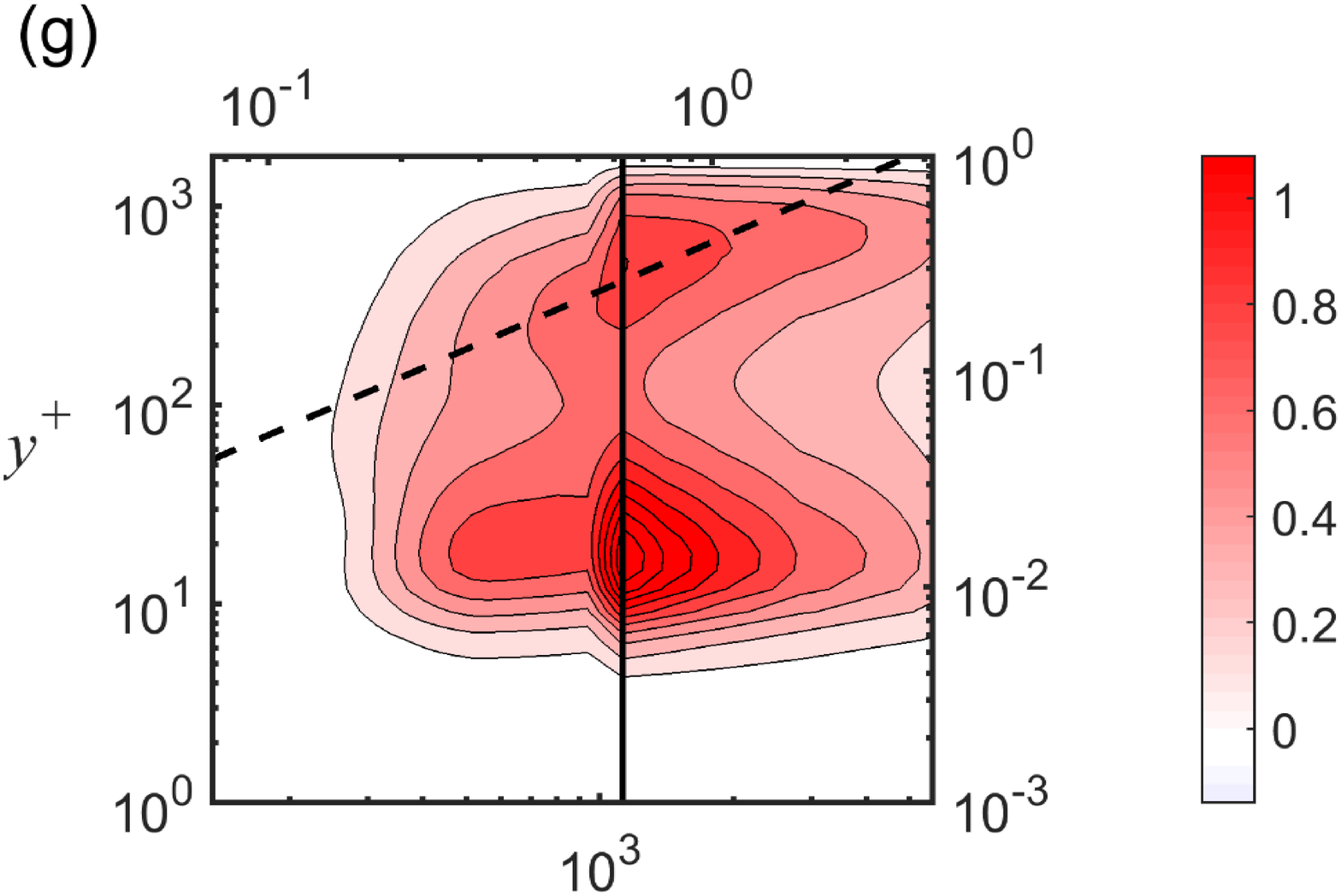}
  \label{5}
\end{subfigure}
\vspace{-0.7cm}
\begin{subfigure}[b]{0.42\textwidth}
  \includegraphics[width=\textwidth]{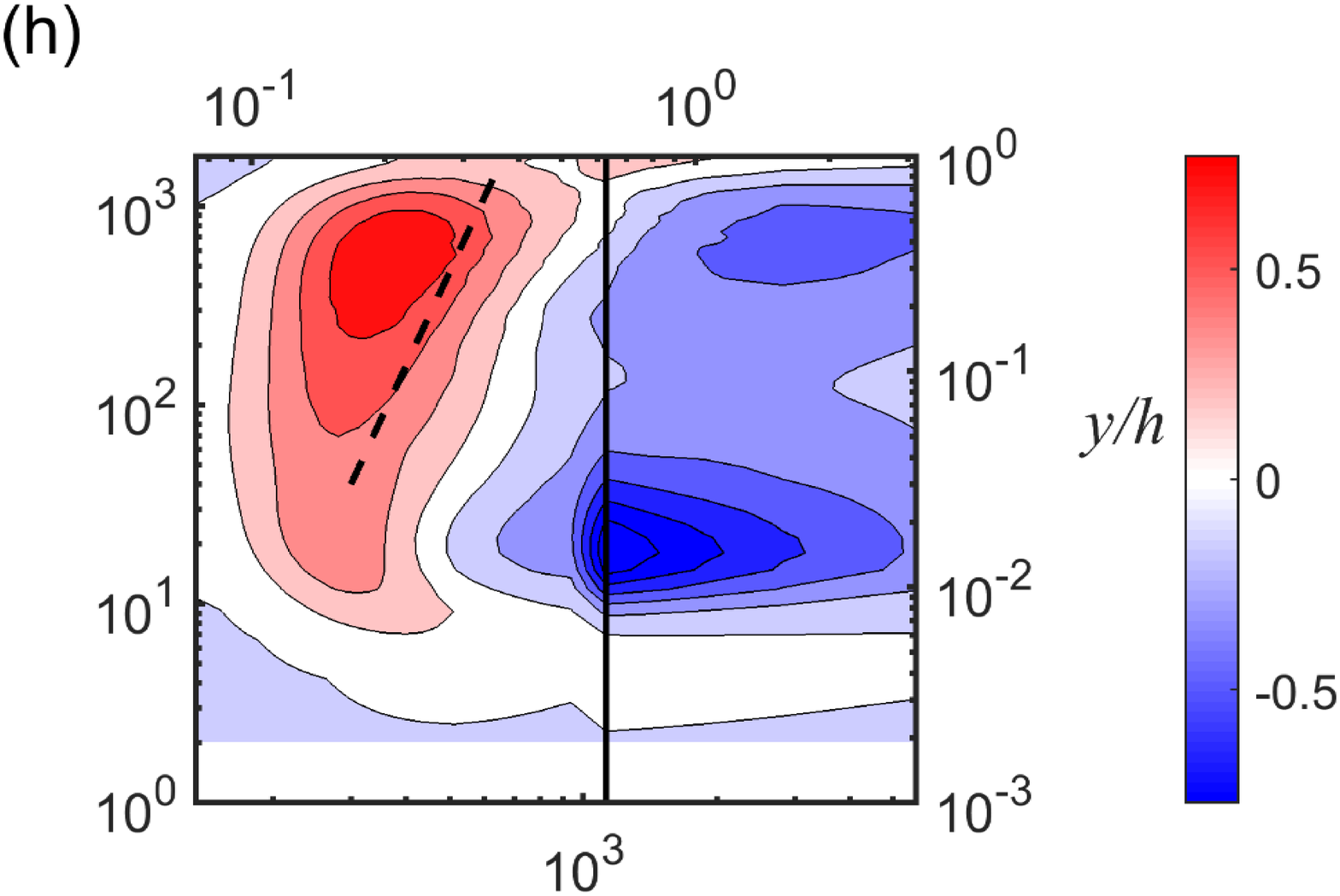}
  \label{6}
\end{subfigure}
\begin{subfigure}[b]{0.42\textwidth}
  \includegraphics[width=\textwidth]{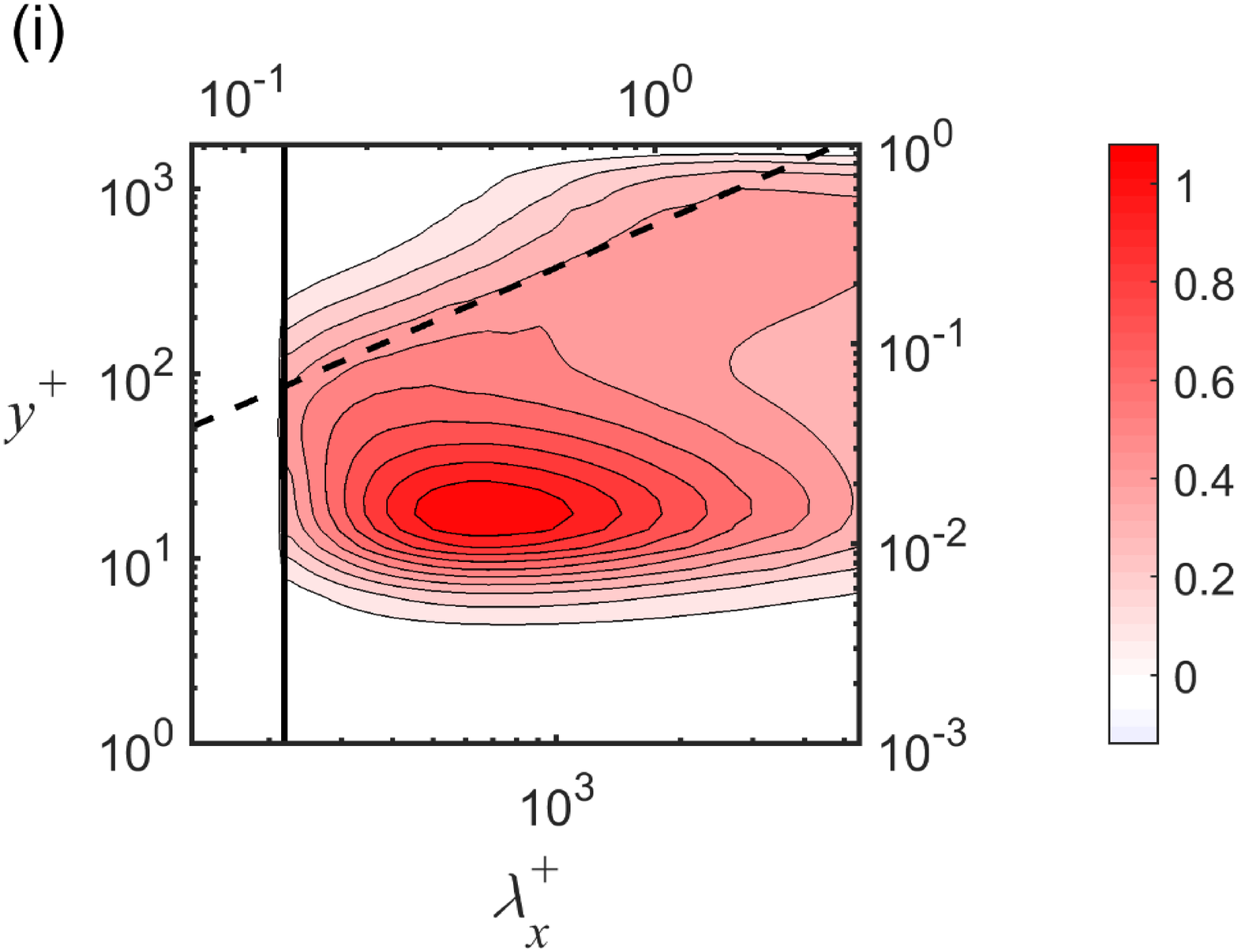}
\end{subfigure}
\begin{subfigure}[b]{0.42\textwidth}
  \includegraphics[width=\textwidth]{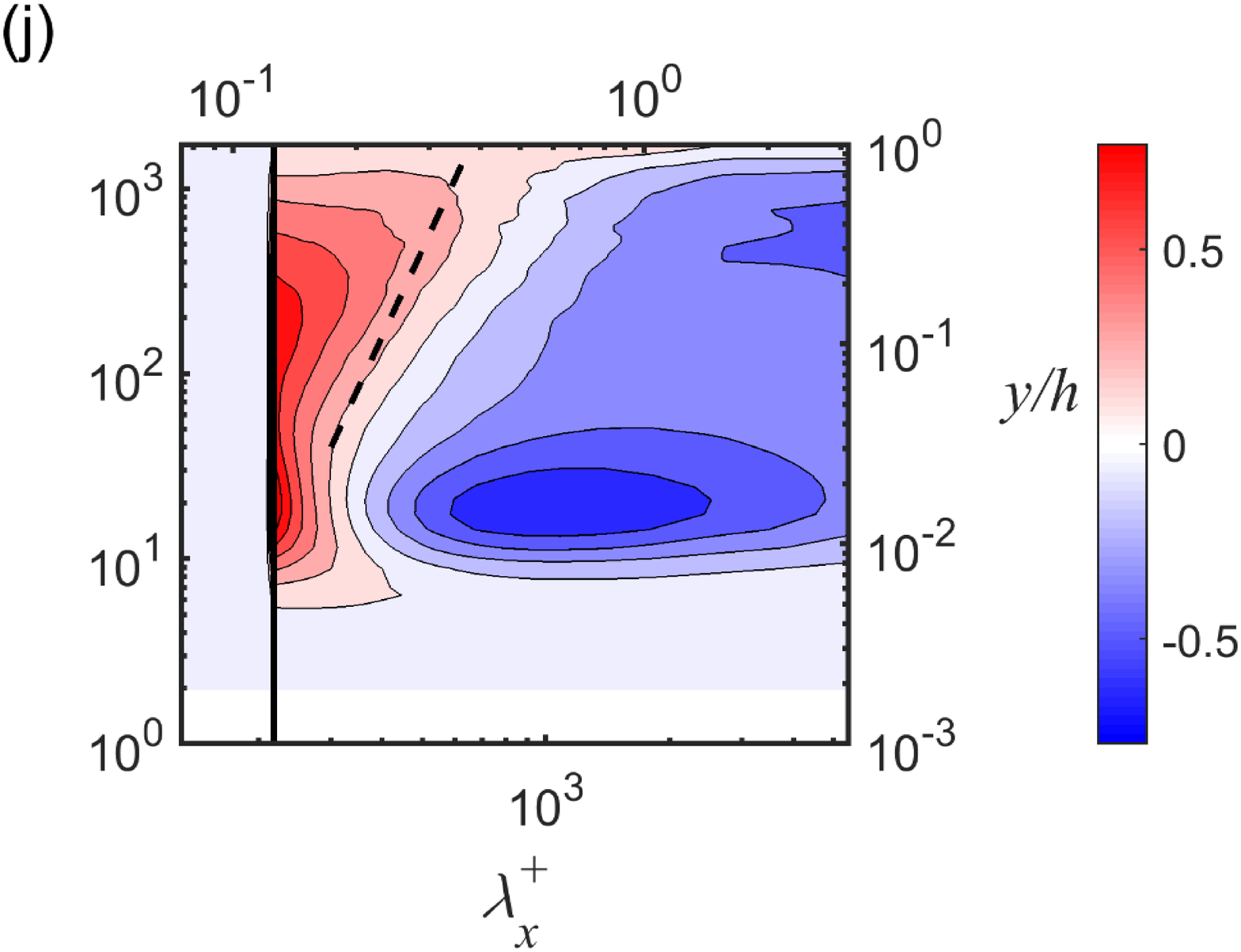}
\end{subfigure}
\end{minipage}
\caption{Premultiplied streamwise wavenumber spectra of production $k_x^+ y^+ \widehat{P}^+(y^+,\lambda_x^+)$ (left column) and turbulent transport $k_x^+ y^+ \widehat{T}_{turb}^+(y^+,\lambda_x^+)$ (right column) for (a,b) LES, (c,d) QL, (e,f) GQL1, (g,h) GQL5 and  (i,j) GQL25 cases. Here, the vertical line represents the streamwise cut-off wavelength ($\lambda_{x,c}$) dividing the $\mathcal{P}_h$- (left) and $\mathcal{P}_l$-subspace  (right) regions.}
\label{fig:xenergy}
\end{figure}

The premultiplied streamwise wavenumber spectra of the energy budget are shown in figure \ref{fig:xenergy}. The spectra of LES show the typical features of energy cascade. However, similarly to the streamwise velocity spectra (figure \ref{fig:xspectra}), the production and turbulent transport spectra do not seem to be aligned along any ridge due to the small streamwise box size employed in this study (for the case with a long streamwise domain, see \citealp{lee19} and \citealp{hwang_lee_2020} where the production and turbulent transport streamwise wavenumber spectra are shown to scale well with the distance from the wall $y$). The production spectra in figure \ref{fig:xenergy}(a) still show a spectral intensity peak at $\lambda_x^+\approx 1000$ and $y^+\approx 20$ and a non-negligible amount of energy is also observed around $y =0.35 \lambda_x$, along which the wall-normal velocity spectra are found to be aligned very well (figure \ref{fig:xspectra}b). The turbulent transport spectra in figure \ref{fig:xenergy}(b) show a region of positive values along the linear ridge corresponding to viscous dissipation ($y=57 \eta$) and a region of negative values corresponding to the streamwise production intensity peak ($\lambda_x^+ \approx 1000$).

As expected, the QL model does not exhibit the typical features of the energy cascade observed in the LES case. In figure \ref{fig:xenergy}(c), the production spectra show no energy intensity for wavelengths below $\lambda_x^+ = 700$, and the two spectra are highly localised. The transport spectra significantly differ to those of the LES case: the negative region of the spectra is displaced to higher wavelengths and reduced to the $y^+ \approx 20$ and $y^+ \approx 500$ locations, while the positive regions appear immediately below and between them -- note that the positive region of the turbulent transport spectra was located at $\lambda_x^+ \lesssim 700$ in the LES case, which has been suppressed in the QL model, and the turbulent transport has reorganized itself to redistribute the energy injected by the streamwise production and lost by the viscous dissipation. Given the linear nature of (\ref{eq:2.6b}), it may not be surprising to see the significantly damaged energy cascade in the streamwise wavenumber space. However, it should be mentioned that the length-scale selection process in the streamwise wavenumber spectra has been understood to be related to the streak instability and/or the related transient growth \cite[]{schoppa02,park11,alizard15,cassinelli17,degiovanetti17}. Indeed, in uniform shear turbulence where the self-sustaining process exists only at single length scale, the intensity distribution of the streamwise turbulent transport spectra of the QL was found to behave consistently with these previous findings \cite[]{hernandez}. In this sense, the behaviour of the streamwise wavenumber spectra of turbulent transport is not entirely expected, and a further discussion on this issue shall be given in \S\ref{sec:sec41}. This behaviour is gradually reverted when $\lambda_{x,c}$ is decreased with the GQL approximations: the GQL1 case already exhibits streamwise spectra reaching down to $\lambda_x^+ \approx 300$, and it goes on for the GQL5 and GQL25 cases. The vertical dashed line in each figure represents the streamwise wavenumber cut-off dividing the $\mathcal{P}_l$- (right) and $\mathcal{P}_h$-subspace (left) groups, and it is therefore clear that by increasing the number of modes allowed to interact nonlinearly (i.e. the GQL cases), the $\mathcal{P}_h$ subspace region of the spectra starts to show a better match to the original LES spectra, developing a healthier cascade in the streamwise direction (GQL5 and GQL25 cases). Once again, a complete lack of energy is observed in the $\mathcal{P}_h$ subspace region of the streamwise wavenumber space for the GQL25 case, as in figure \ref{fig:xspectra}(i,j). 
This issue will be discussed in detail in \S \ref{sec:sec43}.

\subsection{Componentwise energy transport and pressure strain} \label{sec:sec33}
\begin{figure}
\begin{minipage}{\textwidth}
\centering
\begin{subfigure}[b]{0.42\textwidth}
  \includegraphics[width=\textwidth]{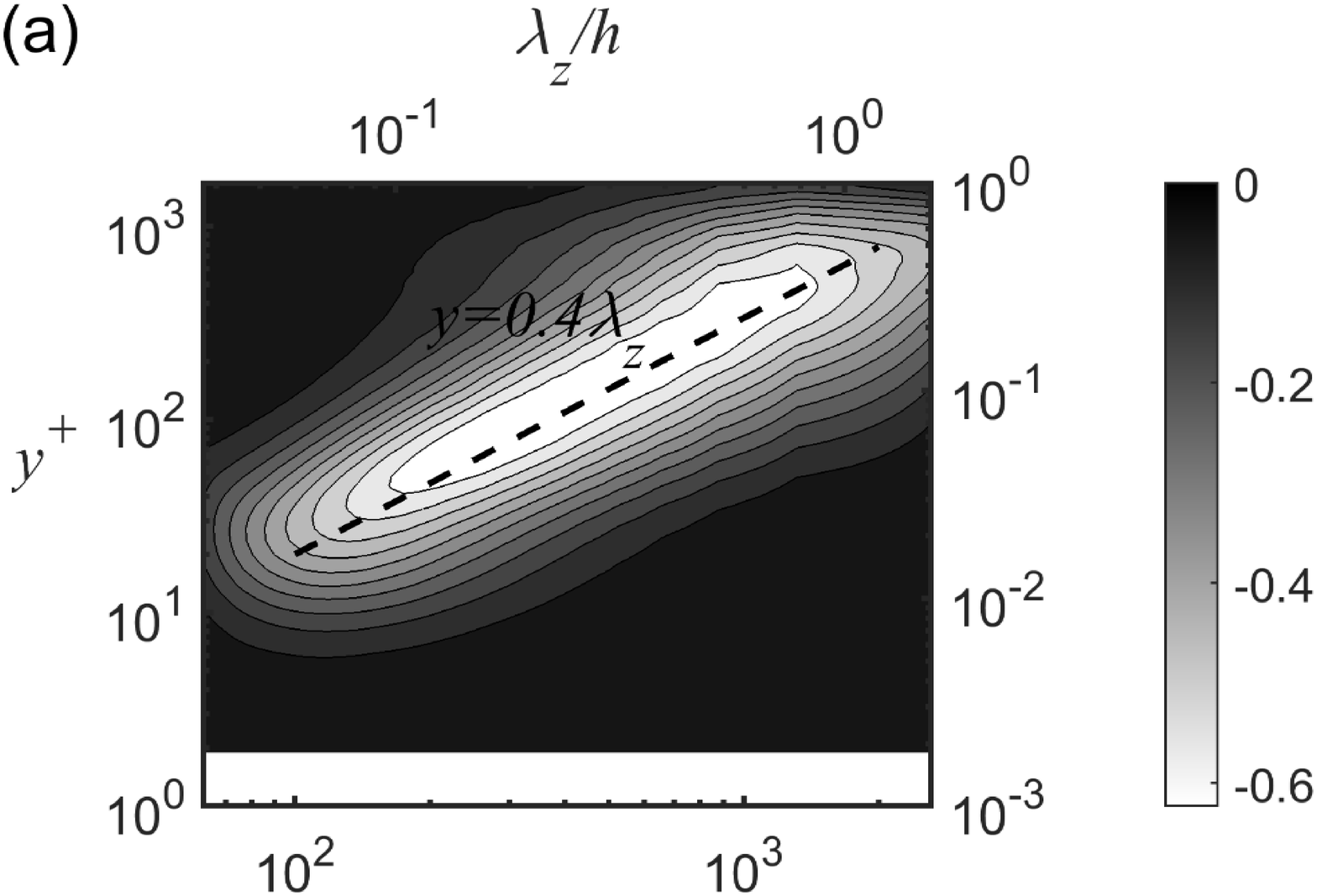}
\label{1}
\end{subfigure}
\vspace{-0.7cm}
\begin{subfigure}[b]{0.42\textwidth}
  \includegraphics[width=\textwidth]{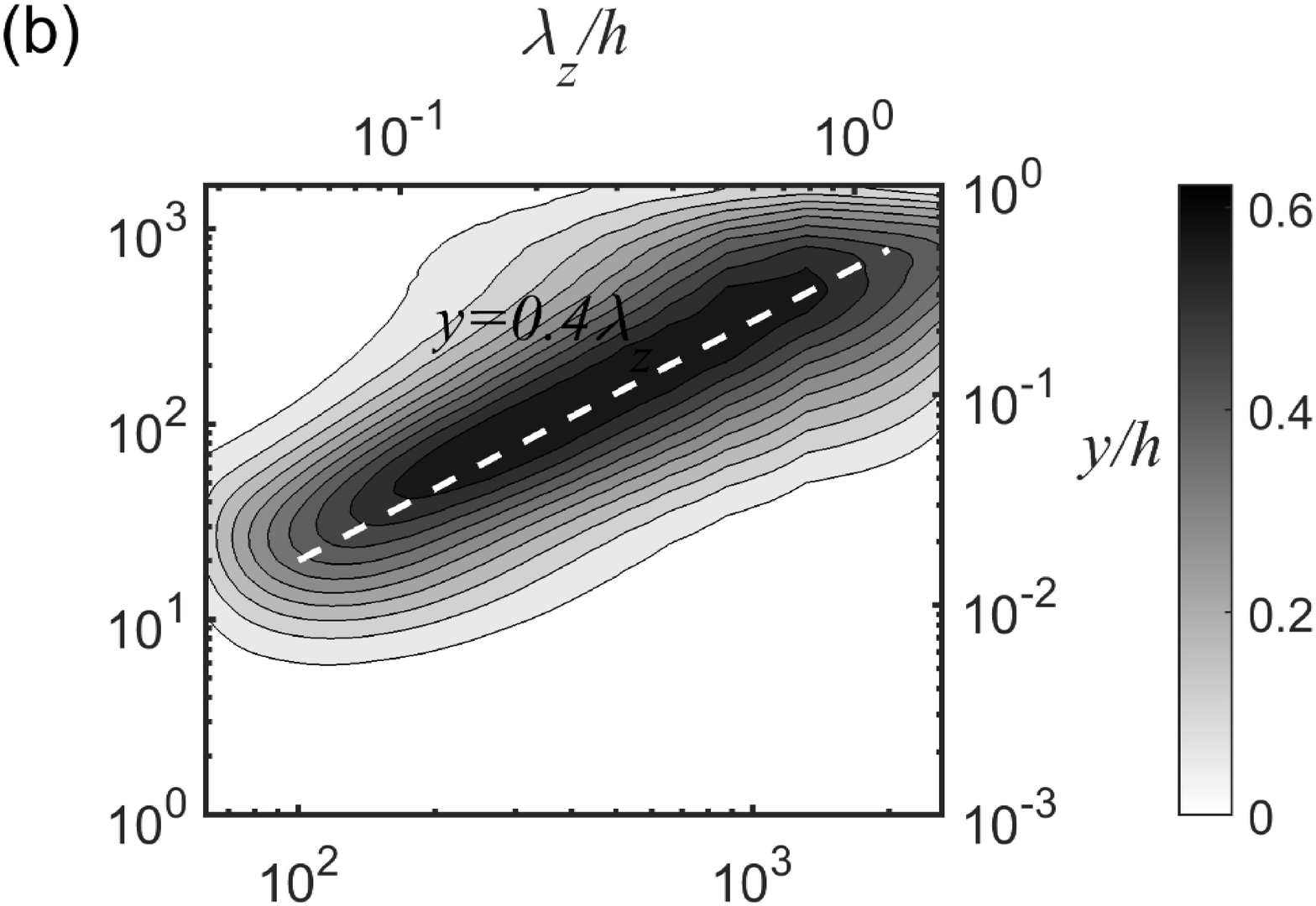}
\label{2}
\end{subfigure}
\begin{subfigure}[b]{0.42\textwidth}
  \includegraphics[width=\textwidth]{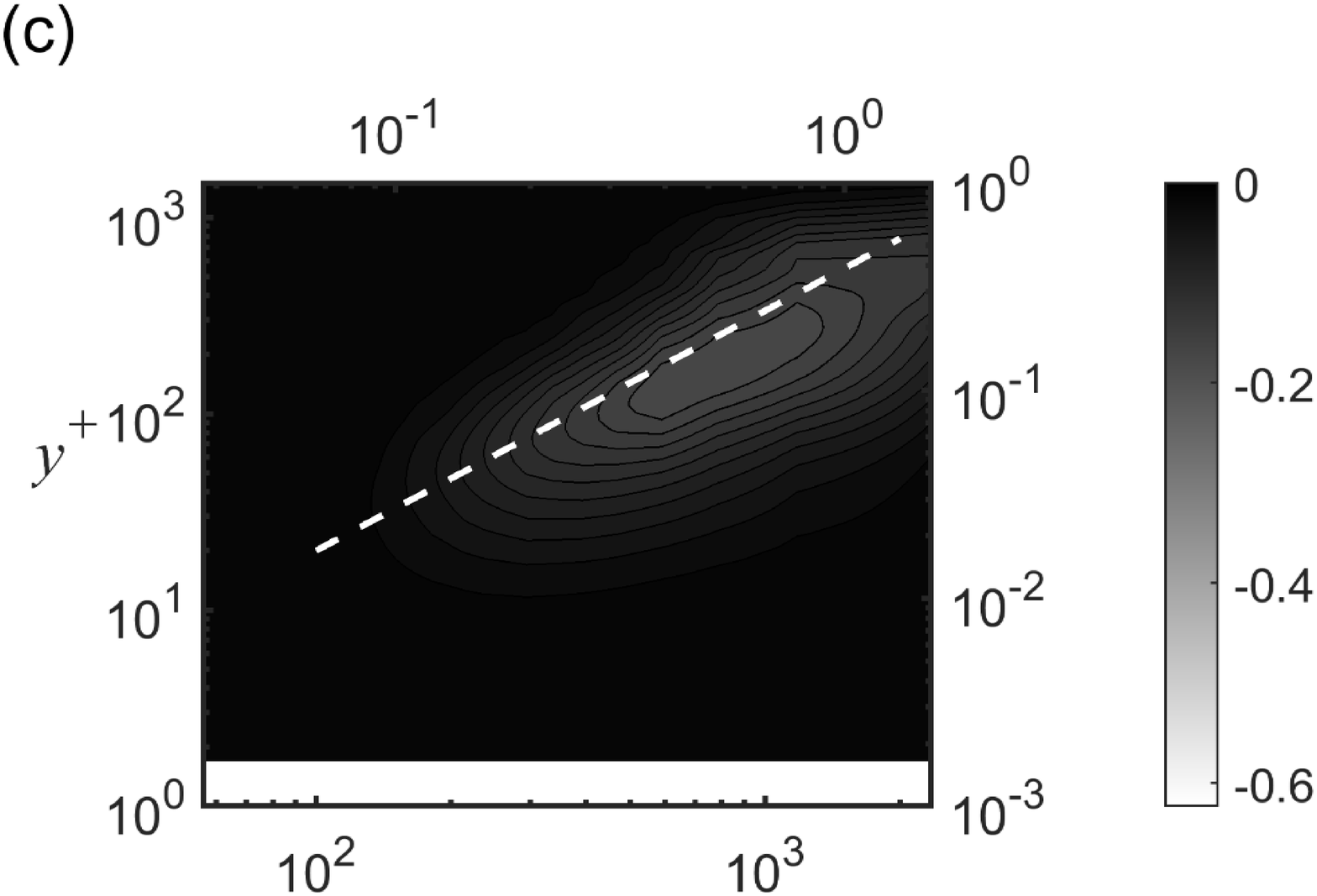}
  \label{3}
\end{subfigure}
\vspace{-0.7cm}
\begin{subfigure}[b]{0.42\textwidth}
  \includegraphics[width=\textwidth]{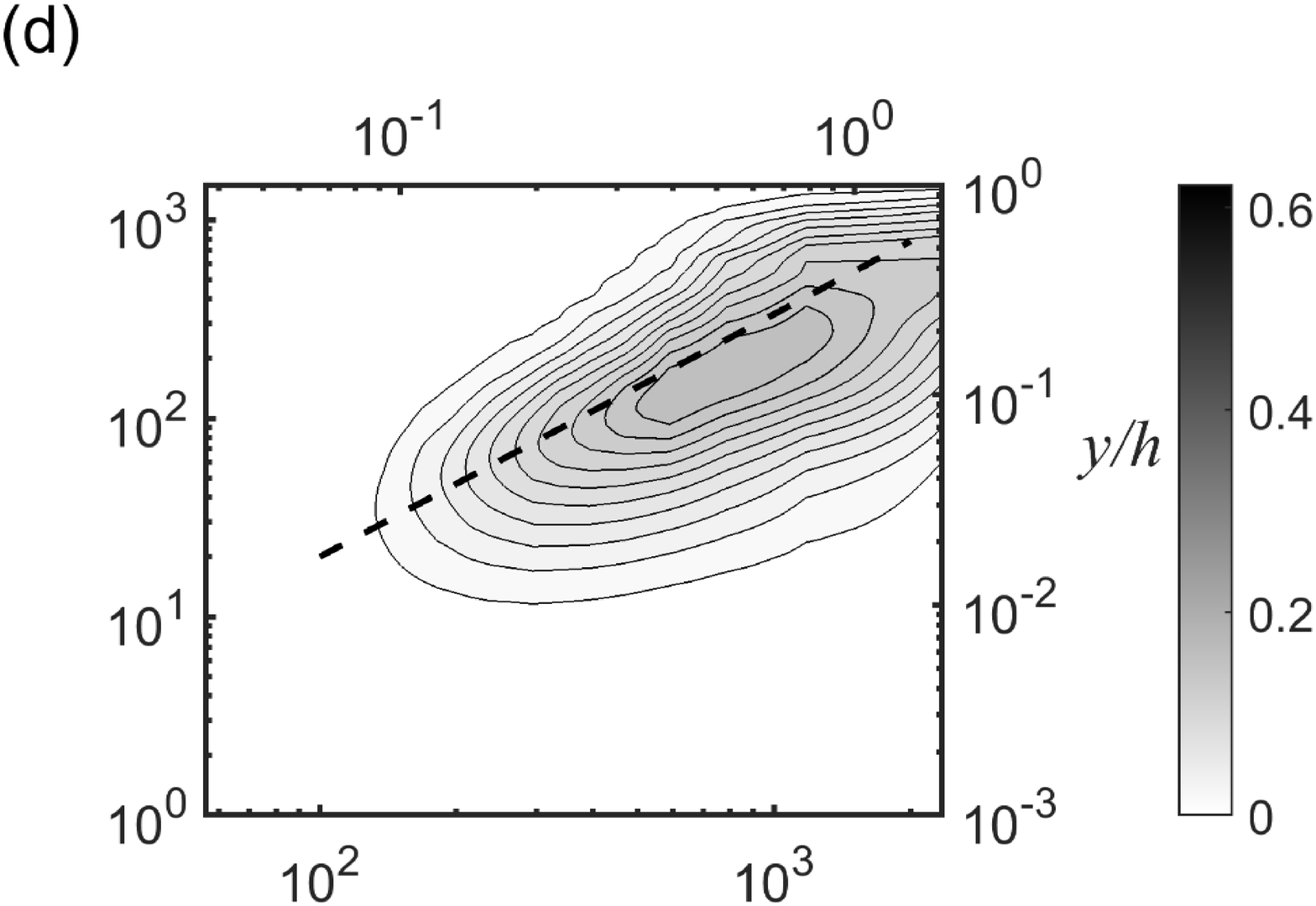}
  \label{4}
\end{subfigure}
\begin{subfigure}[b]{0.42\textwidth}
  \includegraphics[width=\textwidth]{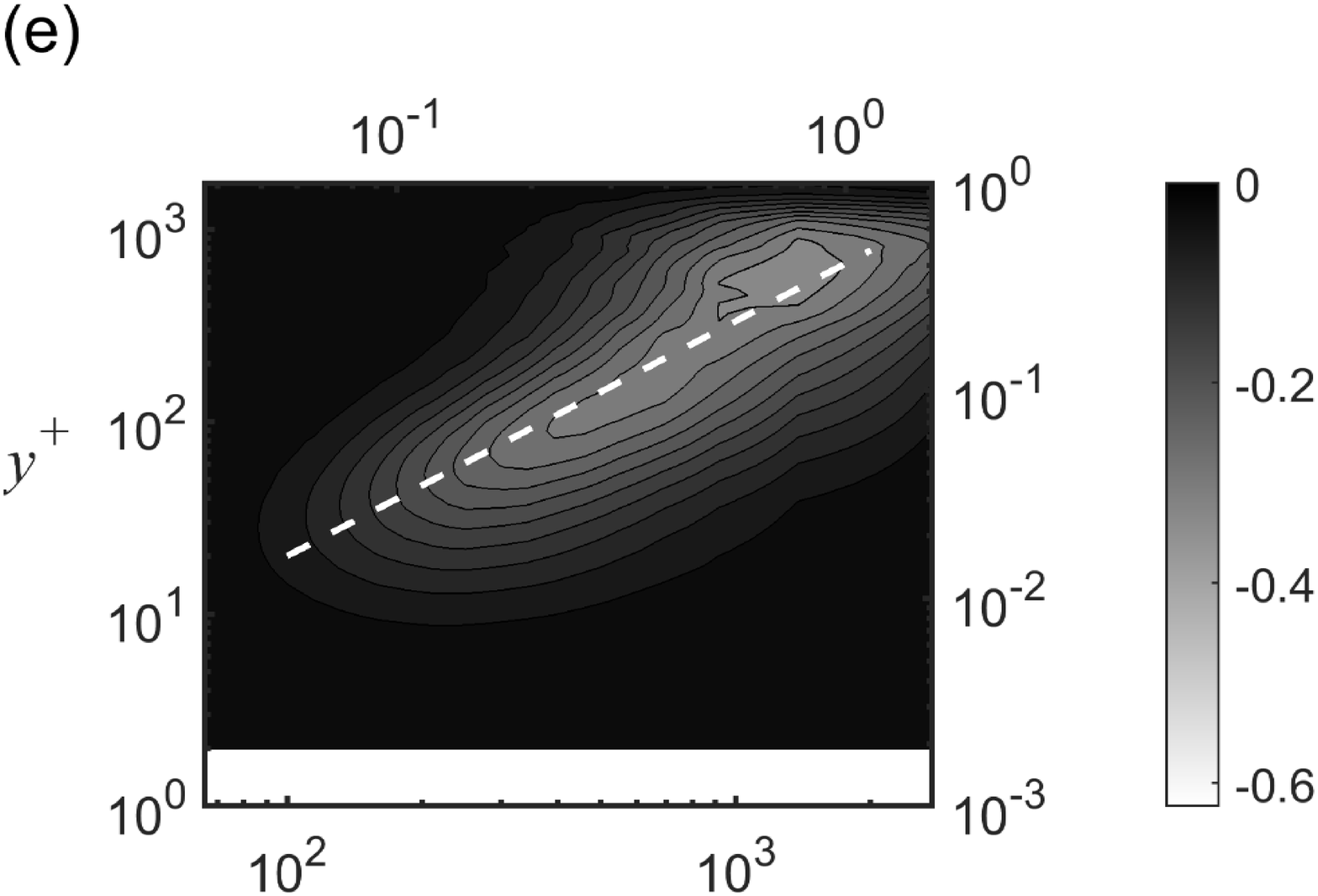}
  \label{5}
\end{subfigure}
\vspace{-0.7cm}
\begin{subfigure}[b]{0.42\textwidth}
  \includegraphics[width=\textwidth]{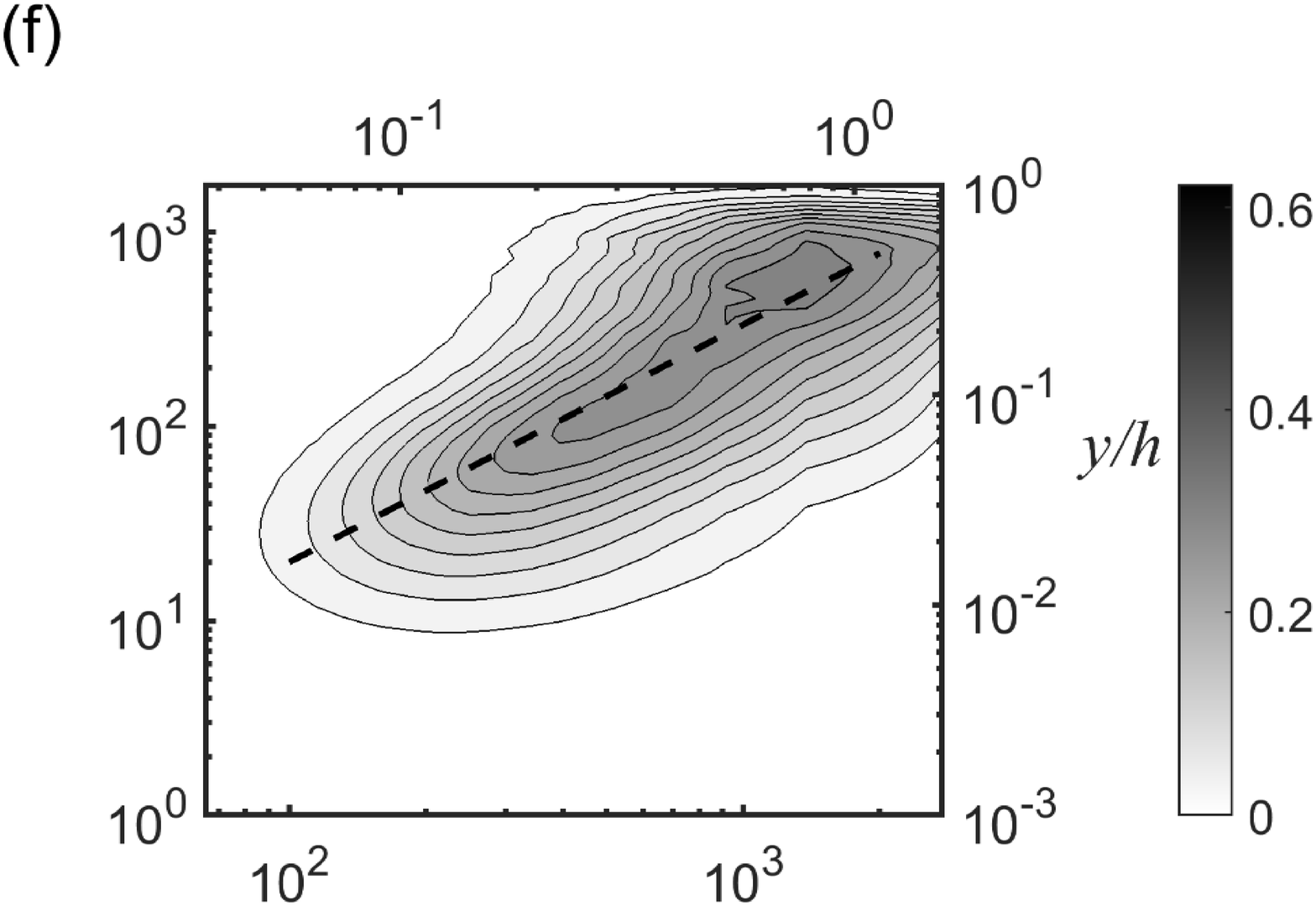}
  \label{6}
\end{subfigure}
\begin{subfigure}[b]{0.42\textwidth}
  \includegraphics[width=\textwidth]{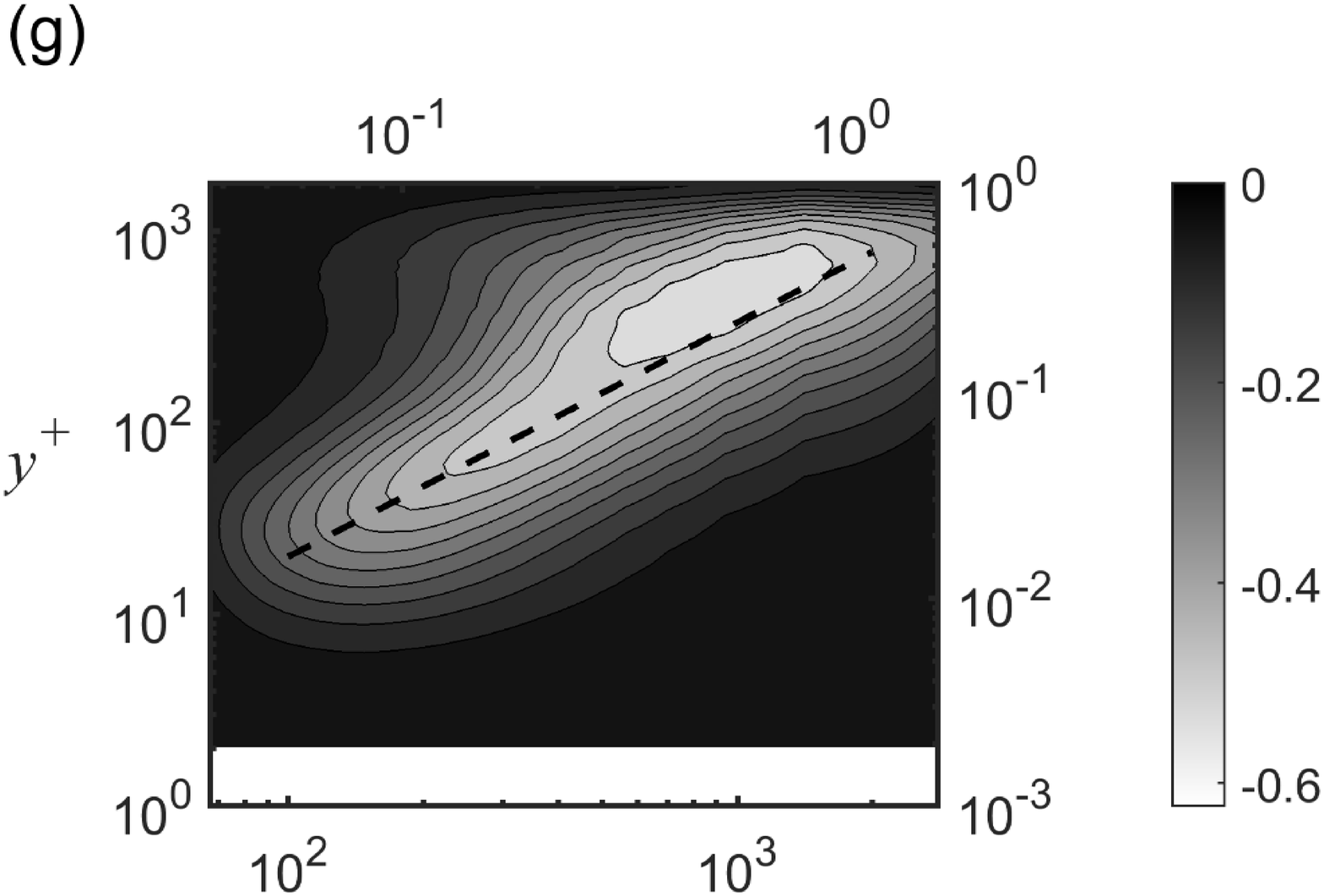}
  \label{5}
\end{subfigure}
\vspace{-0.7cm}
\begin{subfigure}[b]{0.4\textwidth}
  \includegraphics[width=\textwidth]{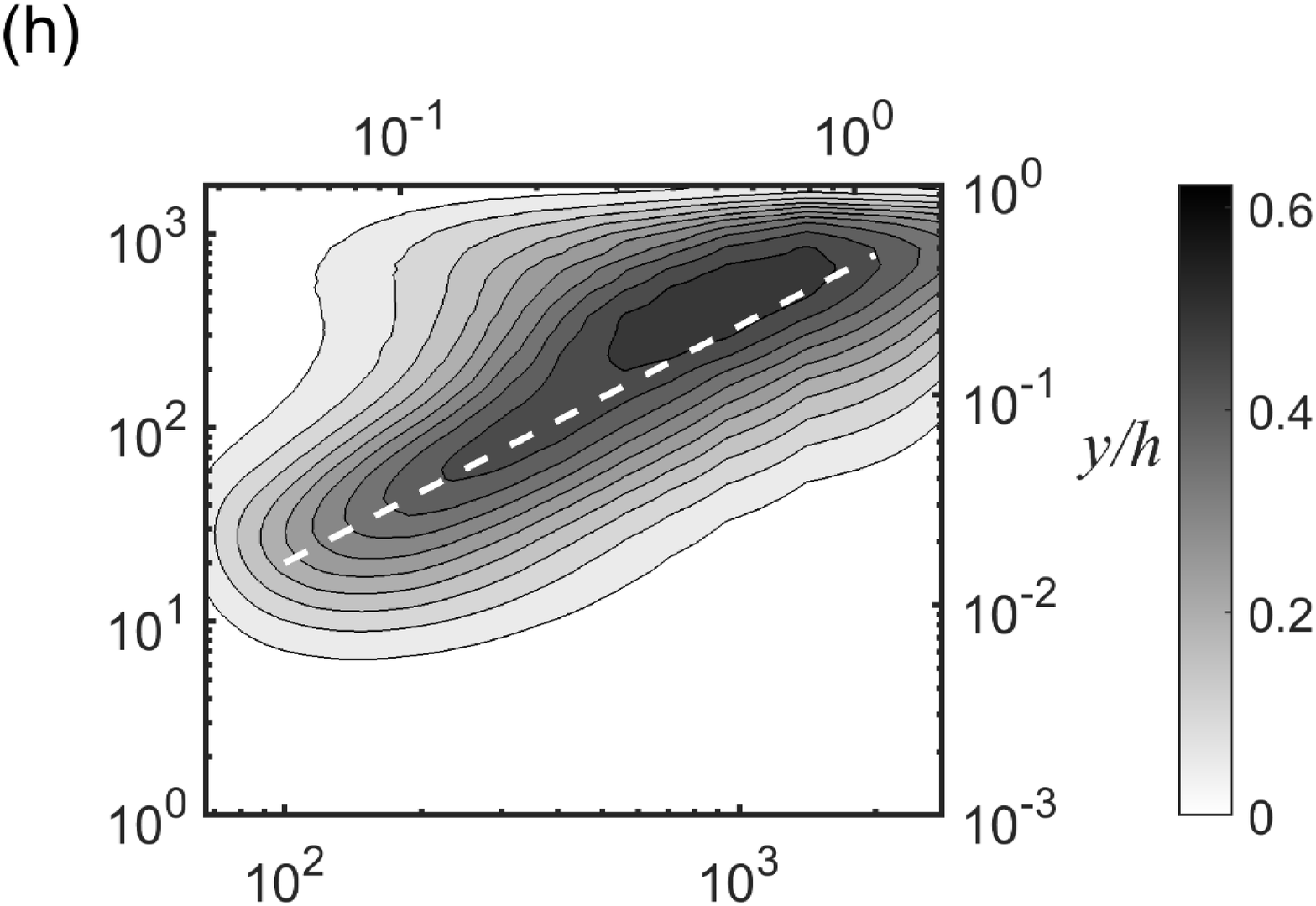}
  \label{6}
\end{subfigure}
\begin{subfigure}[b]{0.42\textwidth}
  \includegraphics[width=\textwidth]{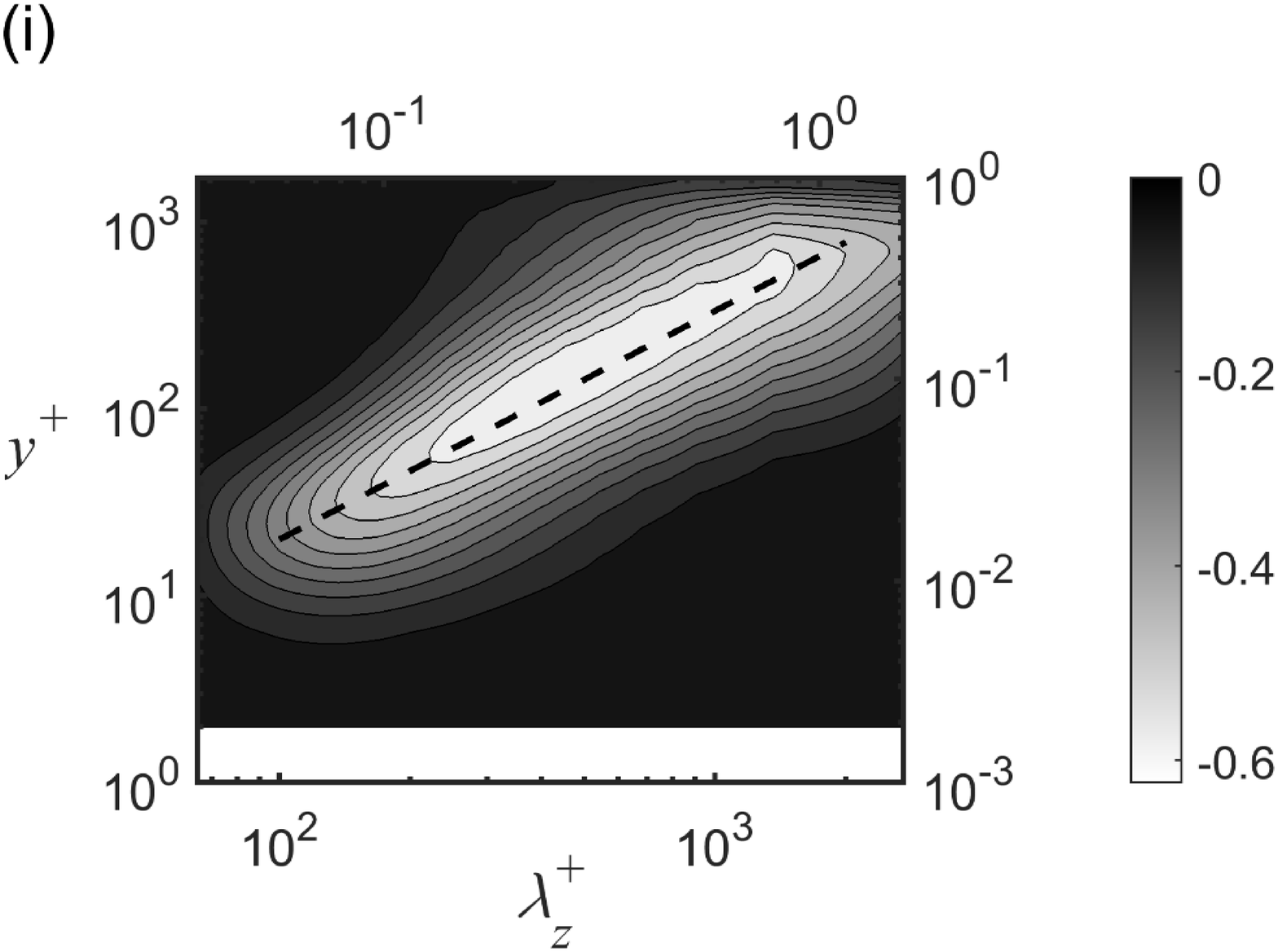}
\end{subfigure}
\begin{subfigure}[b]{0.42\textwidth}
  \includegraphics[width=\textwidth]{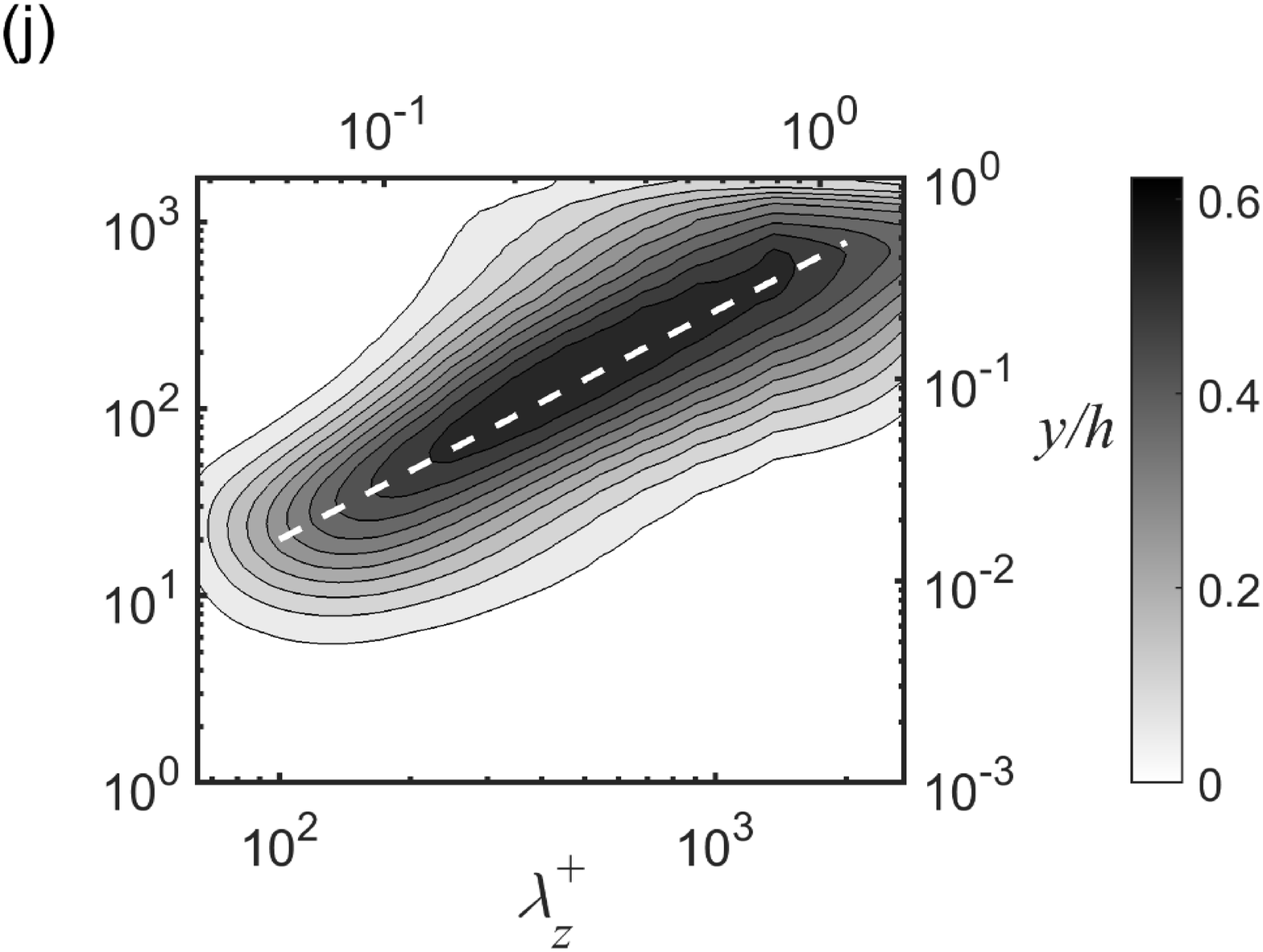}
\end{subfigure}
\end{minipage}
\caption{Premultiplied spanwise wavenumber spectra of $k_z^+ y^+ \widehat{\Pi}_x^+(y^+,\lambda_z^+)$ (left column) and $k_z^+ y^+ \widehat{\Pi}_{yz}^+(y^+,\lambda_z^+)$ (right column) for (a,b) LES, (c,d) QL, (e,f) GQL1, (g,h) GQL5 and (i,j) GQL25 cases.}
\label{fig:piz}
\end{figure}

The pressure-strain spectra are also studied to understand the mechanism of componentwise TKE distribution in the GQL model. Figure \ref{fig:piz} shows the spanwise wavenumber spectra of the pressure strain for LES, QL and GQL model. A negative $\widehat{\Pi}_x$ and mainly positive $\widehat{\Pi}_y$ and $\widehat{\Pi}_z$ (combined into $\widehat{\Pi}_{yz}=\widehat{\Pi}_y + \widehat{\Pi}_z$) can be observed throughout the spanwise scales for all cases. This tendency is consistent with that of \cite{cho18} and the available DNS data \citep{hoyas08,mizuno16}. This means that the energy produced at the streamwise component of the TKE through the production term is distributed to wall-normal and spanwise components via the pressure strain. The spectra also are distributed approximately along the linear ridge $y = 0.4 \lambda_z$. The QL model appears to underestimate the spectral energy intensity of all pressure-strain terms, and their spanwise wavenumber spectra do not span length scales below $\lambda_z^+ \approx 100$. These two features are quickly improved with the application of the GQL approximations: the spectra begin to reach smaller spanwise scales in GQL1 and approach the spectra of LES quantitatively in GQL5. The GQL25 case shows little improvement with respect to the GQL5 case. 

\begin{figure}
\begin{minipage}{\textwidth}
\centering
\begin{subfigure}[b]{0.42\textwidth}
  \includegraphics[width=\textwidth]{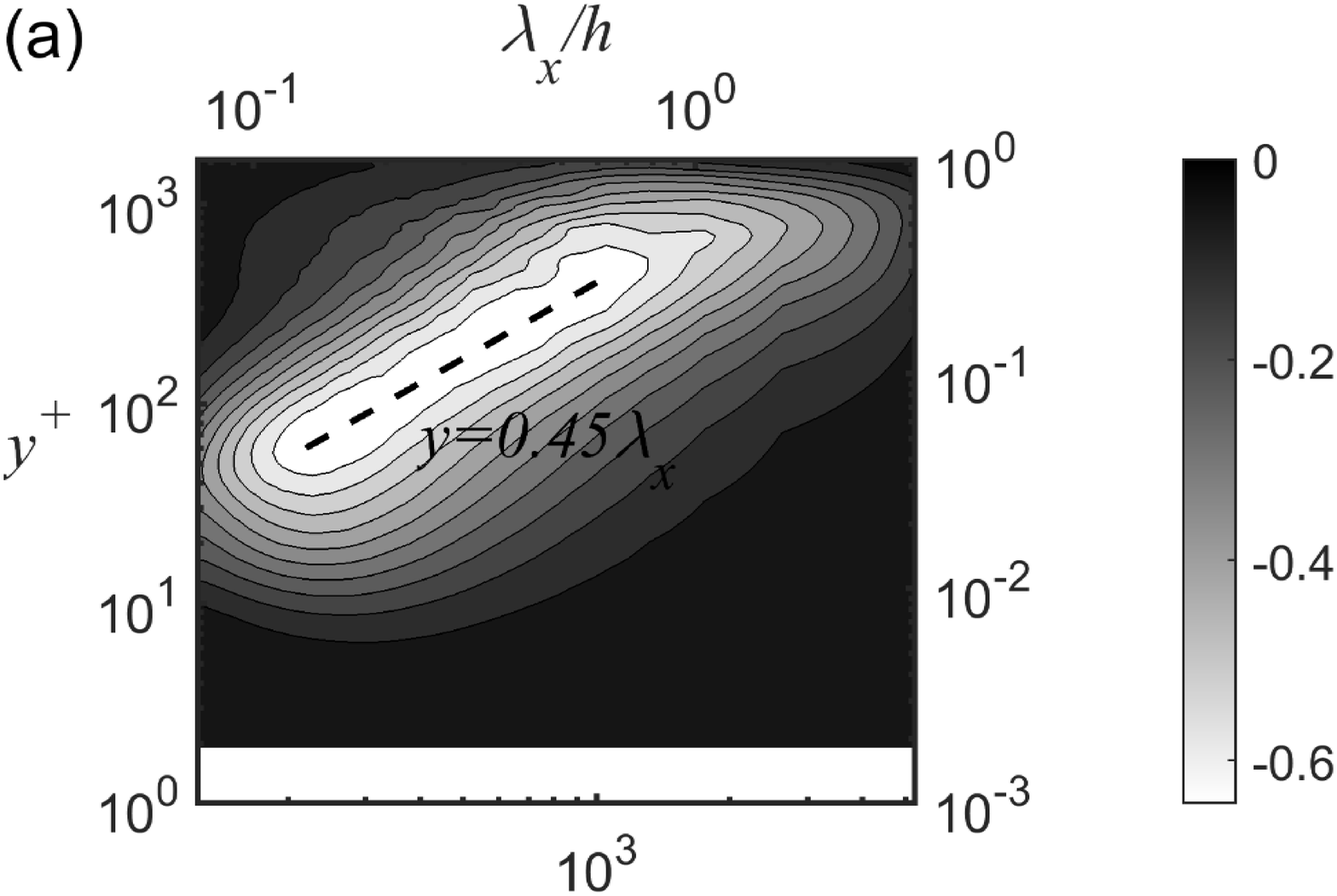}
\label{1}
\end{subfigure}
\vspace{-0.7cm}
\begin{subfigure}[b]{0.42\textwidth}
  \includegraphics[width=\textwidth]{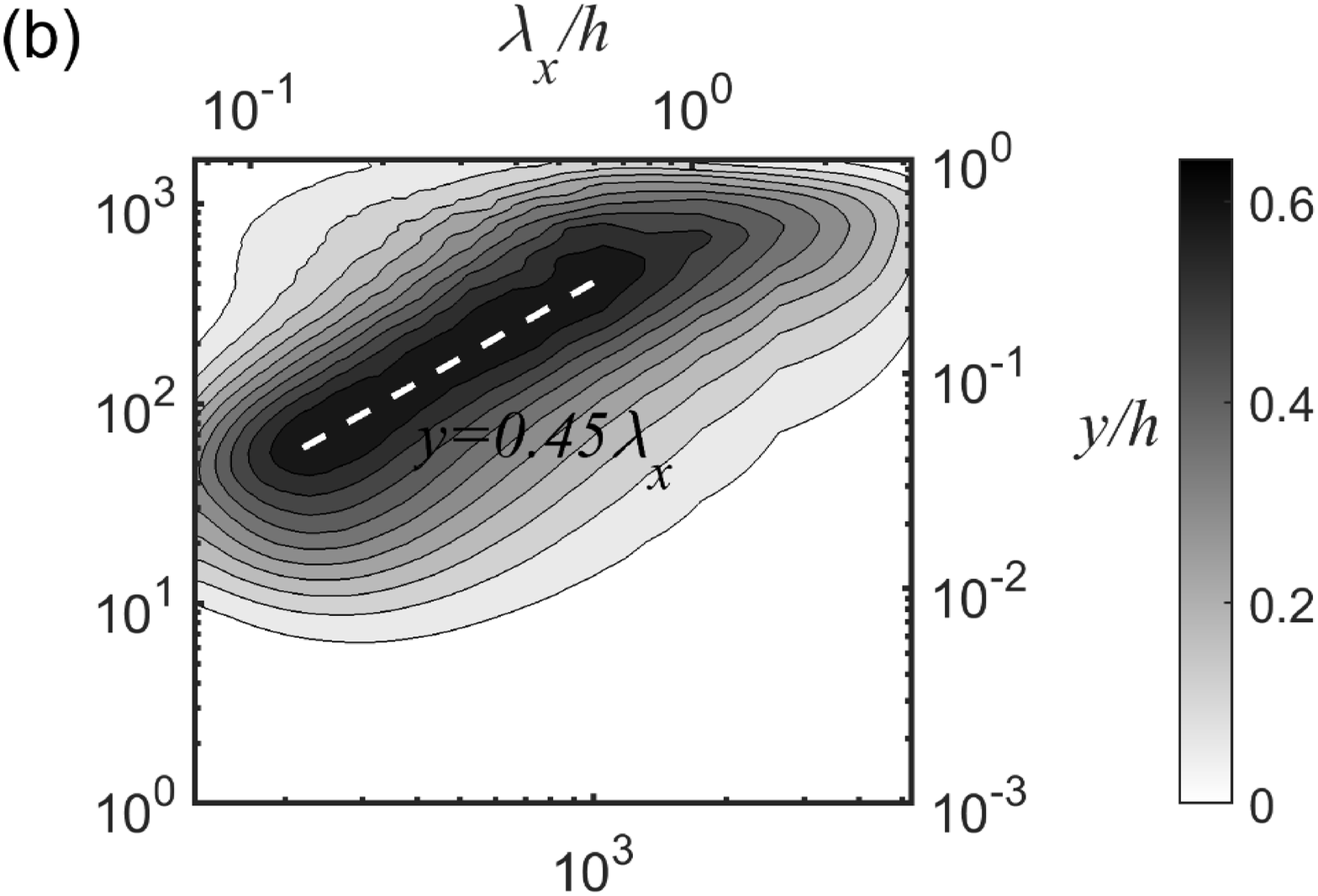}
\label{2}
\end{subfigure}
\begin{subfigure}[b]{0.42\textwidth}
  \includegraphics[width=\textwidth]{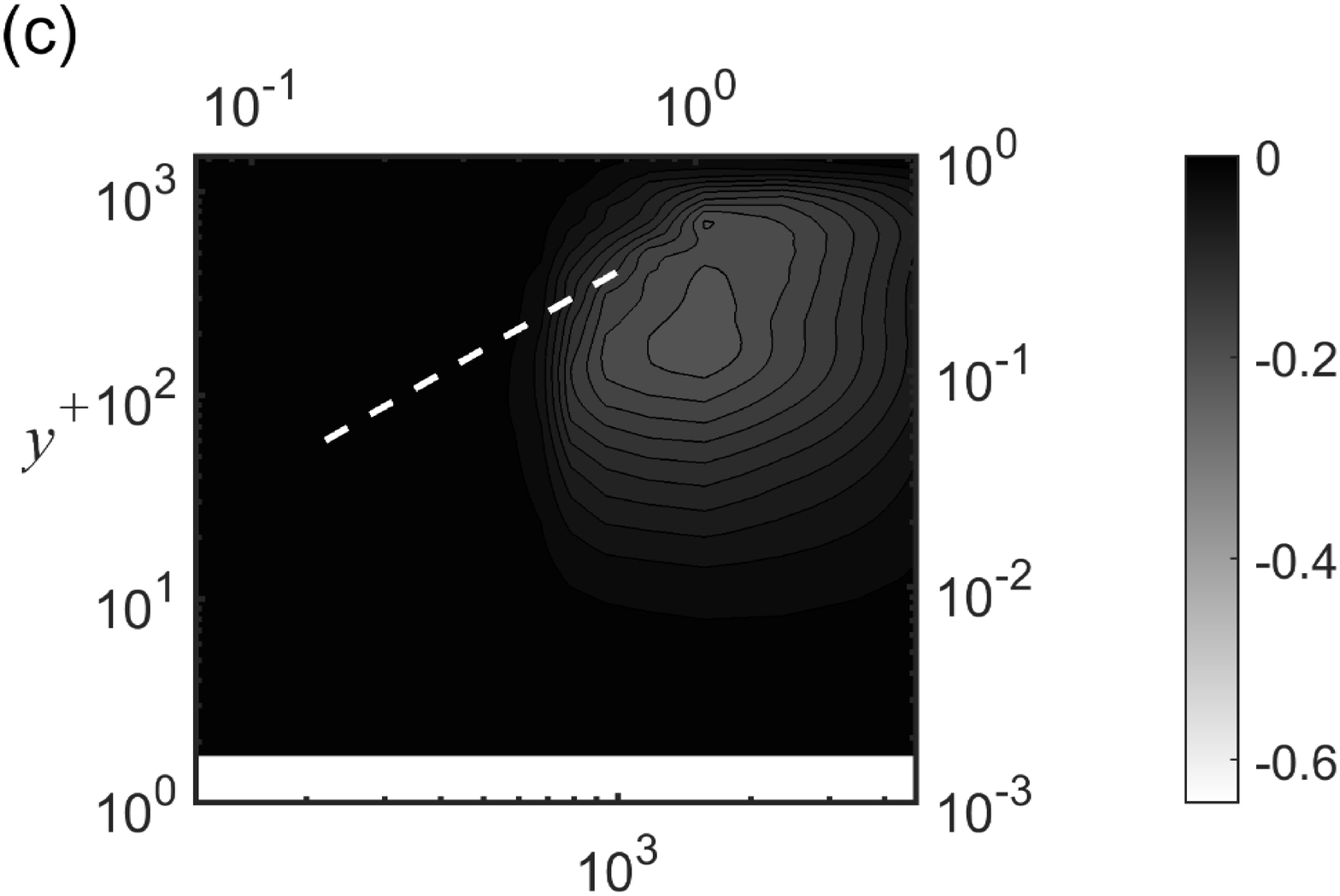}
  \label{3}
\end{subfigure}
\vspace{-0.7cm}
\begin{subfigure}[b]{0.42\textwidth}
  \includegraphics[width=\textwidth]{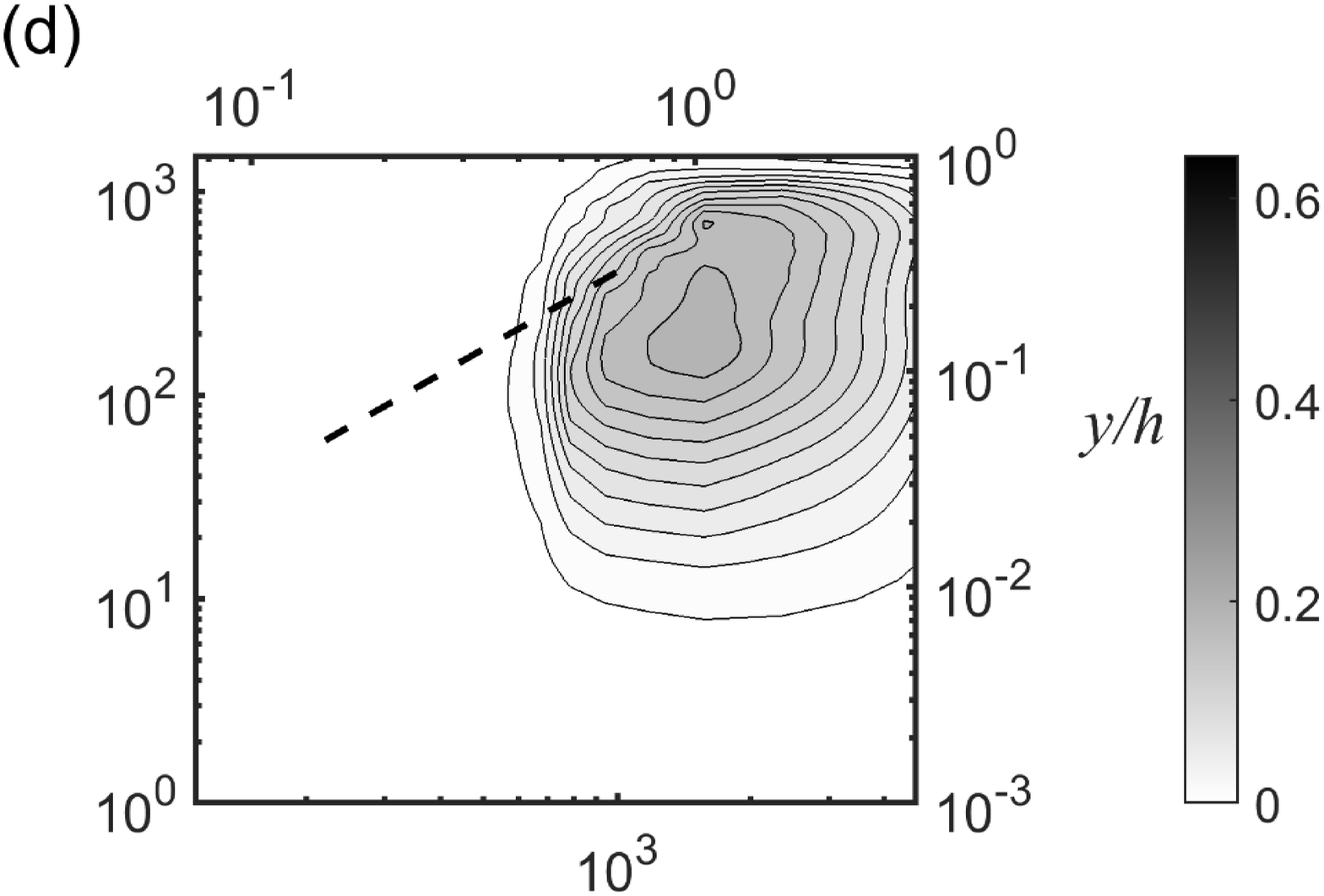}
  \label{4}
\end{subfigure}
\begin{subfigure}[b]{0.42\textwidth}
  \includegraphics[width=\textwidth]{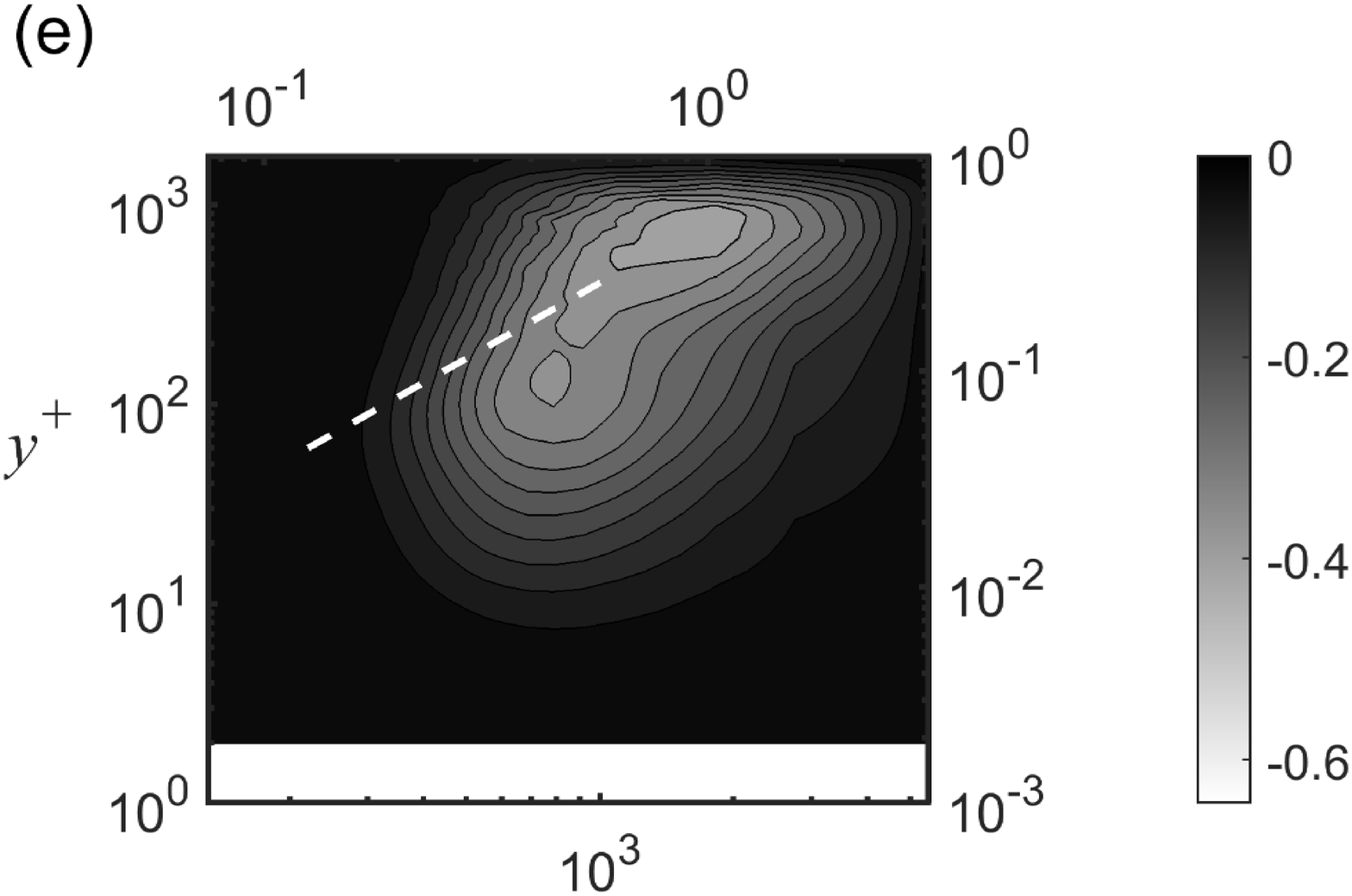}
  \label{5}
\end{subfigure}
\vspace{-0.7cm}
\begin{subfigure}[b]{0.42\textwidth}
  \includegraphics[width=\textwidth]{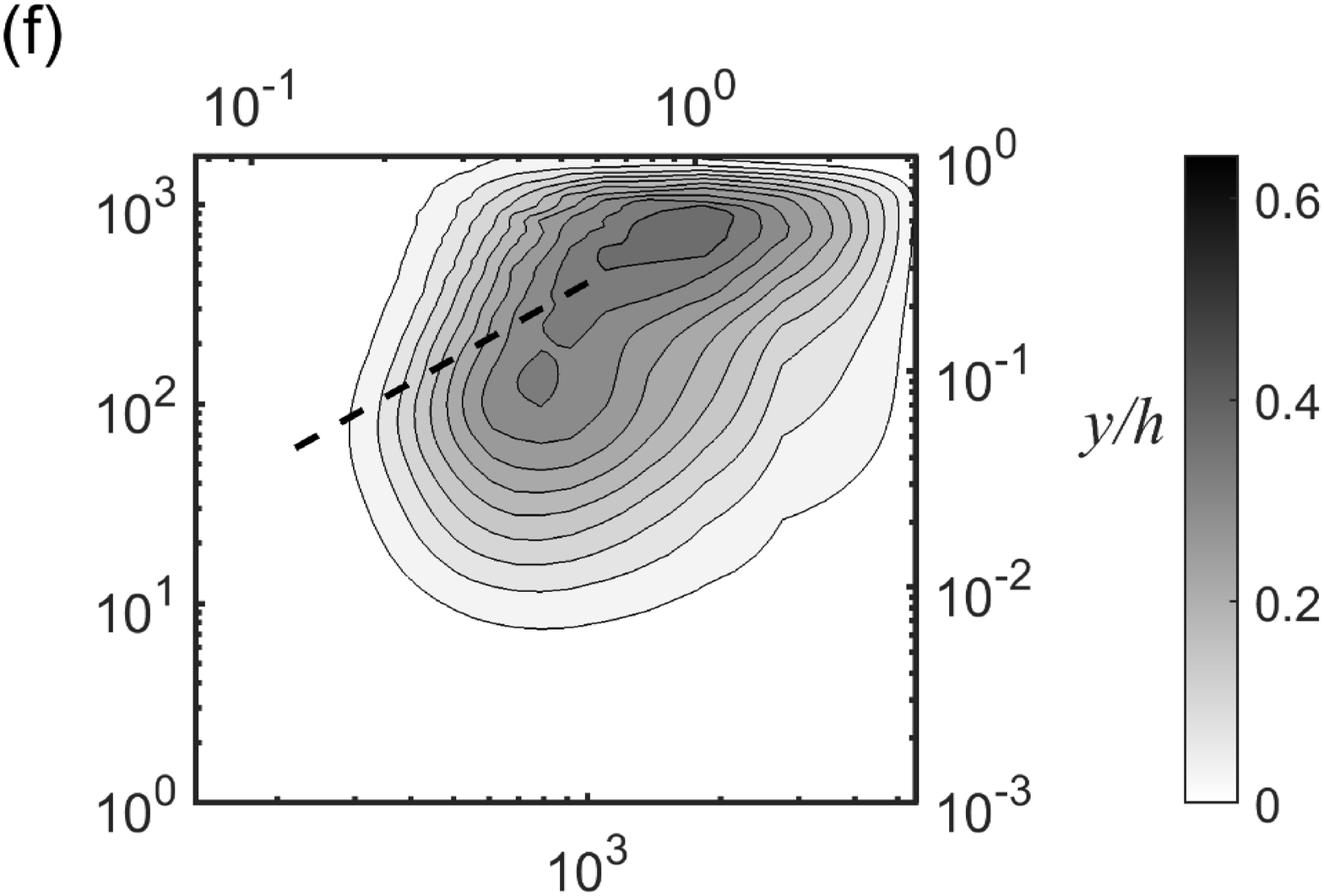}
  \label{6}
\end{subfigure}
\begin{subfigure}[b]{0.42\textwidth}
  \includegraphics[width=\textwidth]{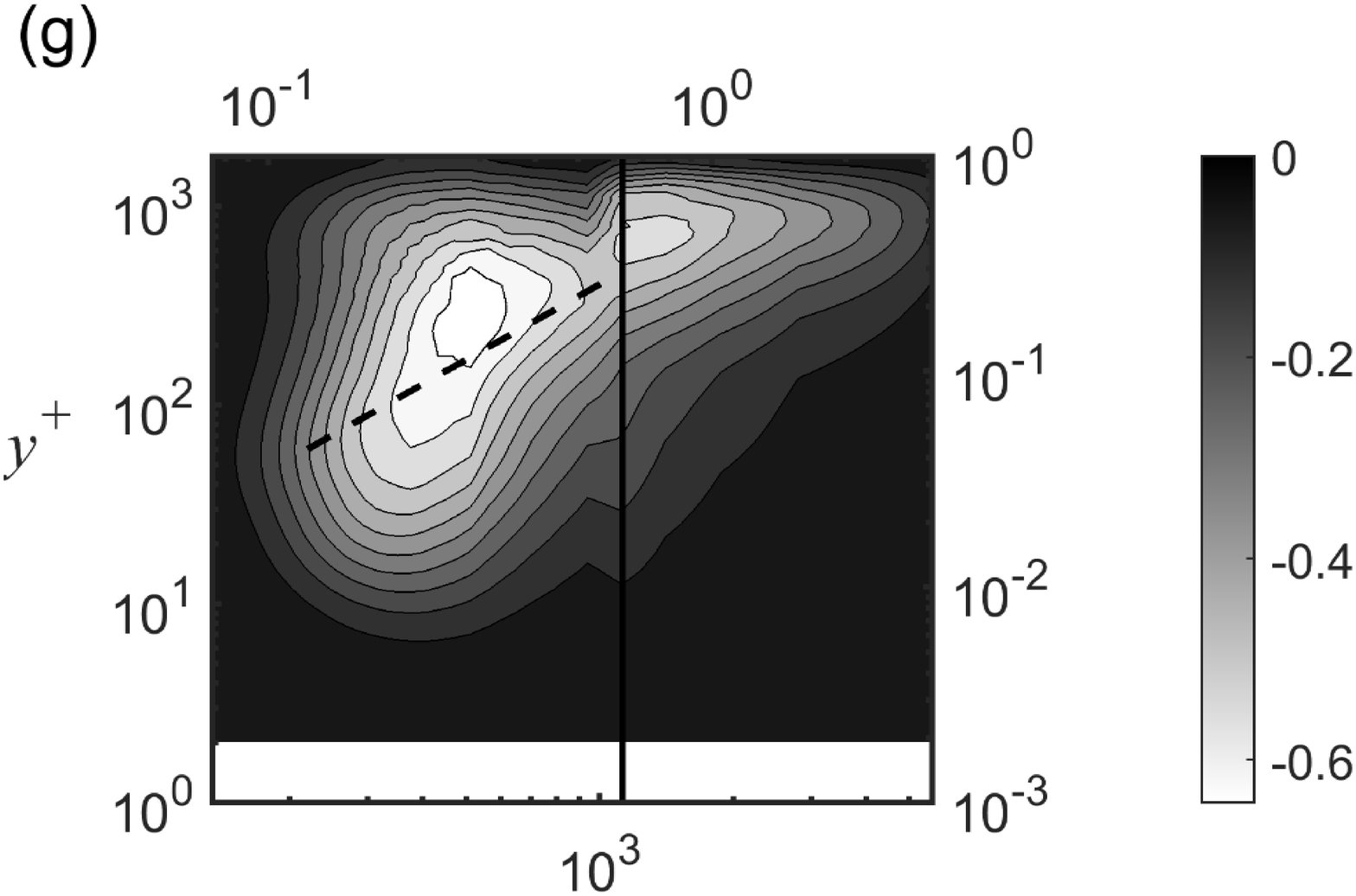}
  \label{5}
\end{subfigure}
\vspace{-0.7cm}
\begin{subfigure}[b]{0.42\textwidth}
  \includegraphics[width=\textwidth]{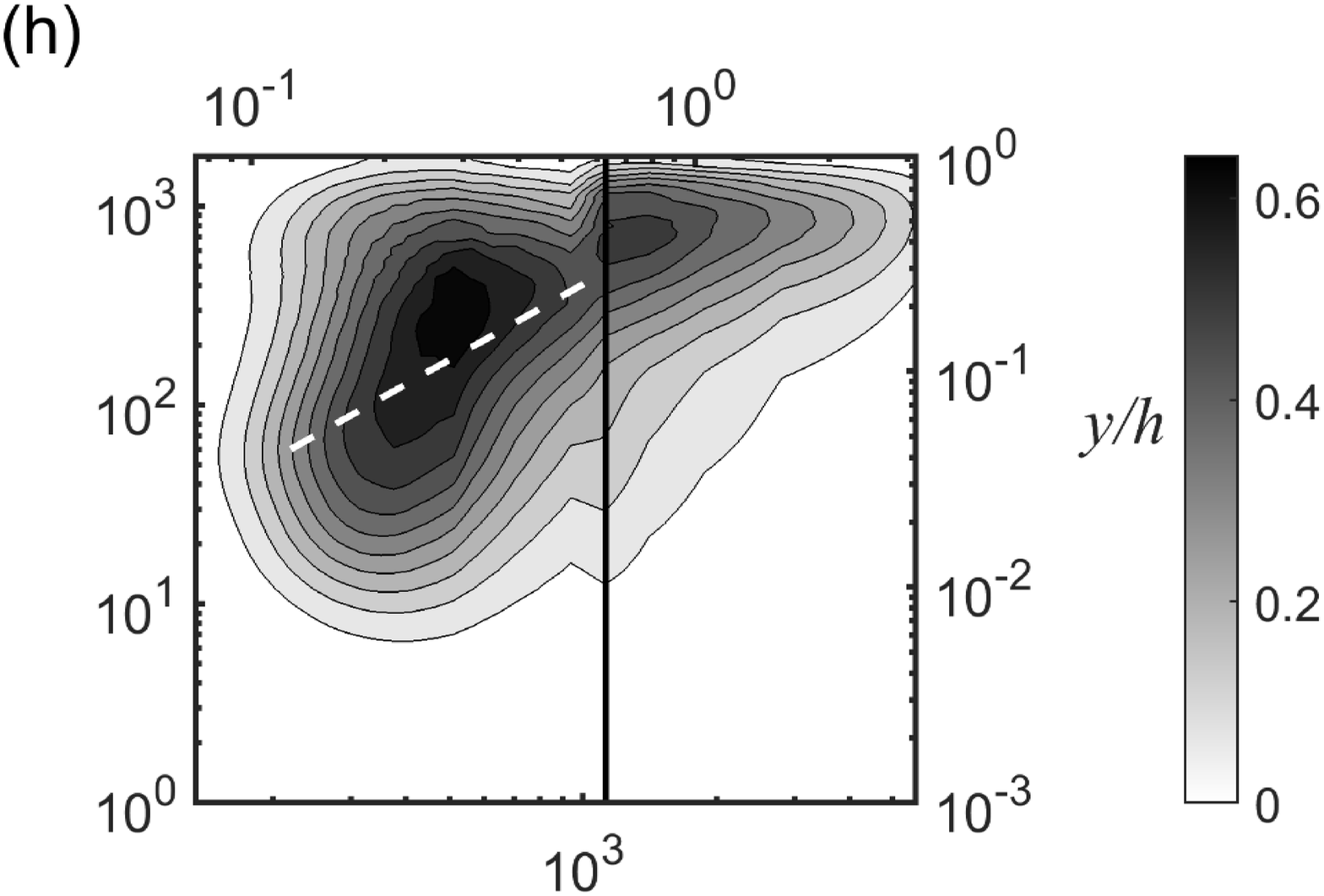}
  \label{6}
\end{subfigure}
\begin{subfigure}[b]{0.42\textwidth}
  \includegraphics[width=\textwidth]{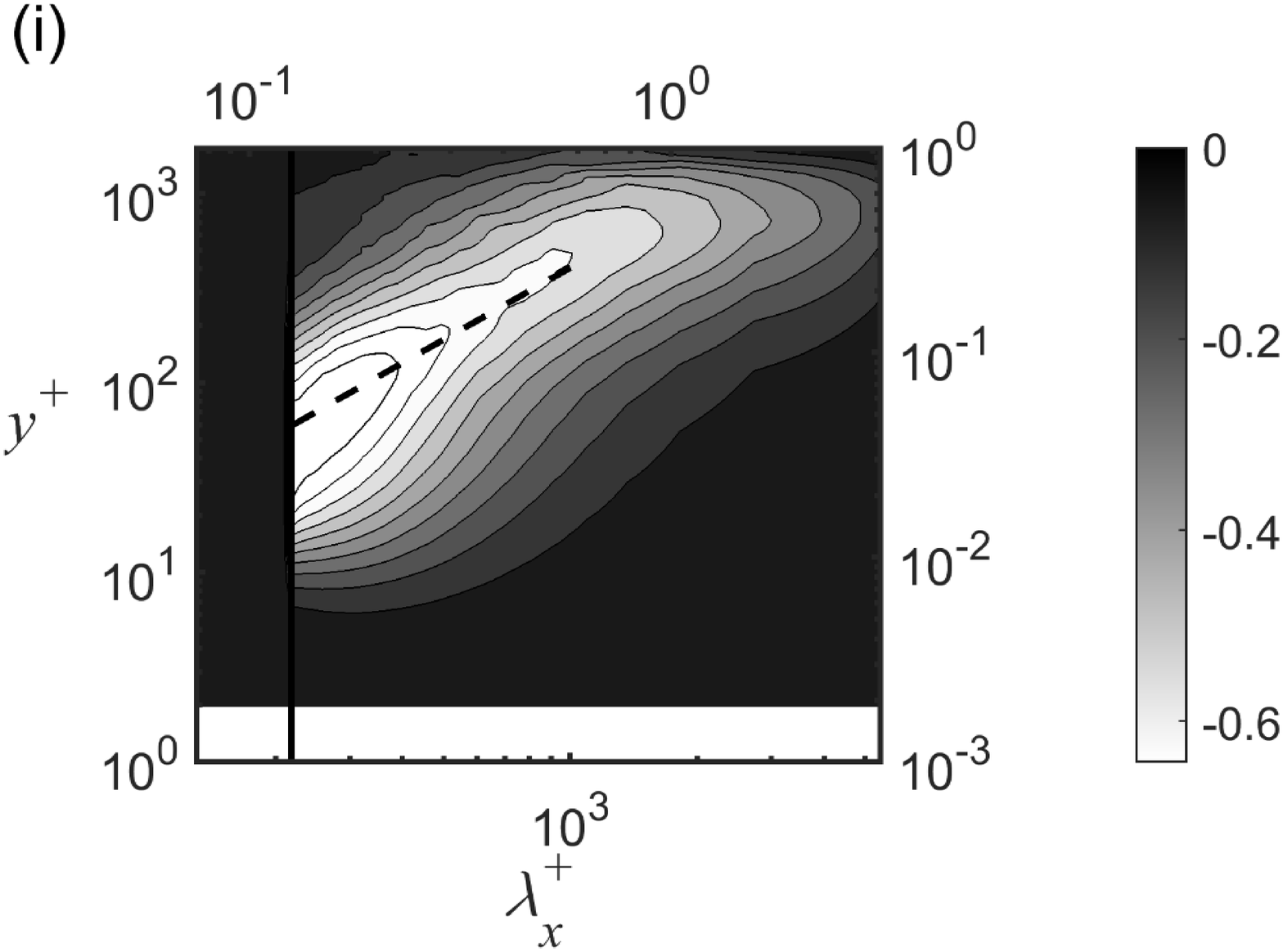}
\end{subfigure}
\begin{subfigure}[b]{0.42\textwidth}
  \includegraphics[width=\textwidth]{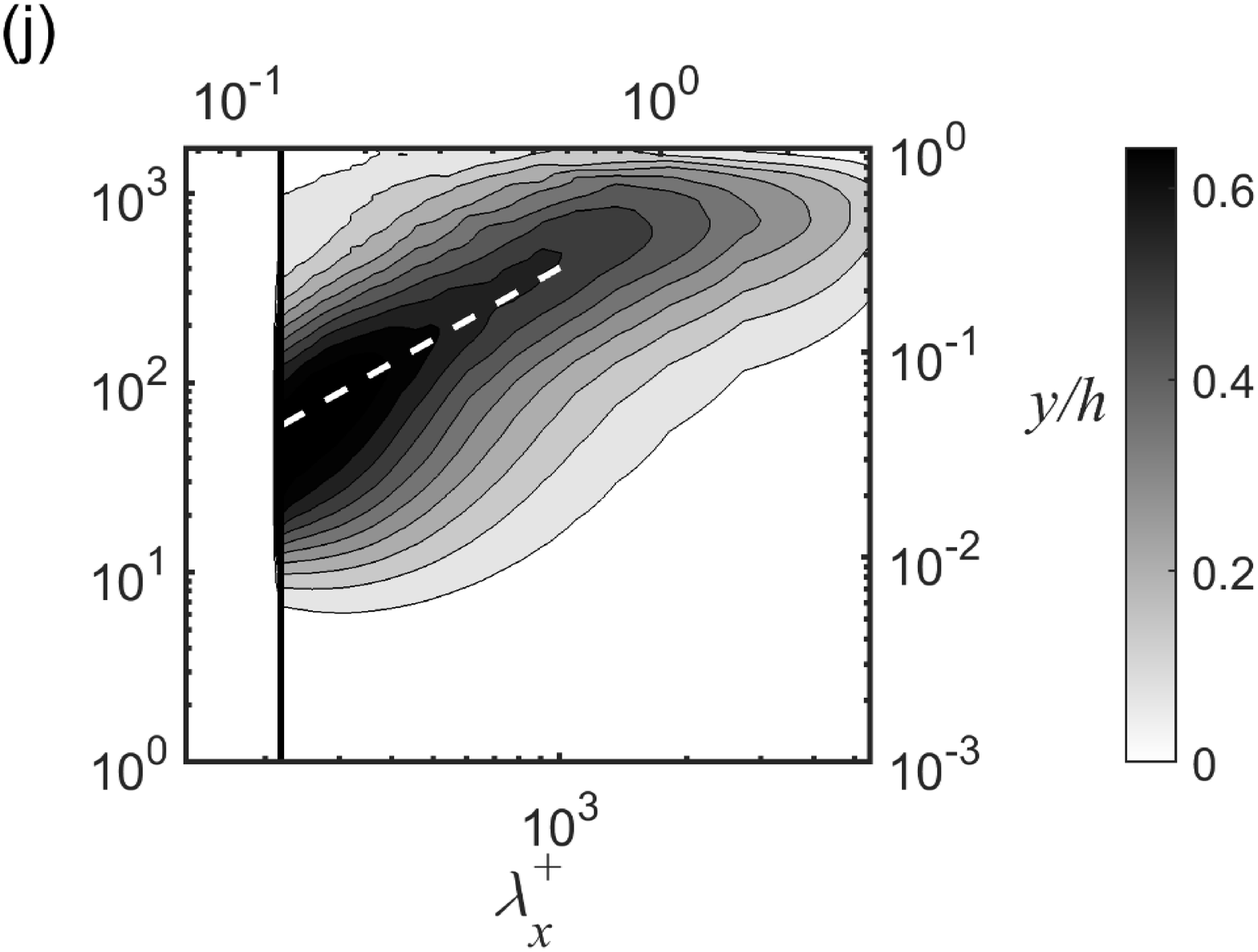}
\end{subfigure}
\end{minipage}
\caption{Premultiplied streamwise wavenumber spectra of $k_x^+ y^+ \widehat{\Pi}_x^+(y^+,\lambda_x^+)$ (left column) and $k_x^+ y^+ \widehat{\Pi}_{yz}^+(y^+,\lambda_x^+)$ (right column) for (a,b) LES, (c,d) QL, (e,f) GQL1,  (g,h) GQL5 and (i,j) GQL25 cases. Here, the vertical line represents the streamwise cut-off wavelength ($\lambda_{x,c}$) dividing the $\mathcal{P}_h$- (left) and $\mathcal{P}_l$-subspace  (right) regions.}
\label{fig:pix}
\end{figure}

The streamwise wavenumber spectra of pressure strain are shown in figure \ref{fig:pix}. Similarly to the velocity spectra, the pressure strain intensity appears to be concentrated for $\lambda_x^+ \gtrsim 700$ in the QL model. The spectra are extended to smaller $\lambda_x$, as $\lambda_{x,c}$ is decreased by the GQL approximations.
A significant improvement is seen in GQL5, whose spectra of the $\mathcal{P}_h$-subspace group reproduces a good portion of the spectra of LES. Once again, the spectra of GQL25 show excellent agreement with those of the LES, except for the phenomenon already observed in figures \ref{fig:xspectra} and \ref{fig:xenergy}: i.e. the complete depletion of energy for the streamwise wavelengths in the $\mathcal{P}_h$-subspace group.

Figure \ref{fig:pix} shows the recovery of the pressure strain transport along the streamwise direction with the GQL approximations. To understand the difference between the QL and the GQL model, we introduce the following equations for pressure fluctuation \cite[]{townsend76,kim89}:
\begin{equation}\label{eq:pres}
    \frac{1}{\rho}\nabla^2{p}^{R}=-2\frac{dU}{dy}\frac{\partial v^\prime}{\partial x} \quad \textrm{and} \quad \frac{1}{\rho}\nabla^2{p}^{S}=-\frac{\partial u_{j}^{\prime}}{\partial x_i}\frac{\partial u_{i}^{\prime}}{\partial x_j},
\end{equation}
where ${p}^{\prime}={p}^{R}+{p}^{S}$, and ${p}^R$ and ${p}^S$ are rapid and slow pressures, respectively. The terms `rapid' and `slow' are derived from the fact that only the rapid part responds immediately to a change imposed on the mean, and the slow part feels the change through nonlinear interactions \cite[]{kim89}. Using the flow decomposition in (\ref{eq:2.3z}) and the projections defined in (\ref{eq:2.4}), (\ref{eq:pres}) can be written as
\begin{subequations}\label{eq:3.6}
\begin{equation}\label{eq:3.6a}
     \frac{1}{\rho}\nabla^2{p_{l}^R}=-2\frac{dU}{dy}\frac{\partial v_l}{\partial x},
\end{equation}
\begin{equation}\label{eq:thel}
     \frac{1}{\rho}\nabla^2{p_l^S}=\mathcal{P}_l\Big[-\frac{\partial u_{l,j}}{\partial x_i}\frac{\partial u_{l,i}}{\partial x_j}\Big]+\mathcal{P}_l\Big[-2\frac{\partial u_{l,j}}{\partial x_i}\frac{\partial u_{h,i}}{\partial x_j}\Big]+\mathcal{P}_l\Big[-\frac{\partial u_{h,j}}{\partial x_i}\frac{\partial u_{h,i}}{\partial x_j}\Big],
\end{equation}
\end{subequations}
in the $\mathcal{P}_l$ subspace and 
\begin{subequations}
\begin{equation}\label{eq:3.7a}
    \frac{1}{\rho}\nabla^2p_h^R=-2\frac{dU}{dy}\frac{\partial v_h}{\partial x},
\end{equation}
\begin{equation}\label{eq:theh}
    \frac{1}{\rho}\nabla^2p_h^S=\mathcal{P}_h\Big[-\frac{\partial u_{l,j}}{\partial x_i}\frac{\partial u_{l,i}}{\partial x_j}\Big]+\mathcal{P}_h\Big[-2\frac{\partial u_{l,j}}{\partial x_i}\frac{\partial u_{h,i}}{\partial x_j}\Big]+\mathcal{P}_h\Big[-\frac{\partial u_{h,j}}{\partial x_i}\frac{\partial u_{h,i}}{\partial x_j}\Big],
\end{equation}
\end{subequations}
in the $\mathcal{P}_h$ subspace. In the QL and GQL models, the first and the last terms in the right-hand side of (\ref{eq:theh}) are absent. In the case of the QL model, this feature and the decomposition of velocity fluctuations (\ref{eq:2.3z}) into a streamwise mean and the remaining fluctuation make $u_{l,j}$ in (\ref{eq:theh}) not vary in the streamwise direction, thus each streamwise Fourier mode of $p_h^S$ is coupled only with that of $u_{h,j}$ at the same wavenumber. Therefore, $p_h^S$ does not play any role in the energy transport between the streamiwse Fourier modes (\citealp{hernandez}). In the GQL model, $u_{l,j}$ is instead allowed to vary in the streamwise direction, which evidently enhances the streamwise-dependent slow pressure generation in the $\mathcal{P}_l$ subspace through (\ref{eq:thel}). Furthermore, in this case, $u_{h,j}$ can now interact with $u_{l,j}$ in a `convolutive' manner in the streamwise wavenumber space for the generation of $p_h^S$, as is indicated by (\ref{eq:theh}). This feature in the GQL model would allow for a much more vigorous pressure strain transport than the QL model. This may explain the substantial improvement of the pressure strain transport only with a small increase in $M_{x,F}$ (e.g. GQL 1 and GQL 5 in figures \ref{fig:piz} and \ref{fig:pix}), rendering the slow pressure in $\mathcal{P}_h$ subspace more important for the streamwise energy transport and energy cascade.

\section{Discussion} \label{sec:sec4}

In this work, the generalized streamwise quasilinear (GQL) approximation in the streamwise direction has been applied to turbulent channel flow at $Re_\tau \approx 1700$. In particular, this study aimed to examine the nonlinear interactions between the energy-containing streamwise waves, which have been understood to originate from the streak instability and/or transient growth mechanisms in the self-sustaining processes at different length scales \citep{park11,alizard15,cassinelli17,degiovanetti17,lozano2021}. This is a direct extension of the previous studies on the QL model \citep{thomas14,thomas15,farrell16,hernandez} using the GQL approximation, as the QL model captures the dynamics of such energy-containing streamwise waves in a minimal manner with the linearised Navier-Stokes equations. The spectral energetics of the QL and GQL models have been studied and compared to those of the LES, with focus on the streamwise nonlinear energy transport to address the efficacy of the models to generate a turbulent state. It has also been found that as the number of streamwise Fourier modes allowed to interact nonlinearly is increased (i.e. GQL1, GQL5 and GQL25 cases), the linear scaling of the spectra with the distance from the wall, which was absent especially in the streamwise spectra of the QL model, is rapidly recovered. The implementation of the GQL approximation, however, has revealed a few points which deserve further discussions: (i) multi-scale behaviour of the QL and GQL models; (ii) dependence of energy transfer to the $\mathcal{P}_h$ subspace on the cut-off wavelength $\lambda_{x,c}$. We will address these points in this section. 

\subsection{Multi-scale dynamics in the QL and GQL models} \label{sec:sec41}




As pointed out in \cref{sec:sec1}, in the QL model, the full nonlinear evolution of streaks generated by a linear mechanism (i.e. the lift-up effect) is captured, but the subsequent streak instability and breakdown processes are approximated by the linearised equations about the streamwise-averaged velocity field. Despite being able of suitably resolving and/or modelling the key structural elements in the self-sustaining process, the size of which varies from the inner to the outer length scales, the following features indicate that the performance of the QL model may not be fully satisfactory (e.g. figures \ref{fig:zspectra}c,d): 1) an excessive energy intensity around particular spanwise wavelengths ($\lambda_z^+ \approx 300-700$); (2) the reduced spectral intensity of the velocity and production at the length scale smaller and larger than these wavelengths. 

\begin{figure*}
\centering
\begin{subfigure}[b]{0.45\textwidth}
\includegraphics[width=\textwidth]{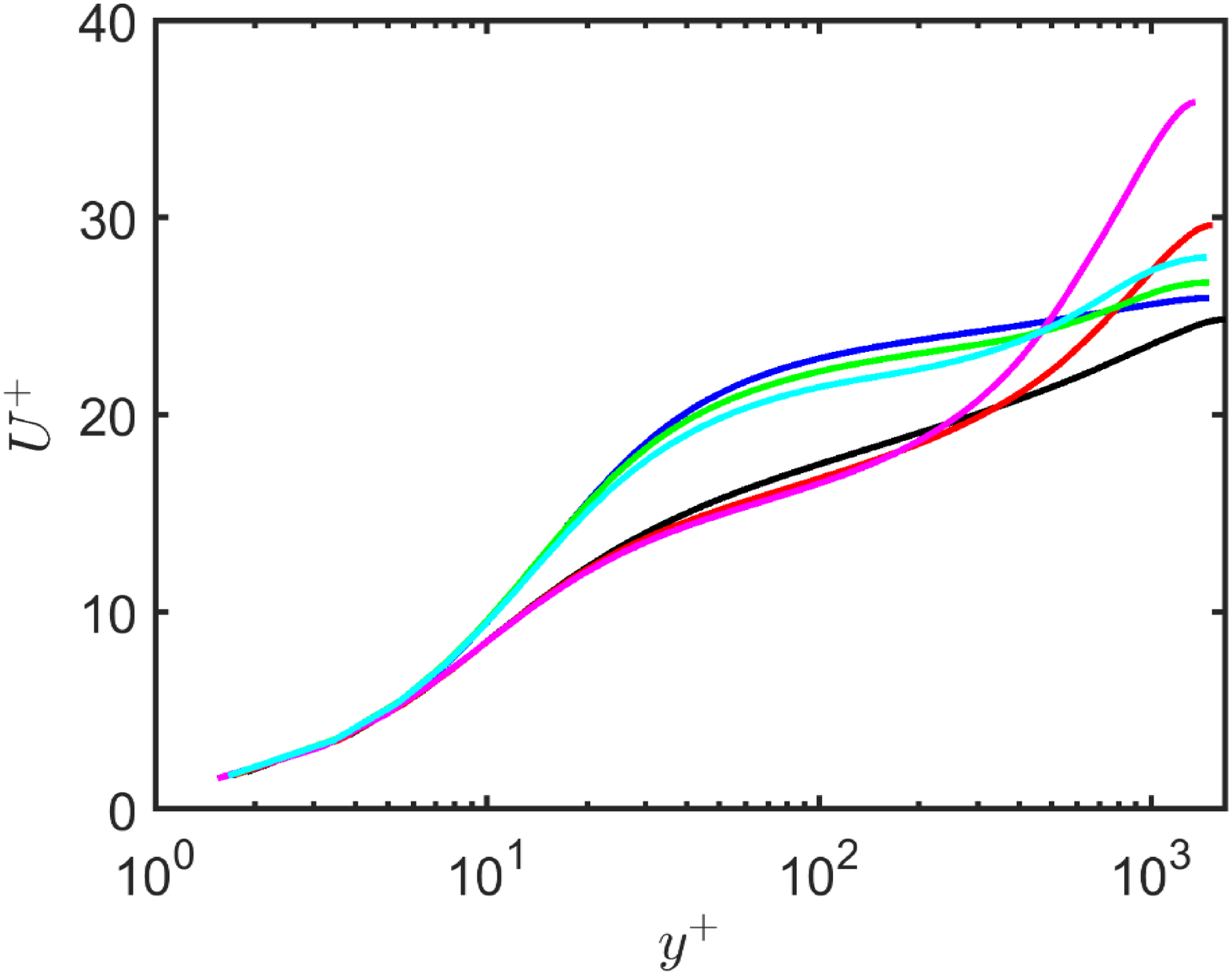}
\caption{$U^+(y^+)$}
\label{fig:ww}
\end{subfigure}
\begin{subfigure}[b]{0.45\textwidth}
\includegraphics[width=\textwidth]{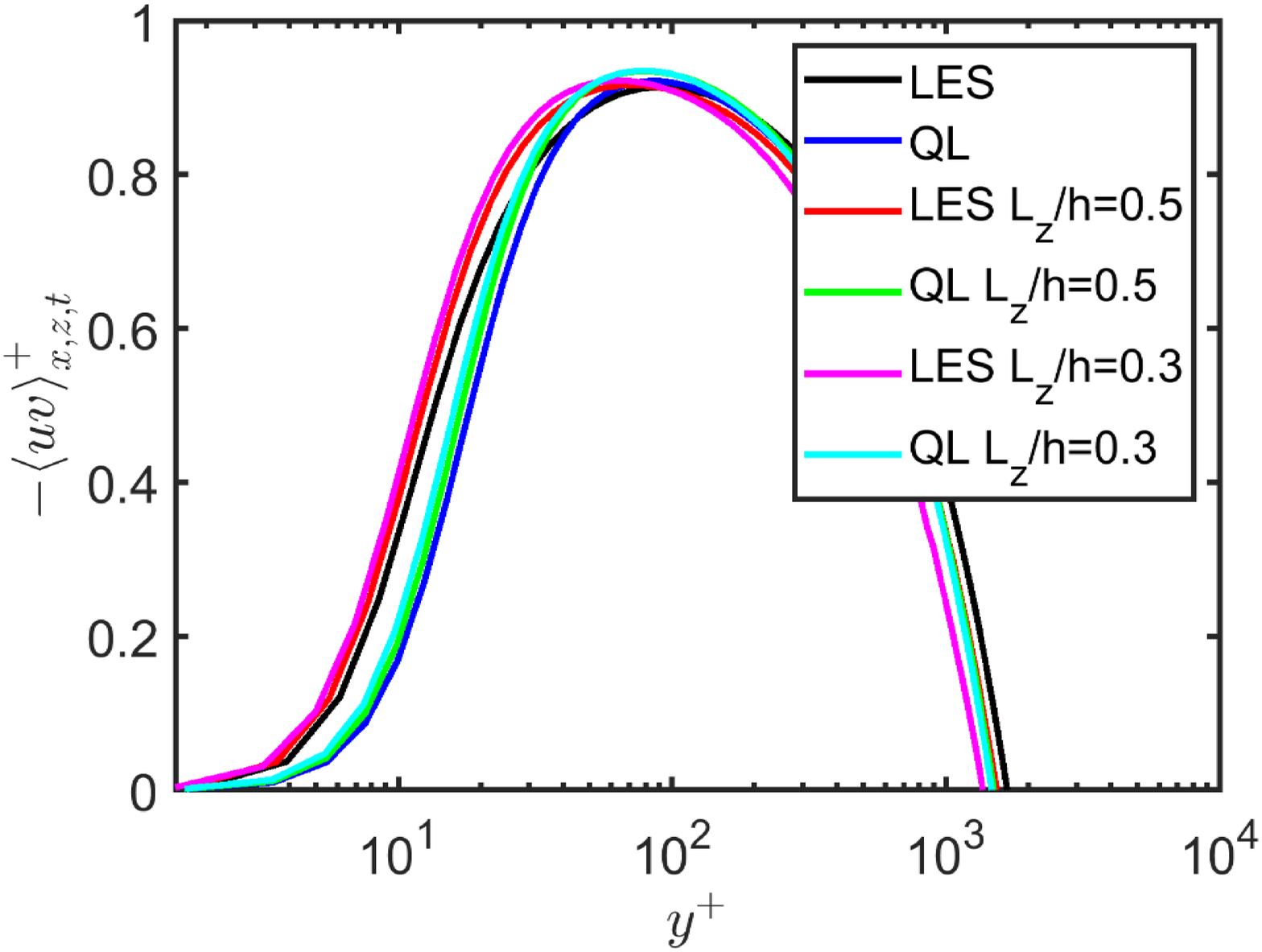}
\caption{$ - \langle u^{\prime} v^{\prime}\rangle_{x,z,t} ^+(y^+)$}
\label{fig:uv}
\end{subfigure}
\caption{First- and second-order turbulence statistics for the LES and the QL model: (a) $U^+(y^+)$; (b) $ - \langle u^{\prime} v^{\prime}\rangle_{x,z,t} ^+(y^+)$. Here, $L_z/h=0.3,0.5, \pi/2$.}
\label{fig:statql}
\end{figure*}

\begin{figure}
\begin{minipage}{\textwidth}
\centering
\begin{subfigure}[b]{0.45\textwidth}
  \includegraphics[width=\textwidth]{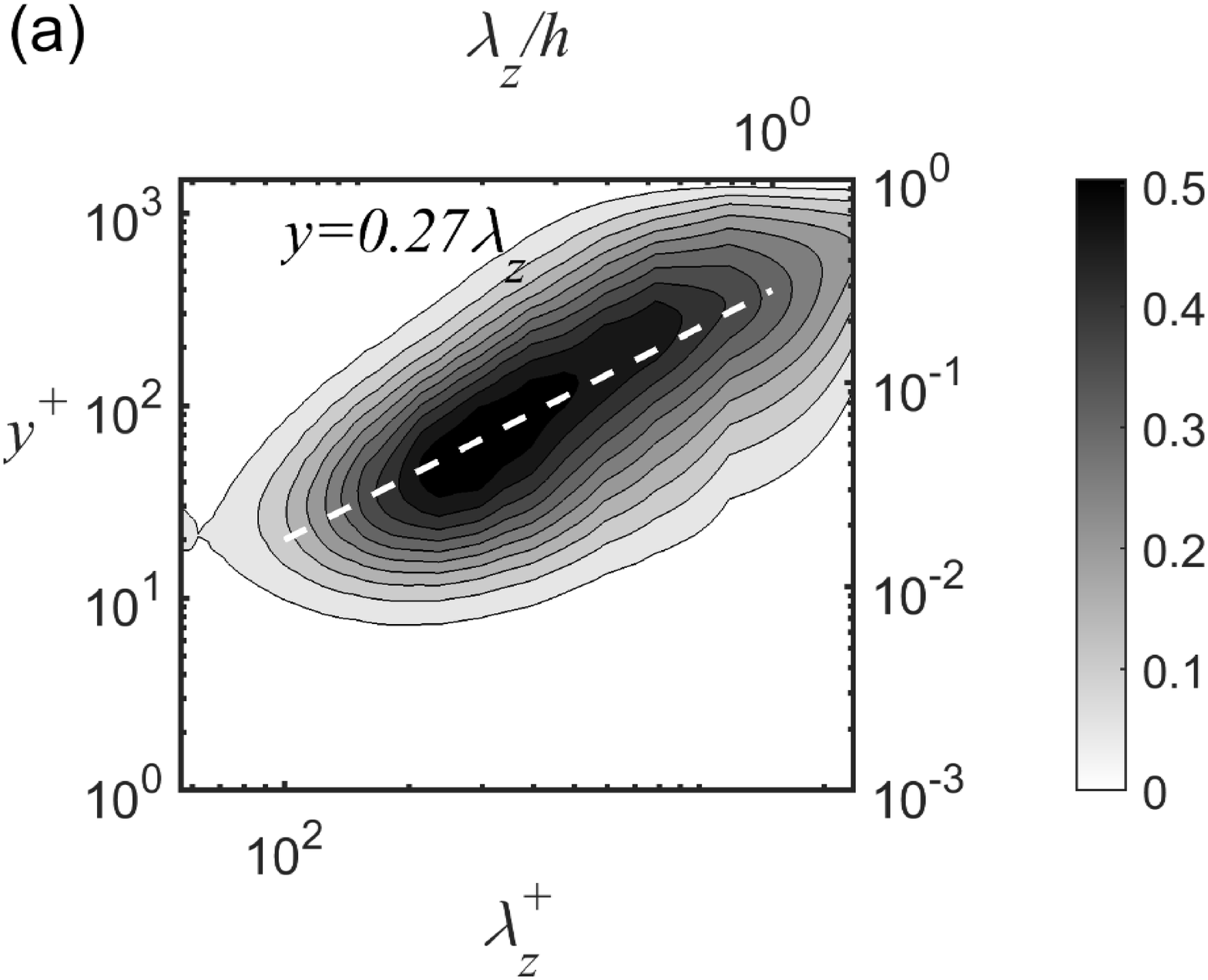}
\label{1}
\end{subfigure}
  \vspace{-0.25cm}
\begin{subfigure}[b]{0.45\textwidth}
  \includegraphics[width=\textwidth]{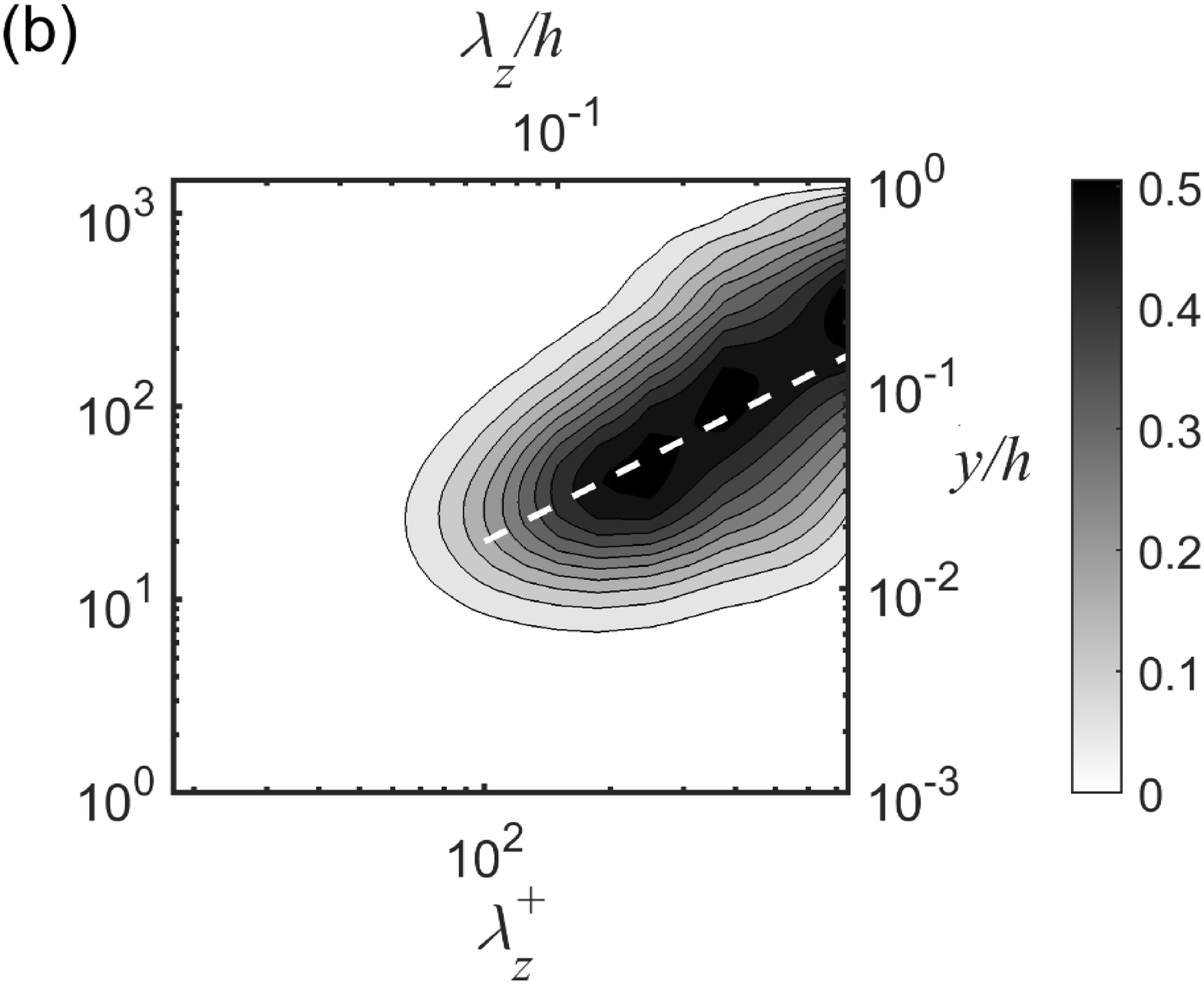}
\label{2}
\end{subfigure}
\begin{subfigure}[b]{0.45\textwidth}
  \includegraphics[width=\textwidth]{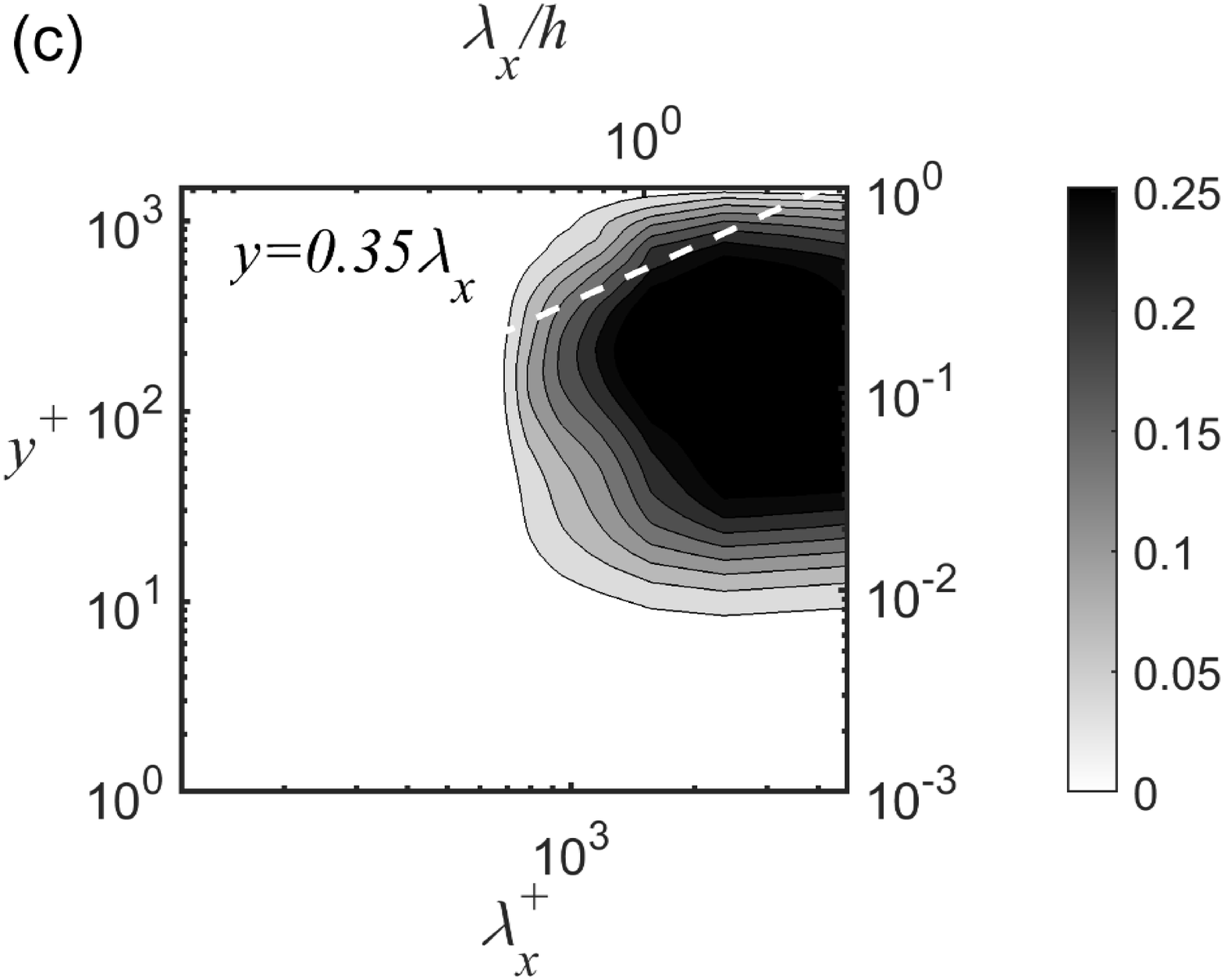}
\end{subfigure}
\begin{subfigure}[b]{0.45\textwidth}
  \includegraphics[width=\textwidth]{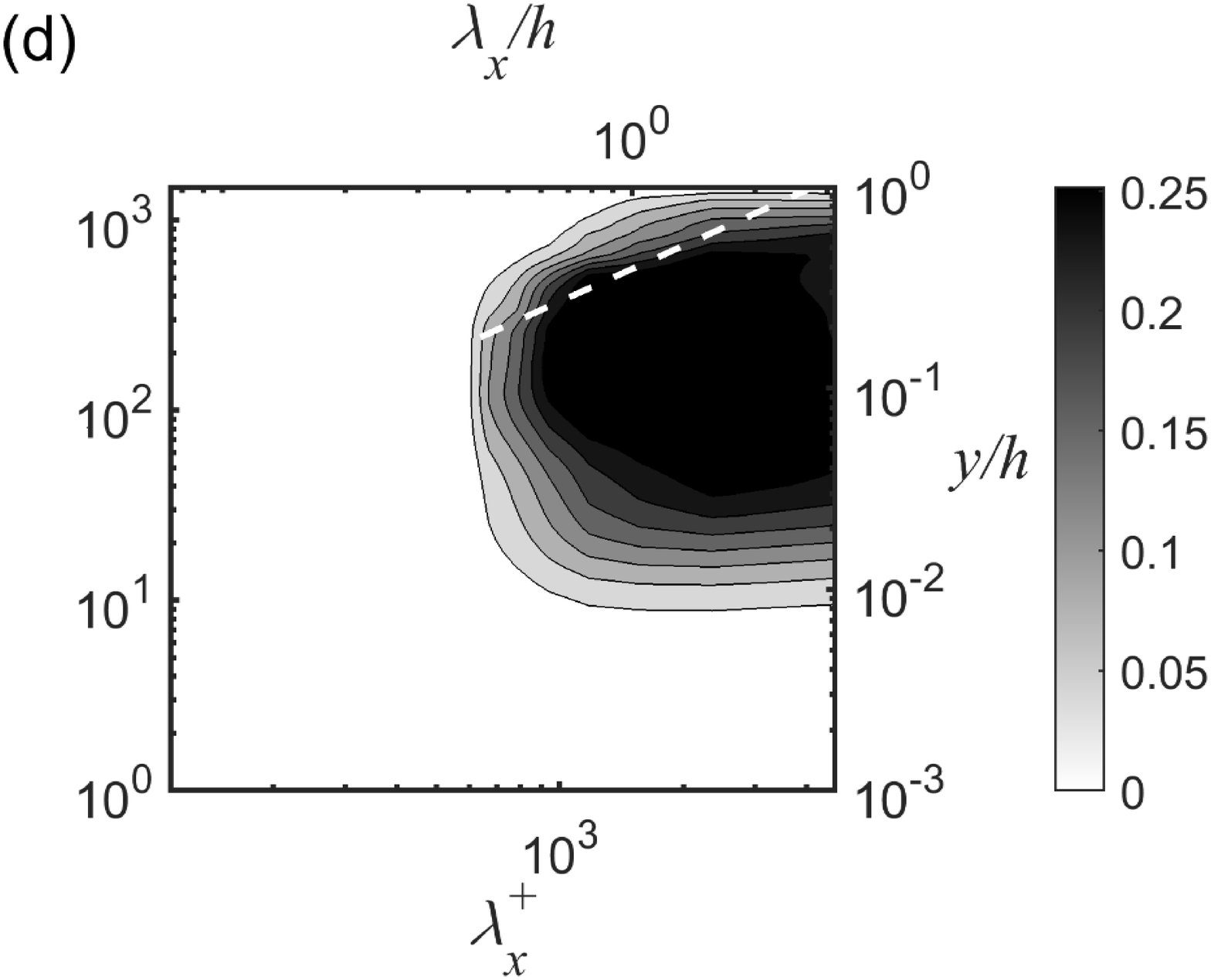}
\end{subfigure}
\end{minipage}
\caption{Premultiplied (a,b) spanwise and (c,d) streamwise spectra of Reynolds shear stress for the QL model: (a,c) $L_z/h=\pi/2$; (b,d) $L_z/h=0.5$.}
\label{fig:spectralz}
\end{figure}

In fact, these observations indicate that the QL model exhibits a considerably reduced multi-scale behaviour. This argument becomes clearer when a systematic reduction of the spanwise box size is carried out in order to examine the multi-scale behaviour of the QL model. Here, the LES and QL cases have been recomputed for different spanwise box size ($L_z/h=0.3,0.5,\pi/2$). Figure \ref{fig:statql} shows the mean velocity and Reynolds shear stress of the simulated cases. It can be observed that the three QL cases have very similar statistics. Although there are small deviations in the mean velocity near the centre of the channel, such deviations are much larger in the full LES cases. Figure \ref{fig:spectralz} shows the spanwise and streamwise wavenumber spectra of Reynolds shear stress of the original QL case ($L_z / h=\pi/2$) and the one recomputed with $L_z / h=0.5$. While the very little difference in the spanwise wavenumber spectra for $\lambda_z/h\leq0.5$ is expected given the existence of the self-sustaining processes at each spanwise length scale \cite[]{hwang15,hwangbengana16}, the very small difference in the streamwise wavenumber spectra confirms that the QL model exhibits a significantly reduced multi-scale behaviour. Here, we note that, owing to equation (\ref{eq:2.9}), there is no direct modification of the production term in the implementation of the QL model -- the QL approximation only changes the form of the nonlinear turbulent transport term. Therefore, this observation implies that the modified fluctuation interaction dynamics by the QL model has subsequently affected the mean-fluctuation interaction described by
\begin{equation}
    \frac{\partial \langle u \rangle_{x,z}}{\partial t}=-\frac{d P_0 }{dx}+\nu \frac{\partial^2 \langle u \rangle_{x,z}}{\partial y^2}-\frac{\partial }{\partial y}\left[\int_{-\infty}^{\infty} \langle{\widehat{u^{\prime}}}^* (k_z)\widehat{v^{\prime}}(k_z)\rangle_x~\mathrm{d}k_z \right],
\end{equation}
where $dP_0/dx$ is the applied mean pressure gradient (note that the eddy viscosity term for the LES is omitted here for convenience). The resulting mean velocity has subsequently affected the fluctuation dynamics, such that the QL model exhibits a reduced multi-scale behaviour across the different energy-containing integral scales. 

It has been shown that the QL approximation to a flow retaining the SSP at single length scale
(i.e. low-Reynolds-number case or uniform shear turbulence case) has a tendency to elevate the production as well as the related wall shear velocity compared to the full simulation (\citealp{hernandez}, and references therein). This is consistent with the elevated energy intensity and the local mean shear rate around a particular spanwise length scale and wall-normal location observed in this study for the QL model ($\lambda_z^+ \simeq 200-700$; figures \ref{fig:zspectra}c,d). However, unlike the low-Reynolds-number or single-scale cases, in the present study where the multi-scale behaviour is prominent due to the high Reynolds number, the elevated energy intensity and production around the particular spanwise length scales lead to a significant disruption in the mean-fluctuation dynamics at other length scales. Indeed, to obtain a logarithmic mean velocity, it is necessary to have approximately self-similar production and transport spectra scaling in $y$ for $\lambda_z \in [200 \delta_\nu,1h]$ (\citealp{hwang_lee_2020}), i.e.
\begin{equation}
    k_z y\hat{P}(k_z,y)\approx f(k_z y).
\end{equation}
However, in the QL model, the elevated production around a particular length scale and wall-normal location ($\lambda_z^+ \approx 300$ and $y^+ \approx 20-30$; figure \ref{fig:zenergy}c) leads to an increase of the mean shear rate at the related wall-normal location ($y^+ \approx 20-30$ in figure \ref{fig:stat}c). Given that the mass flow rate across the channel is constant in the present simulations, this must result in a reduced mean shear rate at some other wall-normal locations. The reduced mean shear rate at those locations would then generate the significantly reduced fluctuations at both small and large scales, as the production there is expected to be reduced. We note that in the case of high-Reynolds-number wall turbulence, all integral scales varying from the viscous inner to the outer one generate turbulent skin friction almost equally (\citealp{degiovanetti16}). This explains why the reduction in skin-friction drag (or mean shear velocity) of the QL model is observed in this case and in \cite{farrell16} unlike the other QL studies at low Reynolds number or uniform shear turbulence.

This behaviour observed in the QL model appears to be very effectively cured in the GQL model. In fact, only a small increase in $M_{x,F}$ (or a decrease in $\lambda_{x,c}$) shows a drastic improvement in the first-order statistics and Reynolds shear stress. As shown in figures \ref{fig:xspectra} and \ref{fig:xenergy}, this is presumably related to the improved energy cascade in the streamwise direction due to the enlarged $\mathcal{P}_l$-subspace group, which more effectively removes the excess of energy at a particular scale (or wavelength) seen in the QL model. This subsequently leads to much more balanced mean-fluctuation interactions at all integral scales like in full LES. This scenario is supported by the wall-normal velocity (figure \ref{fig:zspectra}) and pressure strain (figure \ref{fig:pix}) spectra of the GQL1 and GQL5 cases. In particular, these two spectra of the GQL5 case appear to be well aligned with their linear-scaling ridge with the distance from the wall across the entire range of scales, like those of the LES case, despite the fact that a significant part of the flow (i.e. the $\mathcal{P}_h$-subspace group) is solely obtained by solving the linearised equations. This observation suggests that promoting a balanced mean-fluctuation dynamics across the entire range of scales admitting self-similar production may be the key to a successful prediction at least for the low-order turbulence statistics. In this respect, it is finally worth mentioning the previous work by \cite{bretheim15} where a judicious choice of higher streamwise wavenumber(s) for the $\mathcal{P}_h$-subspace group was shown to improve the mean velocity profile significantly. In their case, such a choice would have led to a reduced fluctuation at the energetic scale (see also \citealp{hernandez} for this issue). Like in our study, it is presumbable that this would subsequently result into a more balanced mean-fluctuation interaction.

\subsection{Energy transfer to the high-wavenumber group} \label{sec:sec43}

In  figures \ref{fig:xspectra}, \ref{fig:xenergy} and \ref{fig:pix}, we have reported that the premultiplied streamwise wavenumber spectra of the GQL cases typically exhibit a large intensity in the $\mathcal{P}_l$ subspace, consistent with \cite{tobias17}. Indeed, as the $\mathcal{P}_l$-subspace group is enlarged by decreasing $\lambda_{x,c}$ (or increasing $M_{x,F}$), the streamwise wavenumber spectra tend to reach smaller wavelengths. It has been claimed that this is the key advantage of the GQL model over the QL model, as the motions in the $\mathcal{P}_h$ subspace can exchange energy through the convolutive interaction (equivalently the non-local interaction in the wavenumber space) with those in the $\mathcal{P}_l$ subspace, and this has been referred to as a `scattering' mechanism \cite[e.g.][]{tobias17}. However, it appears that this scattering mechanism is completely absent when $\lambda_{x,c}$ is sufficiently low (i.e. GQL25 case). In fact, the spectral intensity for $\lambda_x < \lambda_{x,c}$ in this case is zero, which is difficult to understand this observation solely with the proposed `scattering' mechanism. 

To understand this behaviour better, let us consider the equations for the $\mathcal{P}_h$-subspace group. The velocity component $\textbf{u}_h$ is governed by the linear equation (\ref{eq:bb}) whose last two terms $\mathcal{P}_{h}\left[\left(\textbf{u}_{l}\cdot\nabla\right)\textbf{u}_{l}\right]$ and $\mathcal{P}_{h}\left[\left(\textbf{u}_{h}\cdot\nabla\right)\textbf{u}_{h}\right]$ are neglected by the GQL approximation. Since the equation (\ref{eq:bb}) is linear and it has no driving term, (\ref{eq:bb}) can be written for each $k_x$ component as follows:
\begin{equation}\label{lyapu}
\frac{\partial \bold{u}_{h}(k_x)}{\partial t}=\mathcal{L}(\textbf{U}_l, k_x) \bold{u}_{h}(k_x),
\end{equation}
where $\mathcal{L}$ is an autonomous linear operator and $k_x$ is the wavenumber of each streamwise Fourier mode of the $\mathcal{P}_h$-subspace group. For simplicity, let us first assume that $\mathbf{U}_l$ is steady (or time-periodic). In a well-posed QL/GQL model, the flow field should not diverge in time. Therefore, the resulting stable solution to the corresponding QL/GQL model should lead to $\mathbf{u}_h$ in the form of either a non-trivial neutrally-stable leading eigenmode (or Floquet mode) \cite[e.g.][]{malkus56,pausch18} or the trivial solution (i.e. zero). Similarly, when $\textbf{U}_l$ is chaotic, (\ref{lyapu}) becomes the tangent (or linearised) equations to the trajectory $\textbf{U}_l$ in the $\mathcal{P}_h$ subspace. Therefore, any non-trivial solution $\bold{u}_{h}(k_x)$ to (\ref{lyapu}) for each $k_x$ ultimately acquires the structure of the Lyapunov vector associated with the leading Lyapunov exponent of $\mathcal{L}(\textbf{U}_l, k_x)$, which can be obtained as follows:
\begin{equation}
\sigma_{k_x}=\underset{t \rightarrow \infty}{\lim \sup} \frac{\ln || \bold{u}_{h}(k_x) ||}{t},
\end{equation}
where $|| \cdot ||$ is any relevant norm of the velocity field. For the given QL/GQL models to be well-posed (or not blow up), the leading Lyapunov exponent of the linear equations $\mathcal{L}(\textbf{U}_l, k_x)$ should be either negative or zero, meaning that $\bold{u}_{h}(k_x)$ can only decay or be marginally stable, respectively. In the case of the turbulent state produced by the QL/GQL approximation, the leading Lyapunov exponent from (\ref{lyapu}) must be zero (\citealp{farrell12,farrell16}), if (\ref{lyapu}) admits a non-trivial solution. Otherwise, the only possible solution is the trivial solution. 

Given the discussion above, it is now evident that the scattering mechanism proposed by \cite{tobias17} must depend on the nature of the Lyapunov spectrum of (\ref{lyapu}). It is worth mentioning that the leading Lyapunov exponent has been speculated to be proportional to the inverse of the fastest time scale of the system (i.e. the inverse of the Kolmogorov time; see \citealp{ruelle79} and \citealp{crisanti93}). However, more recent evidence suggests that the instability related to the leading Lyapunov exponent grows even faster than the Kolmogorov time scale \citep{mohan17}. It is yet to be clear how the leading Lyapunov exponent would precisely scale, but it appears to be reasonable to assume that it is at least associated with the fastest time scale to some extent.

Now, let us return to the linearised equations (\ref{lyapu}) in the QL/GQL models. As $\lambda_{x,c}$ is decreased (or $M_{x,F}$ is increased), the smallest time scale of the system is expected to be reduced. This is observed in the turbulent transport spectra, where more energy is transferred towards smaller streamwise and spanwise length scales. In other words, the decrease in $\lambda_{x,c}$ could lead the linearised equations for a given streamwise wavenumber in the $\mathcal{P}_h$ subspace to become more unstable before they reach the statistically stationary state. This is consistently seen for the QL, GQL1 and GQL5 cases where the spectra of the $\mathcal{P}_h$-subspace group is extended to smaller streamwise and spanwise wavelengths on decreasing $\lambda_{x,c}$ (figures \ref{fig:zspectra} and \ref{fig:xspectra}). However, if $\lambda_{x,c}$ is too small, (\ref{lyapu}) for the $\mathcal{P}_h$-subspace group is expected to be dominated by viscous dissipation. Therefore, (\ref{lyapu}) may only admit the trivial solution in the $\mathcal{P}_h$ subspace. This explains why the streamwise wavenumber spectra of the GQL25 case (figure \ref{fig:xspectra}) exhibit the trivial solution for $\lambda_x<\lambda_{x,c}$. It is interesting to note that such a streamwise wavelength is $\lambda_{x,c}^+ \approx 200$: i.e. the (smallest) streamwise length scale of the streak instability in the near-wall region \cite[e.g.][]{schoppa02,cassinelli17}. Finally, this suggests that if $\lambda_{x,c}$ is sufficiently small such that the $\mathcal{P}_h$-subspace group only gives the trivial solution, the application of the GQL approximation becomes equivalent to that of a spectral low-pass cut-off filter with the threshold streamwise wavelength $\lambda_{x,c}$. 

\section{Concluding remarks} \label{sec:sec5}

In the present study, we have investigated the spectral energetics of a generalized quasilinear approximation applied to turbulent channel flow at $Re_\tau\simeq 1700$. The focus of the present study is given to its application in the streamwise direction to explore the nonlinear interactions between energy-containing streamwise waves, which have been understood to originate from the streak instability and/or transient growth mechanism in the self-sustaining processes at different length scales. For the GQL approximation, the velocity is decomposed into low and high wavenumber modes, the former of which are solved by considering the full nonlinear equations ($\mathcal{P}_l$-subspace group) whereas the latter are obtained from the linearised equations around the former ($\mathcal{P}_h$-subspace group). The QL case has been found to exhibit the most anisotropic second-order turbulence statistics throughout the entire wavenumber space of the spectra, in agreement with the previous studies \citep{thomas14,farrell16}. Only a small increase in the number of streamwise modes allowed to interact nonlinearly (or the enrichment of the $\mathcal{P}_l$-subspace group; i.e. GQL1 and GQL5 cases) has resulted in a rapid recovery of the scaling of the streamwise wavelengths with the distance from the wall $y$, which was absent in the streamwise spectra of the QL case. These cases also exhibited spectra extending over a wider range and reaching out to smaller scales, when compared to the QL case whose spatial spectra were highly localised in the wavenumber space.

The energetics of the QL and GQL models have been studied using the streamwise and spanwise wavenumber spectral turbulent kinetic energy budget equation. The production spectra of the QL model have been found to be highly localized and the turbulent transport is inhibited in the streamwise direction. Simulations with different spanwise boxes have confirmed that this is related to a considerably reduced multi-scale behaviour of the QL model which originates from both mean-fluctuation and fluctuation-fluctuation dynamics. Finally, it has been proposed that the `scattering mechanism' \cite[]{tobias17}, by which the motions in the $\mathcal{P}_h$ subspace exchange energy through interaction with those in the $\mathcal{P}_l$ subspace, depends on the Lyapunov spectrum of the linearised equations projected to the $\mathcal{P}_h$-subspace group. This has explained the gradual extension of the spectra of the GQL model over a wider range, when the cut-off streamwise wavelenth $\lambda_{x,c}$ for the velocity decomposition in the GQL model is sufficiently large. This is also consistent with the emergence of the trivial solution in the $\mathcal{P}_h$-subspace group, when the cut-off wavelength $\lambda_{x,c}$ is sufficiently small. 

\section*{Acknowledgements}

C. G. H. and Y. H. acknowledge the support of the Leverhulme trust (RPG-123-2019). Q. Y. acknowledges the Natural Science Foundation of China (11802322). Y. H. is also supported by the Engineering and Physical Sciences Research Council (EPSRC; EP/T009365/1) in the UK. 

\section*{Declaration of interest}

The authors report no conflict of interest.

\bibliography{biblio}
\bibliographystyle{jfm}

\end{document}